\documentclass[preprint,12pt]{elsarticle}

\usepackage{amssymb}
\usepackage{amsmath}
\usepackage{graphicx}%
\usepackage{multirow}%
\usepackage[title]{appendix}%
\usepackage{xcolor}%
\usepackage{textcomp}%
\usepackage{manyfoot}%
\usepackage{booktabs}%
\usepackage{algorithm}%
\usepackage{algorithmicx}%
\usepackage{algpseudocode}%
\usepackage{listings}%
\usepackage{caption}
\usepackage{subcaption}
\usepackage[bookmarksnumbered=true]{hyperref}
\usepackage{longtable} % Add this in the preamble
\usepackage{tikz}
\usetikzlibrary{positioning}
\usepackage{geometry}
\journal{arXiv}

\begin{document}

\begin{frontmatter}

\title{Adaptive Physics-Informed System Modeling with Control for Nonlinear Structural System Estimation}


\author[label1]{Biqi Chen} %% Author name
\author[label3]{Chenyu Zhang} %% Author name
\author[label1]{Jun Zhang} %% Author name
\affiliation[label1]{organization={School of Intelligent Civil and Marine Engineering,  Harbin Institute of Technology, Shenzhen},%Department and Organization
           addressline={Taoyuan Street}, 
          city={Shenzhen},
            postcode={518055}, 
            state={Guangdong},
            country={China}}

\affiliation[label3]{organization={Key Lab of Structures Dynamic Behavior and Control of the Ministry of Education,  Harbin Institute of Technology},%Department and Organization
          city={Harbin},
            postcode={150090}, 
            state={Heilongjiang},
            country={China}}
\author[label1,label2]{Ying Wang} %% Author name

\affiliation[label2]{organization={Guangdong Provincial Key Laboratory of Intelligent and Resilient Structures for Civil Engineering,  Harbin Institute of Technology, Shenzhen},%Department and Organization
            addressline={Taoyuan Street}, 
           city={Shenzhen},
            postcode={518055}, 
           state={Guangdong},
           country={China}}
\ead{yingwang@hit.edu.cn}

%% Abstract

\begin{abstract}
Accurately capturing the nonlinear dynamic behavior of structures remains a significant challenge in mechanics and engineering. Traditional physics-based models 
and data-driven approaches often struggle to simultaneously ensure model interpretability, noise robustness, and estimation optimality. To address this issue, this 
paper proposes an Adaptive Physics-Informed System Modeling with Control (APSMC) framework. By integrating Kalman filter-based state estimation with physics-constrained 
proximal gradient optimization, the framework adaptively updates time-varying state-space model parameters while processing real-time input–output data under white noise 
disturbances. Theoretically, this process is equivalent to real-time tracking of the Jacobian matrix of a nonlinear dynamical system.
Within this framework, we leverage the theoretical foundation of stochastic subspace identification to demonstrate that, as observational data accumulates, the 
APSMC algorithm yields state-space model estimates that converge to the theoretically optimal solution. The effectiveness of the proposed framework is validated 
through numerical simulations of a Duffing oscillator and the seismic response of a frame structure, as well as experimental tests on a scaled bridge model. 
Experimental results show that, under noisy conditions, APSMC successfully predicts 19 consecutive 10-second time series using only a single initial 10-second 
segment for model updating, achieving a minimum normalized mean square error (NMSE) of 0.398\%. These findings demonstrate that the APSMC framework not only 
offers superior online identification and denoising performance but also provides a reliable foundation for downstream applications such as structural health 
monitoring, real-time control, adaptive filtering, and system identification.
    \end{abstract}

\begin{graphicalabstract}
We propose an Adaptive Physics-Informed System Modeling with Control (APSMC) framework that integrates stochastic subspace theory, optimal estimation, 
and proximal gradient methods. Physical constraints are embedded into an online optimization scheme, allowing nonlinear dynamics to be reformulated as 
time-varying state-space models. This enables real-time updates of system matrices from sparse and noisy measurements, facilitating the integration of physics-based 
modeling and data-driven learning for digital twin applications in structural systems.

\vspace{1em}

\includegraphics[width=0.95\linewidth]{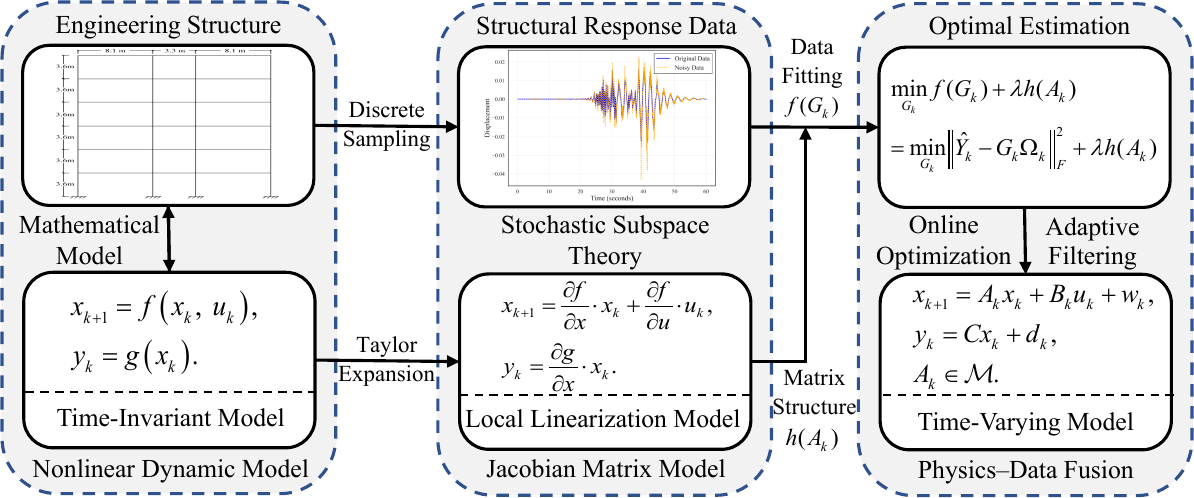}

\vspace{1em}

This framework lies at the intersection of physics-informed machine learning and digital twin technologies, offering a computationally efficient and physically 
consistent approach to online structural system identification under uncertainty.
\end{graphicalabstract}


\begin{highlights}
    \item \textbf{Adaptive Physics-Informed Modeling with Control (APSMC):}  
     A novel physics-based digital twin framework is proposed, integrating real-time adaptive filtering with proximal gradient updates to 
     enable online identification of state-space parameters under arbitrary input excitation.

    \item \textbf{Convergence to Physically Consistent Optimal Estimation:}  
    Within the stochastic subspace identification framework, it is theoretically demonstrated that APSMC incorporates Kalman filter-based optimal state 
    estimation with embedded physical constraints, guaranteeing convergence to a physically consistent optimal solution as data accumulates.

    \item \textbf{Robust Validation through Simulation and Experiment:}  
    The effectiveness of APSMC is validated through simulations of Duffing oscillators, seismic response analyses of frame structures, and scaled bridge 
    experiments, achieving low computational cost and strong generalization performance.
\end{highlights}

%% Keywords

\end{frontmatter}

\section{Introduction}
Nonlinear dynamical systems and partial differential equations (PDEs) are pervasive in the modeling of complex physical systems, including fluid 
mechanics \cite{herrmannDatadrivenResolventAnalysis2021}, structural dynamics \cite{chen5097818adaptive}, and epidemiological modeling \cite{proctor2015discovering}. 
These systems often exhibit complex characteristics such as nonlinearity and multivariable coupling, making their accurate modeling and efficient analysis a highly 
challenging task \cite{khalil2002nonlinear}.
Physics-based numerical solvers, such as the finite element method (FEM), finite difference method (FDM), and finite volume method (FVM), have enabled researchers 
to generate large volumes of high-quality scientific computing data for solving complex dynamical equations \cite{meirovitch1980computational,ferziger2019computational}.
However, when uncertainties exist in boundary conditions, material properties, or external loads, the predictive accuracy of these models often degrades significantly. 

Moreover, the high computational cost of high-fidelity numerical simulations often limits the duration of available simulation data\cite{sharma2020review}.
In contrast, with the rapid advancement of modern sensor technology and the deployment of large-scale monitoring systems, massive amounts of time-series data are continuously 
collected from complex physical systems \cite{zhangOnlineDynamicMode2019,chen2025minimal}. However, such data typically only cover a subset of the system's states, resulting in long time-series datasets 
that are often contaminated with noise \cite{chen5097818adaptive,chen2024online}.
Therefore, efficiently extracting key dynamical features from finite-dimensional noisy data and constructing robust, physics-informed models with high interpretability 
has become a critical and interdisciplinary challenge across multiple scientific and engineering fields \cite{chen2025minimal,kirchdoerfer2016data,de2025automated}.

To address the aforementioned challenges, Koopman Mode Analysis (KMA) has gained significant attention in various disciplines, 
including nonlinear dynamical systems, control, 
and fluid mechanics\cite{mezicAnalysisFluidFlows2013,bruntonModernKoopmanTheory2021,mezic2005spectral}. The core concept of KMA lies in representing nonlinear 
system dynamics within a linear framework, 
offering a new approach for modal decomposition and dynamic 
modeling of high-dimensional nonlinear systems\cite{koopman1931hamiltonian,koopman1932dynamical}. Among these methods, Dynamic Mode Decomposition (DMD) serves as a finite-dimensional approximation of the 
Koopman linear operator\cite{tuDynamicModeDecomposition2014,schmidDynamicModeDecomposition2010}. 
By decomposing high-dimensional time-series data, DMD can extract dominant spatiotemporal modal structures from the system\cite{schmidDynamicModeDecomposition2022}. To further improve the applicability of DMD in nonlinear 
scenarios, Extended DMD (EDMD) was introduced. EDMD projects the original data into a higher-dimensional feature space using user-defined observables, allowing for a more accurate 
approximation of the Koopman linear model\cite{williams2015data}. With the rise of deep learning, neural network models such as autoencoders have 
also been integrated into the DMD framework, enabling the 
adaptive learning of nonlinear observables and further enhancing the capability of Koopman-based modeling in nonlinear 
systems\cite{yeungLearningDeepNeural2017,wu2024koopman}.

Despite the promising potential of the Koopman framework for nonlinear system modeling, obtaining physically interpretable and data-efficient observables from limited measurements 
remains a key challenge in data-driven Koopman operator construction\cite{bruntonModernKoopmanTheory2021}. Meanwhile, Sparse Identification of Nonlinear Dynamics (SINDy) has
 also attracted significant attention in 
recent years\cite{brunton2016discovering}. SINDy leverages sparse optimization techniques to automatically identify governing equations from a candidate function library, providing a compact mathematical 
representation of complex system dynamics. However, SINDy also faces several practical limitations: on the one hand, it heavily depends on the 
completeness of observation data and the accuracy of numerical differentiation\cite{rosafalco2024ekf}; on the other hand, under conditions of limited sensor 
coverage or noisy measurements, the identification accuracy may be significantly 
degraded.  The governing equations identified from finite-dimensional observation data may deviate from the system's true physical mechanisms, thereby compromising the model's 
interpretability and generalization capability.

Although these methods excel at capturing dominant dynamical features, they generally require offline training based on historical data, are sensitive to measurement noise, 
and lack theoretical guarantees for optimal estimation under stochastic disturbances \cite{chen2024online}. As such, they struggle to capture the long-term evolution and 
time-varying characteristics of physical systems. For example, in the field of structural health monitoring (SHM), engineering structures such as bridges, tunnels, and 
urban buildings are continuously subjected to factors such as material aging, fatigue damage, environmental degradation, and extreme events (e.g., earthquakes, typhoons, 
and explosions), leading to gradual changes in their dynamical properties over time \cite{brownjohn2007structural,chen5097818adaptive}. However, existing models, such as Koopman models and neural n
etwork models, are typically parameter-heavy, making them less suitable for real-time tracking of system states.

Against this backdrop, the state-space equation provides a natural framework for physics-data fusion modeling. By embedding physical constraints into the system matrices or 
incorporating physically motivated loss terms during the system identification process, the interpretability of the model can be enhanced while simultaneously improving its 
robustness to noise\cite{baddooPhysicsinformedDynamicMode2021,karniadakis2021physics}. The Adaptive Physics-Informed System Modeling (APSM) method formulates nonlinear 
dynamical systems as stochastic time-varying state-space models, combines 
Kalman filter (KF) for recursive state estimation, and employs proximal gradient optimization to update model parameters while maintaining physical interpretability through 
constraint embedding~\cite{chen5097818adaptive}.

\begin{figure}[h]
\centering
\includegraphics[width=0.9\textwidth]{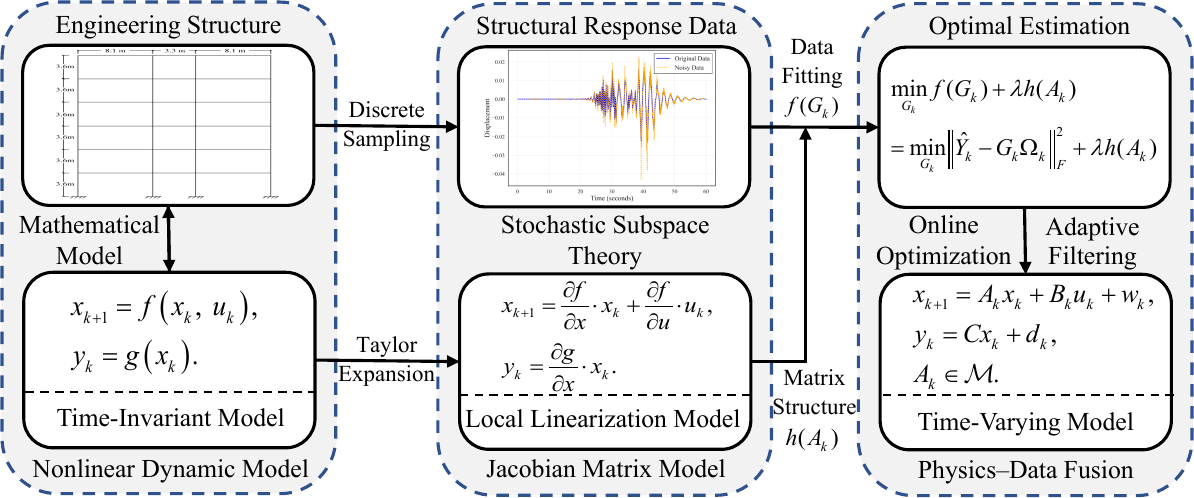}
\caption{Overview of the proposed APSMC framework. For a nonlinear engineering structure governed by $f(x_k, u_k)$ with external input $u_k$, noisy 
response measurements $y_k$ are obtained through discrete sampling. The system is locally linearized into a Jacobian-based time-varying state-space 
representation. Guided by stochastic subspace theory, a physics-informed optimal estimation problem is formulated. By integrating adaptive filtering 
with a convex optimization algorithm constrained by physical priors, the model parameters are updated in real time, enabling accurate and physically consistent system identification.}
\label{fig:ss1}
\end{figure}

This approach has demonstrated strong predictive and modeling performance on real-world acceleration data from the Hangzhou Bay Bridge. However, the original APSM formulation 
lacks a comprehensive theoretical analysis regarding noise robustness, algorithmic convergence, and applicability to broader classes of nonlinear engineering systems, which 
limits its wider adoption. To address these challenges, this study develops a generalized Adaptive Physics-Informed System Modeling with Control (APSMC) framework, extending 
the original APSM method to accommodate stochastic dynamical systems subjected to arbitrary white noise distributions and external excitations.
The key contributions and structure of this paper are summarized as follows:
\begin{itemize}
    \item Section~\ref{SSI_introduction} defines the mathematical model adopted in this study and briefly introduces the theoretical framework of stochastic subspace identification (SSI), along with related prior work.

    \item Within the SSI framework, Section~\ref{SSI_framework1} provides a theoretical proof that, under the assumption of infinite data, the APSM algorithm yields the optimal estimation of time-varying state matrices.

    \item Section~\ref{sec:APSC} further extends the original APSM framework to a general input–output formulation and proposes the APSMC method, which supports arbitrary white noise disturbances and explicit external excitations.

    \item Sections~\ref{numerical} and~\ref{sec:Case} present both numerical simulations and experimental validations, including a nonlinear Duffing oscillator, 
    seismic analysis of a frame structure, and impact testing on a laboratory-scale bridge, to demonstrate the effectiveness of the proposed framework.

\end{itemize}
A concise overview of the proposed APSMC methodology is presented in Figure~\ref{fig:ss1}. An open-source Python implementation is also available 
at \href{https://github.com/Chen861368/Adaptive-Physics-Informed-System-Modeling-with-Control-for-Nonlinear-Structural-Dynamics-Estimation}{GitHub}.

\section{Mathematical Formulation and State-of-the-Art Methods}\label{SSI_introduction}

To support the optimal estimation framework based on stochastic subspace theory presented in Section~\ref{SSI_framework}, and to lay the groundwork for demonstrating 
that the APSMC algorithm achieves the theoretically optimal estimate, this section provides a brief introduction to stochastic subspace theory and defines the mathematical 
models involved in the optimal estimation problem.

First, we introduce the mathematical model of the observed data and noise, and based on this model, we briefly present the Data-Driven Stochastic Subspace 
Identification (SSI-Data) method to demonstrate how state-space models can be estimated from noisy measurement data~\cite{peetersREFERENCEBASEDSTOCHASTICSUBSPACE}.
Finally, a brief overview of the APSM method is provided to prepare the reader for the in-depth discussion in Section~\ref{SSI_framework}.

\subsection{Mathematical Model}
In this section, we present a modified version of the classical state-space model—referred to as the Optimal Estimation Form of the State-Space Model—where the optimal estimates of the state vector are treated as the system's state vectors. First, we begin with the classical state-space model, which assumes that the observed data is generated according to:
\begin{equation}\label{eq_1000}
    x_{k+1} = A x_k + w_k, \quad y_k = C x_k + d_k,
\end{equation}
where \( x_k \in \mathbb{R}^{n} \) is the state vector and \( y_k \in \mathbb{R}^{m} \) is the observed output (typically measured by sensors). Here, \( A \in \mathbb{R}^{n \times n} \) characterizes the system dynamics, while \( C \in \mathbb{R}^{m \times n} \) represents the observation model. The terms \( w_k \in \mathbb{R}^{n} \) and \( d_k \in \mathbb{R}^{m} \) denote the process noise and measurement noise, respectively. In this work, both noise processes are assumed to be zero-mean white noise with ergodic properties.

To express Equation \eqref{eq_1000} in the form of an optimal estimation state vector, we first define what is meant by optimal estimation. 
The estimation problem involves recovering the state sequence \( x_1, x_2, \ldots, x_n \) from the observed data \( y_1, y_2, \ldots, y_n \). 
Under the mean-square error criterion, the minimum error estimator in the Bayesian framework is the conditional expectation:
\begin{equation}\label{eqc_7}%\eqref{eqc_7} 
    \mathbb{E}\left[x_k \mid y_1, y_2, \ldots, y_n\right] 
    = \underset{g\left(y_1, y_2, \ldots, y_n\right)}{\mathrm{argmin}} \; \mathbb{E}\left[\left(x_k - g\left(y_1, y_2, \ldots, y_n\right)\right)^2\right],
\end{equation}
for \( k \in \{1, 2, \ldots, n\} \). We then define the optimal estimate as the conditional expectation:
\begin{equation}
    \hat{x}_{k|n} = \mathbb{E}\left[x_k \mid y_1, y_2, \ldots, y_n\right].
\end{equation}
This formulation, which utilizes all available observation data, is commonly referred to as smoothing. Alternatively, if only the 
data up to the current time \( k \) are used for estimation, the process is known as filtering, and the estimate becomes:
\begin{equation}\label{eqc_1}%\eqref{eqc_1} 
    \hat{x}_{k|k}= \mathbb{E} \left[ x_k \mid  y_1, y_2, \ldots, y_k \right ].
\end{equation}
It is worth noting that the renowned KF is a special case of equation \eqref{eqc_1} under the assumptions of linear dynamics and Gaussian noise, 
with recursive estimation. For more general noise distributions or nonlinear dynamic models, alternative Bayesian filters, such as the Unscented Kalman Filter\cite{wan2000unscented} 
or the Particle Filter, may be employed.

These adaptive filters typically combine predictions from the system's dynamic model with corrections derived from real-time measurements. In many civil engineering 
applications, due to the large scale of the structures, external excitations are often difficult to monitor comprehensively and are therefore typically assumed to be 
white noise. Accordingly, we further assume that:
\begin{equation}
    w_k \sim \mathcal{N}(0, W_k), \quad d_k \sim \mathcal{N}(0, D_k),
\end{equation}
with \(W_k \in \mathbb{R}^{n \times n}\) and \(D_k \in \mathbb{R}^{m \times m}\) as the respective covariance matrices.
Based on the above linear state-space model \eqref{eq_1000} and the assumption of Gaussian white noise, stochastic subspace identification theory typically 
employs the KF framework to analyze the noise in the data. However, the following discussion process and framework also apply to more general noise 
distributions and nonlinear dynamic model types.

In summary, the Bayesian filtering process can be decomposed into the prediction step and the correction step. This adaptive adjustment process in the KF, 
involving posterior estimation, can be expressed in the optimal estimation form of the state-space model, based on the analysis in \cite{chen2025principal}:
\begin{equation}\label{eqc_4}%\eqref{eqc_4} 
    \hat{x}_{k+1|k} = A\hat{x}_{k|k-1} + K_k e_k, \quad y_k = C \hat{x}_{k|k-1} + e_k,
\end{equation}
where the innovation term is defined as \( e_k = y_k - \hat{y}_{k|k-1} \). Here, \( e_k \) is assumed to be white noise, and the Kalman gain matrix \( K_k \) links 
the process noise with the observation noise.
More generally, for any \( i \in \mathbb{Z}^+ \), the future output \( y_{k+i} \) can be expressed as\cite{chen2025principal}:
\begin{equation}\label{eqc_2}
    y_{k+i} = CA^i \hat{x}_{k|k-1} + \sum_{j=0}^{i-1} CA^{i-1-j} K_{k+j} e_{k+j} + e_{k+i}.
\end{equation}

\subsection{Noise Modeling and Data-Driven Stochastic Subspace Identification Method}

For the convenience of subsequent analysis, this section first models the noise in the monitoring data based on the previously discussed 
equation \eqref{eqc_4}, and briefly introduces the SSI-Data method\cite{peetersREFERENCEBASEDSTOCHASTICSUBSPACE}. 
We begin by forming two Hankel matrices from the observed output data:
\begin{equation}
    Y_p = \begin{bmatrix}
        y_{0} & y_{1} & \cdots & y_{j-1} \\ 
        y_{1} & y_{2} & \cdots & y_{j} \\ 
        \vdots & \vdots & \ddots & \vdots \\ 
        y_{i-1} & y_{i} & \cdots & y_{i+j-2}
        \end{bmatrix}, \quad Y_f = \begin{bmatrix}
    y_i & y_{i+1} & \cdots & y_{i+j-1} \\ 
    y_{i+1} & y_{i+2} & \cdots & y_{i+j} \\ 
    \vdots & \vdots & \ddots & \vdots \\ 
    y_{2i-1} & y_{2i} & \cdots & y_{2i+j-2}
    \end{bmatrix}.
\end{equation}
with \(Y_p, Y_f \in \mathbb{R}^{im \times j}\). These matrices capture the past and future outputs of the system, respectively.
Based on equation \eqref{eqc_2}, any column of the Hankel matrix can be written in a compact form as:
\begin{equation}
    \underbrace{
\begin{bmatrix}
y_k \\
y_{k+1} \\
\vdots \\
y_{k+i-1}
\end{bmatrix}}_{y_i[k]}
=
\underbrace{
\begin{bmatrix}
C \\
CA \\
\vdots \\
CA^{i-1}
\end{bmatrix}
}_{O_i}
\hat{x}_k
+
\underbrace{
\begin{bmatrix}
I & 0 & 0 & \cdots & 0 \\
CK_k & I & 0 & \cdots & 0 \\
\vdots & \vdots & \vdots & \ddots & \vdots \\
CA^{i-2} K_k & CA^{i-3} K_{k+1} & \cdots & CK_{k+i-2} & I
\end{bmatrix}
}_{F_i}
\underbrace{
\begin{bmatrix}
e_k\\
e_{k+1} \\
\vdots \\
e_{k+i-1}
\end{bmatrix}}_{e_i[k]}.
\end{equation}
Here, \(y_i[k] \in \mathbb{R}^{im}\) represents the observed output vector from \(y_k\) to \(y_{k+i-1}\), \(\hat{x}_k\) (a shorthand for \(\hat{x}_{k|k-1}\)) is the 
optimal estimation state, \(O_i \in \mathbb{R}^{im \times n}\) is the extended observability matrix, and \(F_i \in \mathbb{R}^{im \times im}\) captures the influence 
of noise, with \(e_i[k] \in \mathbb{R}^{im}\) denoting the noise components.
For convenience, we organize the state estimates and noise components over the data window into the matrices:
\begin{equation}
    \hat{X}_k = 
    \begin{bmatrix}
    \hat{x}_k & \hat{x}_{k+1} & \hat{x}_{k+2} & \cdots & \hat{x}_{k+j-1}
    \end{bmatrix}, \quad 
    E_k = 
    \begin{bmatrix}
    e_i[k] & e_i[k+1] & \cdots & e_i[k+j-1]
    \end{bmatrix}. 
\end{equation}
with \(\hat{X}_k \in \mathbb{R}^{n \times j}\) and \(E_k \in \mathbb{R}^{im \times j}\). Under the KF based state estimation, the output Hankel matrix can be expressed as\cite{chen2025principal}:
\begin{equation}
    Y_f = O_i \hat{X}_i + F_i E_i = (O_i T)(T^{-1}\hat{X}_i) + F_i E_i.
\end{equation}
where \( T \in \mathbb{R}^{n \times n} \) represents an invertible transformation. This decomposition indicates that 
the term \(O_i \hat{X}_i\) represents the signal components, while \(F_i E_i\) captures the noise.
Although these two components are unknown, the signal subspace \(\text{span}( O_i \hat{X}_i)\) and the noise subspace \(\text{span}(F_i E_i)\) are statistically 
independent. Leveraging this relationship, the instrumental variable (IV) \(\Psi^\top \in \mathbb{R}^{j \times p}\) is introduced, which can be applied to eliminate 
the noise term \(F_i E_i\), satisfying the following constraints:
\begin{equation}
    \lim_{j \to \infty} \frac{1}{j} (F_iE_0) \Psi^\top = 0, \quad \text{and} \quad \text{rank}\left(\lim_{j \to \infty} \frac{1}{j} O_i \hat{X}_0 \Psi^\top\right) = n. 
\end{equation}
These constraints ensure that the subspace \(\text{span}(\Psi^\top)\) is statistically orthogonal to the noise subspace \(\text{span}(F_iE_i)\), and 
that \(\text{span}(\Psi^\top)\) captures the full information of the signal subspace \(\text{span}(O_i \hat{X}_i)\). Leveraging these properties, the 
SSI-Data method utilizes the projection matrix \(\Pi_{Y_p}\) to eliminate the noise and derive the following relationship:
\begin{equation}\label{eqc_{9}}%\eqref{eqc_{9}} 
    \lim_{j \to \infty} \frac{1}{j} Y_f \Pi_{Y_p} = O_i \hat{X}_i, \quad \text{with} \quad \Pi_{Y_p} \doteq  Y_p^\top (Y_pY_p^\top)^\dagger Y_p,
\end{equation}
where $\dagger$ denotes the Moore–Penrose pseudoinverse, $\doteq$ indicates a definition, and $\Pi_{Y_p}$ projects the row space of the future outputs onto that of the past outputs.
However, in practical applications, since \(j\) is finite, this expression typically contains noise and can be written as:
\begin{equation}\label{eqc_6}%\eqref{eqc_6} 
    \frac{1}{j} Y_f \Pi_{Y_p}  = O_i \hat{X}_0 + O_j(\varepsilon),
\end{equation}
where \(O_j(\varepsilon)\) represents residual noise that diminishes as \(j \to \infty\). 
To minimize the influence of \(O_j(\varepsilon)\), the SSI-Data method performs SVD on \(\frac{1}{j} Y_f \Pi_{Y_p}\) as follows:
\begin{equation}
    \frac{1}{j}Y_fY_p^\top = U \Sigma V^\top =
    \begin{bmatrix}
    U_1 & U_2
    \end{bmatrix}
    \begin{bmatrix}
    \Sigma_1 & 0 \\ 
    0 & 0
    \end{bmatrix}
    \begin{bmatrix}
    V_1^\top \\ 
    V_2^\top
    \end{bmatrix}
    = U_1 \Sigma_1 V_1^\top.  
\end{equation}
Here, \( \Sigma_1 \in \mathbb{R}^{n \times n} \) is a diagonal matrix containing the \( n \) largest singular values, 
while \( U_1 \in \mathbb{R}^{im \times n} \) and \( V_1^\top \in \mathbb{R}^{n \times j} \) are the corresponding left and right singular matrices.
Consequently, based on the above analysis, the SSI-Data method obtains the approximate extended observability matrix and the state sequence, which are expressed as follows:
\begin{equation}
    O_i \thickapprox U_1 S_1^{1/2}, \quad \hat{X}_i \thickapprox  S_1^{1/2}V_1^\top.
\end{equation}
Due to the influence of noise \(O_j(\varepsilon)\), the approximate symbol is used here. With the estimated state sequence \(\hat{X}_i\), we construct:
\begin{equation}
    \hat{X}_0 \thickapprox \begin{bmatrix} \hat{x}_i & \hat{x}_{i+1} & \cdots & \hat{x}_{i+j-2} \end{bmatrix}, \quad
    \hat{X}_1 \thickapprox \begin{bmatrix} \hat{x}_{i+1} & \hat{x}_{i+2} & \cdots & \hat{x}_{i+j-1} \end{bmatrix}. 
\end{equation}
Using the observed data \(Y_0\), the system matrices \(A\) and \(C\) are then estimated from the linear relationship:
\begin{equation}
    \begin{bmatrix}
        \hat{X}_1 \\
        Y_0
        \end{bmatrix}
        =
        \begin{bmatrix}
        A \\
        C
        \end{bmatrix}
        \hat{X}_0 + \begin{bmatrix}
        \rho_w \\
        \rho_v
        \end{bmatrix}, \quad Y_0 = \begin{bmatrix} y_i & y_{i+1} & \cdots & y_{i+j-2} \end{bmatrix}, 
\end{equation}
where \(\rho_w\) and \(\rho_v\) are residual terms. A least-squares solution gives:
\begin{equation}\label{eqc_5}%\eqref{eqc_5} 
    \begin{bmatrix}
        A \\
        C
        \end{bmatrix} =
        \begin{bmatrix}
        \hat{X}_1 \\
        Y_0
        \end{bmatrix}
        \hat{X}_0^\dagger.
\end{equation}
Here, \(\hat{X}_0^\dagger\) represents the Moore-Penrose pseudoinverse of \(\hat{X}_0\). Since the matrix \(\hat{X}_0\) is typically not square, the 
pseudoinverse provides an optimal fit solution to solve the overdetermined system. In theory, with a sufficiently large dataset, the residual noise \(O_j(\varepsilon)\) in 
equation \eqref{eqc_6} tends to zero, yielding an optimal estimation of the state-space model. Consequently, the modal parameter estimates will also be optimal. 

However, in practice, computational constraints limit the sample size \(j\), and \(O_j(\varepsilon)\) may remain non-negligible. As a result, the estimated 
matrices \(A\) and \(C\) obtained using the SSI-Data method are still affected by noise, preventing the method from achieving its theoretical optimal performance.

\subsection{Adaptive Physics-Informed System Modeling}

To address the issue that traditional system identification algorithms struggle to leverage massive monitoring data due to computational resource limitations, 
we proposed the APSM method in \cite{chen5097818adaptive}, which adaptively and in real-time corrects system parameters. 
This allows the method to operate on sufficiently large datasets, resulting in more accurate state-space model estimates.

The method was validated in the paper 
using measured acceleration data from the Hangzhou Bay Bridge, achieving the best state estimation results compared to existing classical time-domain methods.
In the APSM framework, the engineering structure is represented by a time-varying discrete stochastic state-space model \( A_k \), with physical constraints imposed on \( A_k \) as follows:
\begin{align}\label{eqc_{10}}%\eqref{eqc_{10}} 
    x_{k+1} &= A_k x_k  + w_k, \\
    y_k &= Cx_k + d_k, \\
    A_k &\in \mathcal{M}.
  \end{align}
Here, \( A_k \in \mathcal{M} \) indicates that the system matrix \( A_k \) is 
constrained within a set \( \mathcal{M} \) that satisfies specific physical 
properties. Some commonly used physical constraints and methods for incorporating them can be found in \ref{physics_constrain}.
Although, in real-world monitoring systems, only the monitoring 
data \( y_1, y_2, \ldots, y_{k+1} \) are available, the state 
vector \( \hat{x}_i \) can be reconstructed from \( y_k \) using the KF. Thus, based on the monitoring data, the following data matrices can be incrementally constructed:
\begin{equation}
    \hat{X}_k = \begin{bmatrix}
        | & | & & | \\
        \hat{x}_1 & \hat{x}_2 & \cdots & \hat{x}_k \\
        | & | & & |
    \end{bmatrix}, \quad 
    \hat{Y}_k = \begin{bmatrix}
        | & | & & | \\
        \hat{x}_2 & \hat{x}_3 & \cdots & \hat{x}_{k+1} \\
        | & | & & |
    \end{bmatrix}.
\end{equation}
Building upon this, the APSM algorithm employs an online proximal gradient convex optimization method to identify the 
system matrix \( A_k \) by solving the following optimization problem: 
\begin{equation}\label{eqc_{11}}%\eqref{eqc_{11}} 
    A_k = \underset{A_k}{\mathrm{argmin}} \left\|\hat{Y}_k - A_k\hat{X}_k\right\|_F^2 + \lambda h(A_k) 
\end{equation}
Here, $\|\cdot\|_F$ denotes the Frobenius norm, $h(A_k)$ is a regularization term that imposes physical constraints, and $\lambda$ is a Lagrange multiplier that balances data fidelity and conformity to the prescribed physical laws. Based on the Taylor expansion, $A_k$ can be regarded as the Jacobian matrix of the nonlinear 
dynamical system at each time step \cite{chen5097818adaptive}. Detailed 
algorithmic steps are provided in Algorithm \ref{algo:opdmd}. An introduction to the proximal gradient descent method can be found in \ref{proximal_gradient}.
\begin{algorithm}  
    \caption{Adaptive Physics-Informed System Modeling Algorithm}
    \label{algo:opdmd}
    \begin{algorithmic}[1]
    \Require System matrices \( A_{k-1} \), monitoring data \( y_k \), state vector \( \hat{x}_{k-1} \), measurement matrix \( C \), noise covariance matrices \( P_{k-1} \), \( Q \), \( R \), and step sizes \( \{t_k\} \).
    
    \State \textbf{Step 1}: Use the Kalman Filter to compute \( \hat{x}_{k} \) and \( P_{k} \).
    
    \State \textbf{Step 2}: Compute the gradient of the differentiable part of the cost function at \( A_{k-1} \):
    \[
        f(A) \doteq  \left\| \hat{x}_{k} - A \hat{x}_{k-1} \right\|_2^2
    \]
    \[
        \nabla f(A_{k-1}) = -2 \left( \hat{x}_{k} - A_{k-1} \hat{x}_{k-1} \right) \hat{x}_{k-1}^\top
    \]
    \State \textbf{Step 3}: Update \( A_{k-1} \) using the online proximal gradient step:
    \[
        A_k = \text{prox}_{t_k h} \left( A_{k-1} - t_k \nabla f(A_{k-1}) \right)
    \]
    
    \State \textbf{Where} the proximal mapping \( \text{prox}_t(v) \) is defined as:
    \[
        \text{prox}_{t h}(v) = \arg\min_z \left\{ \frac{1}{2t} \|v - z\|_2^2 + h(z) \right\}
    \]
    
    \State \textbf{Return}: Updated states \( \hat{x}_{k} \), noise covariance matrix \( P_{k} \), and system matrix \( A_k \).
    \end{algorithmic}
\end{algorithm}

The APSM algorithm integrates residual-based updates with physical information to adaptively correct the parameters of the initial system matrix \( A_0 \) at each time step. 
The system matrices identified through this method possess clear physical significance, providing valuable insights into the dynamic properties of the structural system.

\section{An Adaptive Physics-Informed System Modeling with Control Framework}\label{SSI_framework}

In this section, we first provide a theoretical justification for the optimality of the APSM algorithm in state-space model estimation. Based on these insights, we extend the framework to accommodate more 
general noise conditions and input-influenced scenarios, leading to the proposed APSMC algorithm.

\subsection{Adaptive Physics-Informed System Modeling and Optimal Estimation}\label{SSI_framework1}

Although the APSM algorithm has demonstrated promising results in practice, it does not explicitly incorporate noise modeling, making it difficult to 
assess its statistical performance~\cite{chen5097818adaptive}. In this section, we analyze the statistical properties of the estimated state-space model 
based on the theoretical foundations of stochastic subspace identification.
Following the structure of equation~\eqref{eqc_7}, the optimal estimator within a Bayesian framework can be defined. Specifically, the minimum-error 
estimator of \( A \) is the conditional expectation:
\begin{equation}\label{eqc_{8}}%\eqref{eqc_{8}} 
    \mathbb{E}\left[A \mid y_1, y_2, \ldots, y_n\right] 
    = \underset{g\left(y_1, y_2, \ldots, y_n\right)}{\mathrm{argmin}} \; \mathbb{E}\left[\left(A - g\left(y_1, y_2, \ldots, y_n\right)\right)^2\right],
\end{equation}
where the expectation $\mathbb{E}[\cdot]$ is taken with respect to the posterior distribution $p(A \mid y_1, y_2, \ldots, y_n)$.
In contrast, under a classical statistical framework where \( A \) is treated as deterministic, equation~\eqref{eqc_{8}} lacks a well-defined optimal 
solution. Therefore, in general, directly extracting a state-space model from noisy data \( y_k \) 
is a highly challenging task.

\begin{figure}[h]
    \centering
    \includegraphics[width=0.9\textwidth]{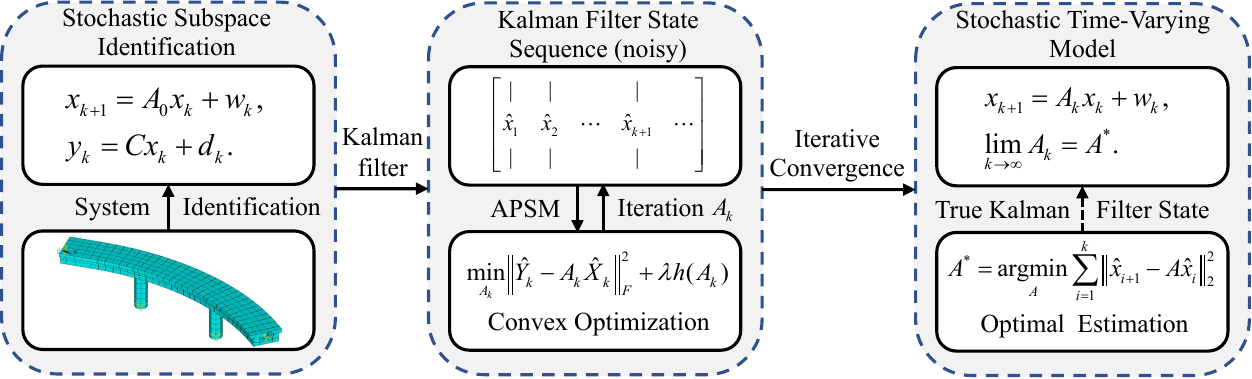}
    \caption{Illustration of the APSM framework for optimal system matrix estimation.}
    \label{fig:ss12}
\end{figure}

However, as shown in equation \eqref{eqc_{9}}, when an infinite amount of monitoring data is available, the noise 
component is entirely eliminated, leaving only the signal subspace term $O_i \hat{X}_i$.
In this ideal case, if the true Kalman-filtered state sequence \( \hat{X}_i \) were known, the optimal estimate of the system matrix \( A^* \) could be defined as:
\begin{equation}
    A^* = \underset{A}{\mathrm{argmin}} \sum_{i=1}^{k} \left\| \hat{x}_{i+1} - A \hat{x}_i \right\|^2_2 =\hat{Y}_k \hat{X}_k^{\dagger} 
\end{equation}
This reformulates the original estimation problem into one of obtaining an accurate Kalman-filtered state sequence. However, due to the presence of the noise 
term \( O_j(\varepsilon) \) in equation~\eqref{eqc_6}, the SSI-Data method can only approximate the state sequence via SVD, resulting in residual noise 
in the system matrix estimate. Therefore, the initial estimate \( A_0 \) obtained via SSI-Data can be expressed as:
\begin{equation}
    A_0 = A^* + \Delta A_0 
\end{equation}
where \( \Delta A_0 \) represents the estimation error introduced by noise, limited data, and model approximation.
Despite this, the SSI-derived model \( A_0 \) can still serve as a suitable initialization for the APSM algorithm. Since the Kalman filter minimizes 
the mean square error, each update yields a 
state estimate \( \hat{x}_k \) that is more accurate than the original SSI-based sequence:
\begin{equation}\label{eqc_{12}}%\eqref{eqc_{12}} 
    \hat{x}_1, \quad \hat{x}_2, \quad \hat{x}_3, \quad \cdots, \quad \hat{x}_{i}, \quad \cdots
\end{equation}
Using these optimized state estimates, we can apply the proximal gradient algorithm to solve equation \eqref{eqc_{11}}, progressively refining 
the initial model \( A_0 \), continuously improving the model's accuracy. This results in the following:
\begin{equation}
    A_1=A^* + \Delta A_1, \quad A_2=A^* + \Delta A_2,\quad \cdots,\quad A_{i}=A^* + \Delta A_{i},\quad \cdots
\end{equation}
Since the objective function in equation~\eqref{eqc_{11}} is convex, and each update follows a convex optimization scheme, the algorithm is 
guaranteed to converge to the global optimum based on the true Kalman-filtered state sequence. Ultimately, as time progresses, \( \lim_{i \to \infty} \Delta A_{i} = 0 \), 
completely reducing the influence of noise interference \( \Delta A \) on the system matrix \( A_k \). 
The above process is illustrated in Figure~\ref{fig:ss12}.

While this study demonstrates the use of the SSI-Data method to construct the initial model $A_0$, other techniques—such as the Eigensystem Realization 
Algorithm (ERA)~~\cite{juang1985eigensystem}, N4SID, and MOESP~~\cite{van2012subspace,viberg1995subspace}—can also be employed. Empirical results indicate 
that subsequent optimization using APSM achieves similarly high performance regardless of the initial model construction method.

\subsection{An Adaptive Physics-Informed System Modeling with Control Framework}\label{sec:APSC}

Although it has been previously shown that APSM can achieve optimal estimation, this is only applicable to systems without external inputs or systems that assume
 white noise excitation. In this section, we will further consider the APSMC framework. It assumes the presence 
 of process noise \( w_k \in \mathbb{R}^n \) and measurement noise \( d_k \in \mathbb{R}^m \), both of which may follow arbitrary distributions of white noise, 
 and that the system may include external excitation inputs. Based on this assumption, equation \eqref{eqc_{10}} can be rewritten as:
\begin{align}
    x_{k+1} &= A_k x_k + B_k u_k + w_k, \\
    y_k &= C x_k + d_k, \\
    A_k &\in \mathcal{M}.
\end{align}
Here, \( B_k \in \mathbb{R}^{n \times l} \) represents the external control input matrix, and \( u_k \in \mathbb{R}^l \) is the external control or excitation vector. 
At this point, we can consider two cases:
\begin{enumerate}
    \item \textbf{\( B_k \) is known}. In this case, the APSM algorithm can be applied directly, with the non-stochastic term \( B_k u_k \) simply being incorporated 
    into the calculation at each time step.
    \item \textbf{\( B_k \) is unknown}. Since many time-domain algorithms can estimate both \( A \) and \( B \) from data, we assume that the initial system 
    matrices \( A_0 \) and \( B_0 \) have already been obtained.
\end{enumerate}
Given the initial \( A_0 \) and \( B_0 \), algorithms such as the KF or particle filter can be run, and the following relationship can be written:
\begin{equation}
    \underbrace{\begin{bmatrix}
        | & | & & | \\
        \hat{x}_2 & \hat{x}_3 & \cdots & \hat{x}_{k+1} \\
        | & | & & |
    \end{bmatrix}}_{\hat{Y}_k} = A \underbrace{\begin{bmatrix}
        | & | & & | \\
        \hat{x}_1 & \hat{x}_2 & \cdots & \hat{x}_k \\
        | & | & & |
    \end{bmatrix}}_{\hat{X}_k}+ B \underbrace{\begin{bmatrix}
        | & | & & | \\
        u_1 & u_2 & \cdots & u_k \\
        | & | & & |
    \end{bmatrix}}_{\Upsilon_k}
\end{equation}
The above equation can be simplified as:
\begin{equation}
    \hat{Y}_k = G \Omega_k, \quad G  \doteq \begin{bmatrix} A & B \end{bmatrix}, \quad \Omega_k  \doteq \begin{bmatrix} \hat{X}_k & \Upsilon_k \end{bmatrix}^\top.
\end{equation}
It is worth noting that the column vectors in \( \hat{X}_k \) may not necessarily represent the KF estimates. They actually correspond to adaptive filters that can 
yield optimal estimates, depending on the nature of the noise. Therefore, we could define the optimal estimate of the system matrix \( A^* \) as:
\begin{align}\label{eqc_{13}}%\eqref{eqc_{13}} 
    \begin{bmatrix} A^* &  B^*  \end{bmatrix}  &= \underset{A,B}{\mathrm{argmin}} \sum_{i=1}^{k} \left\| \hat{x}_{i+1} - A \hat{x}_i - B u_k \right\|^2_2 \\
    &= \underset{G}{\mathrm{argmin}} \left\|  \hat{Y}_k - G\Omega_k \right\|_F^2 = \hat{Y}_k \Omega_k^{\dagger}
\end{align}
It is important to note that when there is correlation between the columns of the matrix \( \Omega_k \) in the above convex optimization problem, the problem may 
not be strictly convex. In this case, although an optimal solution can still be found, these solutions may not be unique. The term \( \Omega_k^{\dagger} \) guarantees 
that the solution with the minimum Frobenius norm will be obtained. 

\begin{figure}[h]
\centering
\includegraphics[width=0.9\textwidth]{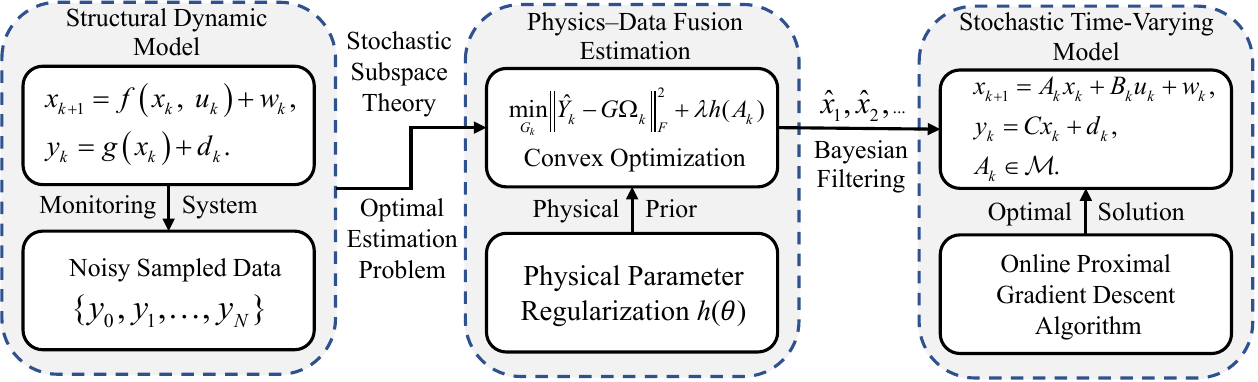}
\caption{Illustration of the convex optimization formulation within the APSMC framework.}
\label{fig:ss2}
\end{figure}

To discuss how to solve the above problem online and allow the system matrices \( A_k \) and \( B_k \) to be adaptively updated based on real-time sampling data, 
we modify the purely data-driven loss function as follows:
\begin{align}\label{eqc_{14}}%\eqref{eqc_{14}} 
    A_k, B_k &= \underset{G_k}{\mathrm{argmin}} \, f(G_k) + \lambda h(A_k) \\
    &= \underset{G_k}{\mathrm{argmin}} \left\|\hat{Y}_k - G_k\Omega_k\right\|_F^2 + \lambda h(A_k) ,\quad  G_k \doteq  \begin{bmatrix} A_k & B_k \end{bmatrix}
\end{align}
where \( f(G_k) \) represents the differentiable part, and \( h(A_k) \) represents the physical constraints on \( A_k \), which may be non-differentiable.
A visual representation of the overall physics–data fusion estimation framework is provided in Figure~\ref{fig:ss2}.

To solve this using the online proximal gradient descent algorithm, we first compute the gradient of the differentiable part \( f(G_k) \). The gradient computation proceeds as follows:
\begin{align}
    \nabla_G \left\|  \hat{Y}_k - G\Omega_k \right\|_F^2 &= \nabla_G \mathrm{Tr} \left[ (\hat{Y}_k - G \Omega_k)^\top (\hat{Y}_k - G \Omega_k) \right] \\
    &=-2 (\hat{Y}_k - G \Omega_k)\Omega_k^\top \\
    &=-2 \left( \hat{Y}_k - A \hat{X}_k - B \Upsilon_k \right) \begin{bmatrix} \hat{X}_k^\top & \Upsilon_k^\top \end{bmatrix}
\end{align}
To satisfy the physical constraints, \(A_k \in \mathcal{M}\) or the conditions imposed by \(h(A_k)\), we can update \( G_{k-1} \) using the online proximal gradient step:
\begin{equation}
    G_k = \text{prox}_{t_k h} \left( G_{k-1} - t_k \nabla f(G_{k-1}) \right),\quad    
\end{equation}
For the given objective function, the proximal mapping \( \text{prox}_{t_k h}(v) \) essentially solves an optimization problem that balances the distance between the current value \( v \) and the 
regularization term \( h(z) \), thus yielding the optimal solution that satisfies the physical constraints. 
This entire process for solving equation \eqref{eqc_{14}} online can be summarized in Algorithm \ref{algo:1opdmd}.
\begin{algorithm}[H]
    \caption{Adaptive Physics-Informed System Modeling with Control Framework}
    \label{algo:1opdmd}
    \begin{algorithmic}[1]
    \Require  Previous system matrices \( A_{k-1} \), \( B_{k-1} \); monitoring output \( y_k \); control input \( u_k \); estimated 
    state \( \hat{x}_{k-1} \); measurement matrix \( C \); covariance matrices \( P_{k-1} \), \( Q \), \( R \); and step size sequence \( \{t_k\} \).
    
    \State \textbf{Step 1}: Select an appropriate adaptive filter to compute \( \hat{x}_{k|k} \) and the noise covariance matrices \( P_{k|k} \).
    
    \State \textbf{Step 2}: Define the cost function for the data terms as follows:
    \[
        f(G) \doteq  \left\| \hat{x}_{k} - A \hat{x}_{k-1} - B u_{k-1} \right\|_2^2 = \left\| \hat{x}_{k} - G r_{k-1} \right\|_2^2
    \]
    
    \State \textbf{Step 3}: Compute the gradient of the cost function with respect to \( G_{k-1} \):
    \[
        \nabla f(G_{k-1}) = -2 \left( \hat{x}_{k} - A_{k-1} \hat{x}_{k-1} - B_{k-1} u_{k-1} \right) 
        \begin{bmatrix} \hat{x}_{k-1}^\top & u_{k-1}^\top \end{bmatrix}
    \]
    
    \State \textbf{Step 4}: Update \( G_{k-1} \) using the online proximal gradient step:
    \[
        \begin{bmatrix} A_k &  B_k \end{bmatrix} = G_k = \text{prox}_{t_kh} \left( G_{k-1} - t_k \nabla f(G_{k-1}) \right)
    \]
    
    \State \textbf{Return}: Updated  \( \hat{x}_{k} \),  \( P_{k} \), and system matrices \( A_k \), \( B_k \).
    \end{algorithmic}
\end{algorithm}

The choice of step size $t_k$ can be constant or determined through   variations of the gradient descent 
algorithm designed to expedite convergence, such as Variance Reduction\cite{johnson2013accelerating}, 
Acceleration and Momentum\cite{nitanda2014stochastic,qian1999momentum,liu2020accelerating}, 
and Adaptive Step Sizes\cite{duchi2011adaptive,kingma2014adam}.

\begin{figure}[htbp]
    \centering
    \resizebox{0.9\textwidth}{!}{ 
    \begin{tikzpicture}[node distance=1cm and 1cm, auto]
      \usetikzlibrary{positioning}
      \tikzstyle{block} = [rectangle, draw, text centered, fill=blue!20, minimum height=2em, minimum width=3em]
      \tikzstyle{data} = [draw, rectangle, fill=green!20, minimum width=3em, minimum height=2em]
      \tikzstyle{input} = [draw, rectangle, fill=red!20, minimum width=3em, minimum height=2em]
  
      \node [block, align=center] (Ak) {
        $\begin{aligned}
          x_{k+1} &= A_k x_k + B_k u_k + w_k, \\
          y_k &= Cx_k + d_k.
        \end{aligned}$
      };
      \node [input, below=of Ak] (yvalues) {\((A_{k},\ y_{k+1}, \ u_k)\)};
      \node [block, right=of Ak] (algorithm) {\(\left\| \hat{x}_{k+1} -A_{k} \hat{x}_{k} - B_{k} u_{k}\right\|^2_2 + \lambda h(A_k)\)};
      \node [data, above=of algorithm] (xkplus) {\((\hat{x}_{k},\hat{x}_{k+1})\)};
      \node [data, right=of algorithm] (Akplus) {\(G_{k+1}\)};
  
      \node [input] at (yvalues -| algorithm) (proximal) {Proximal mapping $\text{prox}_{th}(v)$};
  
      \draw [-latex] (yvalues) -- (Ak);
      \draw [-latex] (proximal) -- (algorithm);
      \draw [-latex] (xkplus) -- (algorithm);
      \draw [-latex] (Ak) |- (xkplus);
      \draw [-latex] (algorithm) -- (Akplus);
    \end{tikzpicture}
    }
    \caption{A schematic representation of the APSMC algorithm process.}
    \label{fig:algorithm_process}
  \end{figure}
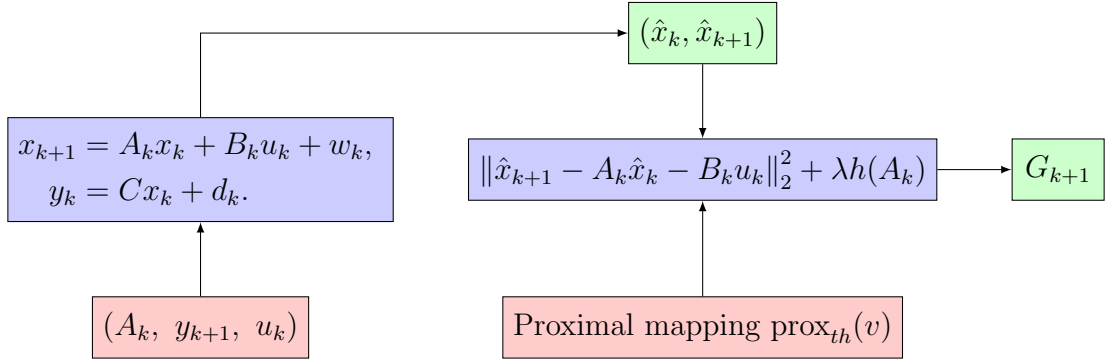

 By repeatedly executing this process, the system 
matrices \( A_k \) and \( B_k \) are adaptively updated at each time step, ensuring that the model not only fits the real-time data but also adheres to the physical 
constraints. The APSMC algorithm process is illustrated in Figure \ref{fig:algorithm_process}.
 Some common proximal mappings of physical 
constraints \( h(A_k) \) can be found in \cite{chen2024online}.
  
\section{Numerical Validation of the Proposed Framework}\label{numerical}

\subsection{Duffing Oscillator}
\begin{figure}[!ht]
    \centering
    \begin{subfigure}{0.29\textwidth}
        \centering
        \includegraphics[width=\textwidth]{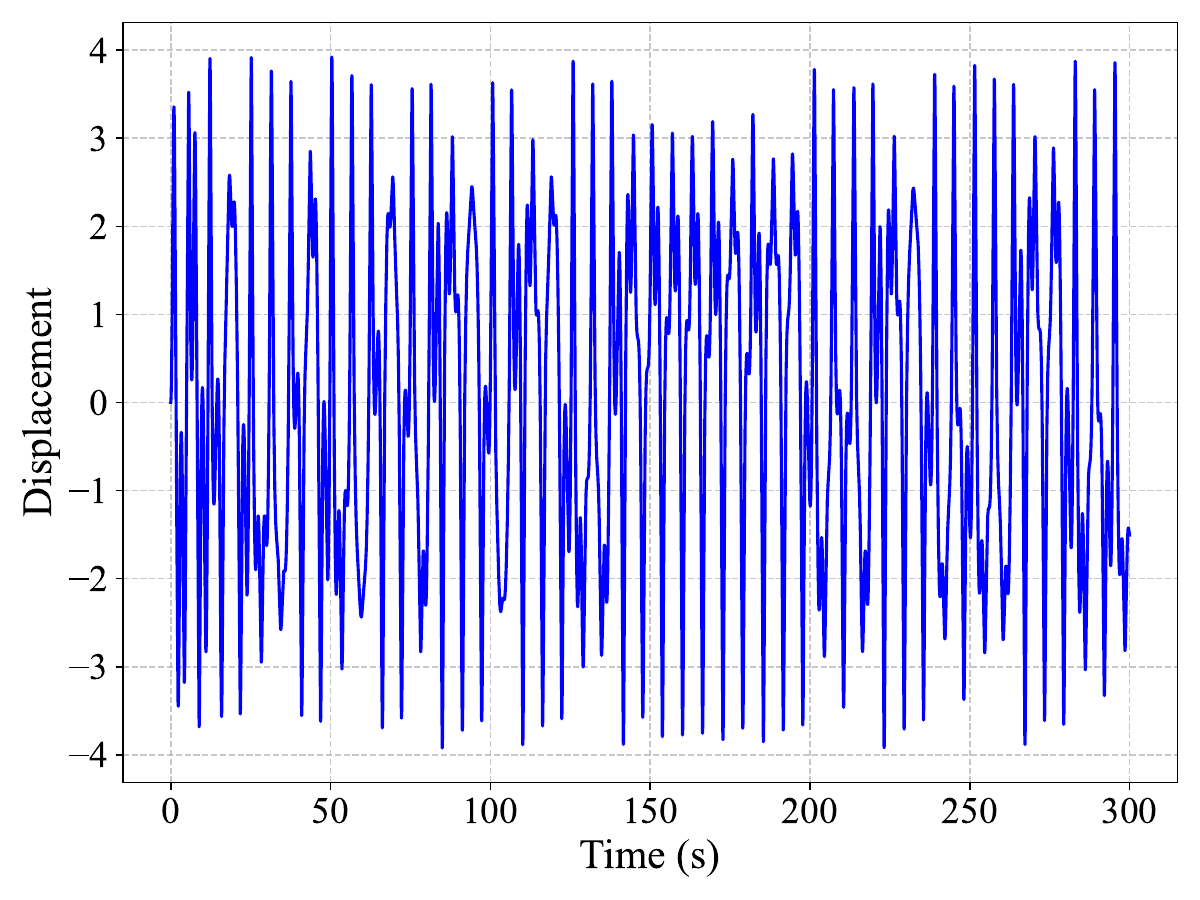}
        \caption{Displacement \( x_1(t) \)}
    \end{subfigure}
    \hspace{1em}% Space between figures
    \begin{subfigure}{0.29\textwidth}
        \centering
        \includegraphics[width=\textwidth]{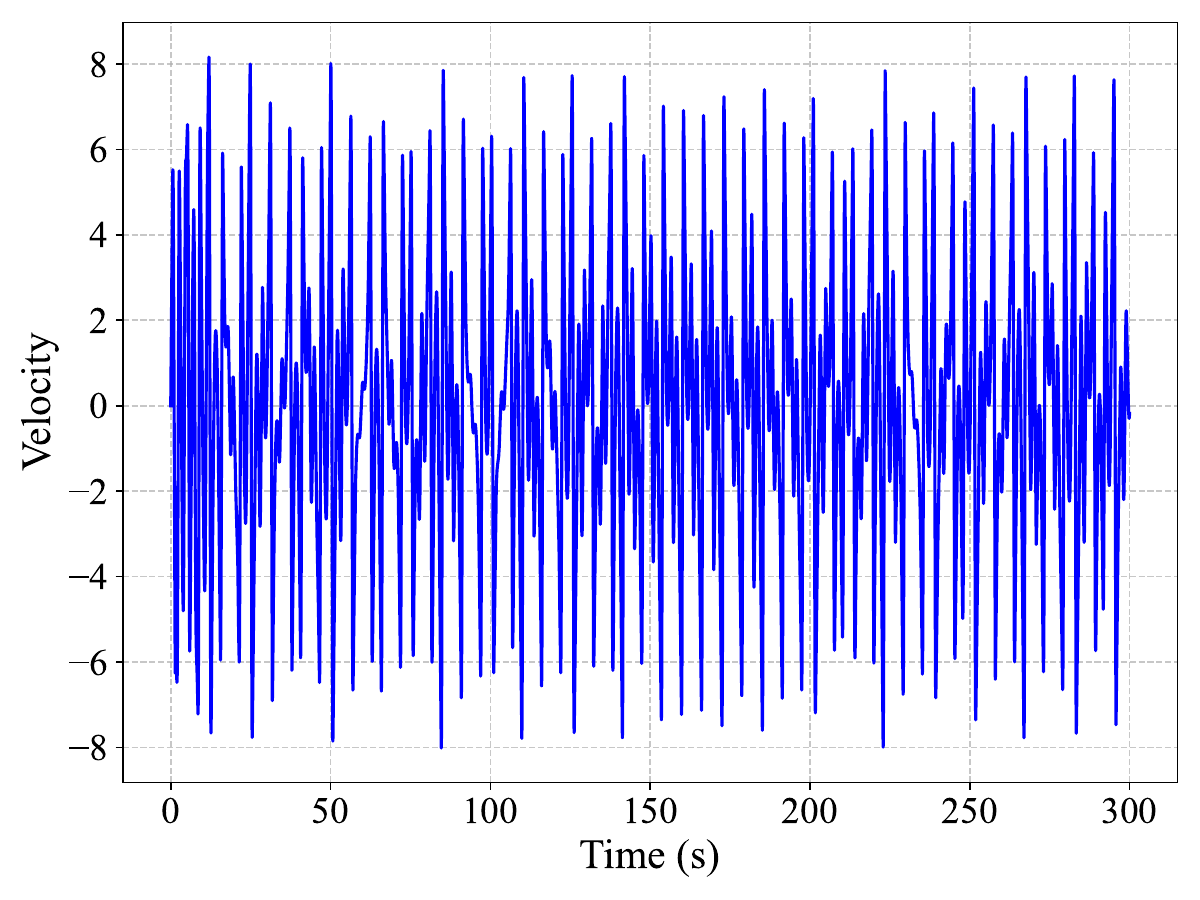}
        \caption{Velocity \( x_2(t) \)}
    \end{subfigure}
    \hspace{1em}% Space between figures
    \begin{subfigure}{0.29\textwidth}
        \centering
        \includegraphics[width=\textwidth]{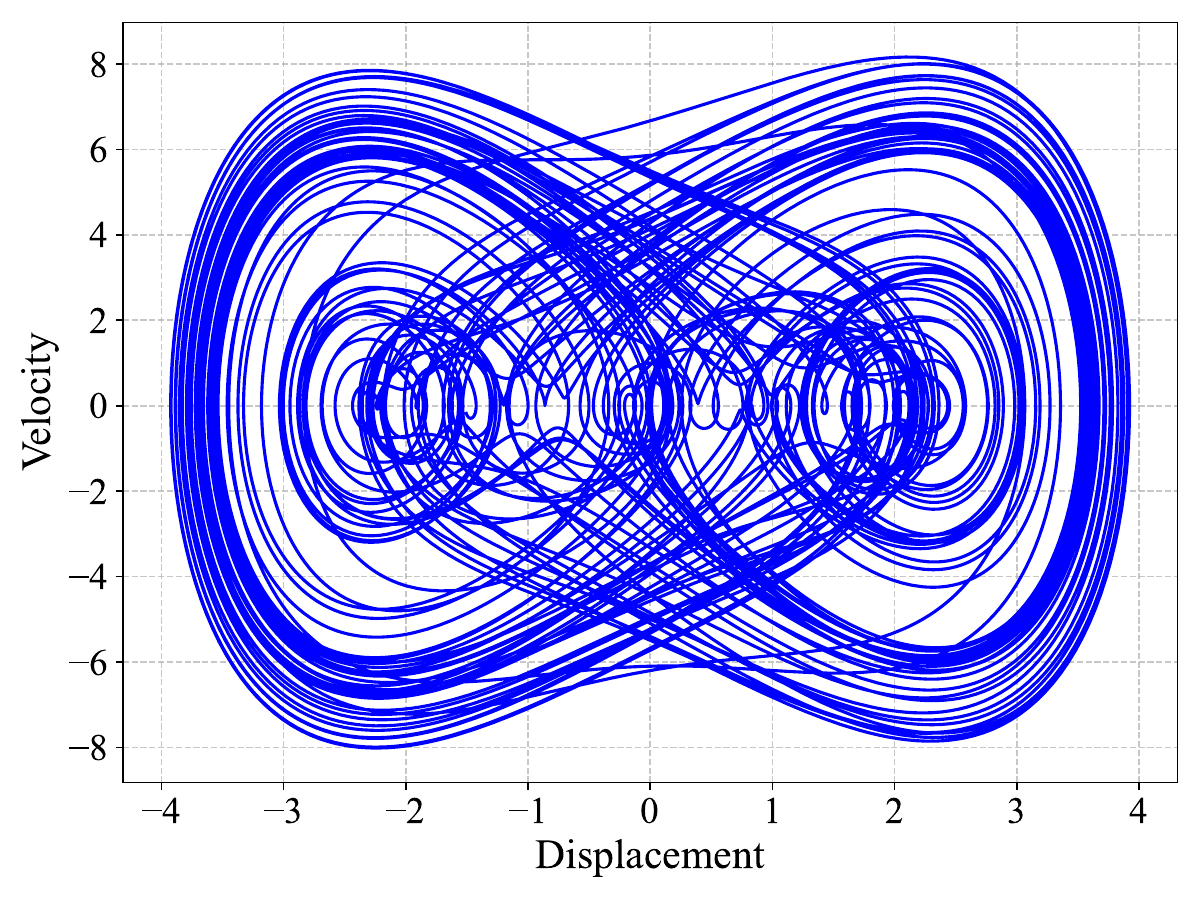}
        \caption{Phase space attractor}
    \end{subfigure}
    \caption{The displacement \( x_1(t) \), velocity \( x_2(t) \), and phase space attractor of the Duffing oscillator.}
    \label{fig:1}
\end{figure}

The Duffing oscillator is a classical nonlinear system that exhibits rich dynamic behaviors—including periodic motion, bifurcations, 
and chaos—depending on its parameters. It models a mass–spring–damper system with a nonlinear restoring force, governed by:
\begin{equation}\label{eqc_{15}}%\eqref{eqc_{15}} 
\ddot{x} + c \dot{x} + \alpha x + \beta x^3 = F \cos(\omega t)
\end{equation}
where \( c \) is the damping coefficient, \( \alpha \) and \( \beta \) are the linear and nonlinear stiffness parameters, \( F \) is 
the forcing amplitude, and \( \omega \) is its frequency.
Introducing the state vector \( \mathbf{x} = [x_1, x_2]^\top = [x, \dot{x}]^\top \), the system can be written in first-order state-space form:
\begin{equation}
    \mathbf{\dot{x}} =   f(\mathbf{x})   \Longleftrightarrow  \begin{bmatrix} \dot{x}_1 \\ \dot{x}_2 \end{bmatrix} = 
    \begin{bmatrix} x_2 \\  \alpha x_1 - \beta x_1^3 - c x_2 \end{bmatrix} +\begin{bmatrix}
        1 & 0 \\
        0 & 1
    \end{bmatrix} F \cos(\omega t)
\end{equation}
where \( f(\mathbf{x}) \) denotes the system’s nonlinear vector field.
For numerical validation, the parameters are set as:
\begin{equation}
    c = 0.1, \quad \alpha = 1, \quad \beta = 1, \quad F = 10, \quad \omega = 1
\end{equation}
These values ensure that the system exhibits both linear and nonlinear behaviors under the influence of periodic external forcing. 
With initial conditions \( x_1(0) = 0 \) and \( x_2(0) = 0 \), the system is integrated over 300 seconds with a time step of 0.01 using 
the \texttt{odeint} solver from the SciPy library. The resulting displacement \( x_1(t) \) and velocity \( x_2(t) \) are shown in Figure~\ref{fig:1}.

To validate the effectiveness and physical interpretability of the system matrix identified by the APSMC algorithm, we 
compare it with the theoretical Jacobian matrix of the Duffing system.
The Jacobian matrix \( J(\mathbf{x}) \) is defined as the partial derivative of the system's vector field \( f(\mathbf{x}) \) with respect to the state variables \( \mathbf{x} \):
\begin{equation}\label{eqc_{16}}%\eqref{eqc_{16}} 
    J(\mathbf{x}) = \begin{bmatrix} \frac{\partial f_1}{\partial x_1} & \frac{\partial f_1}{\partial x_2} \\ \frac{\partial f_2}{\partial x_1} & \frac{\partial f_2}{\partial x_2} \end{bmatrix} 
    = \begin{bmatrix}
        0 & 1 \\
        \alpha - 3\beta x_1^2 & -c
    \end{bmatrix} = \begin{bmatrix}
        0 & 1 \\
        1 - 3 x_1^2 & -0.1
    \end{bmatrix}
\end{equation}
Accordingly, the nonlinear system can be represented as a continuous time-varying linear system:
\begin{equation}
    \begin{bmatrix} \dot{x}_1 \\ \dot{x}_2 \end{bmatrix} = \begin{bmatrix}
        0 & 1 \\
        1 - 3 x_1^2 & -0.1
    \end{bmatrix} \begin{bmatrix} x_1 \\ x_2 \end{bmatrix} + \begin{bmatrix}
        1 & 0 \\
        0 & 1
    \end{bmatrix} F \cos(\omega t)
\end{equation}
Applying a discrete-to-continuous transformation (e.g., bilinear transformation or matrix exponentiation), the system can be expressed in discrete form as:
\begin{equation}
    x_{k+1} = J(x_k) \cdot x_k + B_k \cdot F \cos(\omega t_k) \Longleftrightarrow \mathbf{\dot{x}} = J(\mathbf{x}) \cdot \mathbf{x} + B \cdot F \cos(\omega t)
\end{equation}
where \( B \) is the identity matrix, and \( J(x_k) \), \( B_k \) denote the system matrices after discretization. Here, \( x_{k+1} \) and \( F \cos(\omega t_k)  \) are the 
discrete-time state and input vectors, respectively.

\subsubsection{Impact of Physical Constraints in APSMC}
\begin{figure}[!ht]
    \centering
    \vspace{1em} % Space between the two rows of figures
    \begin{minipage}{\textwidth}
        \centering
        \begin{subfigure}{0.405\textwidth}
            \centering
            \includegraphics[width=\textwidth]{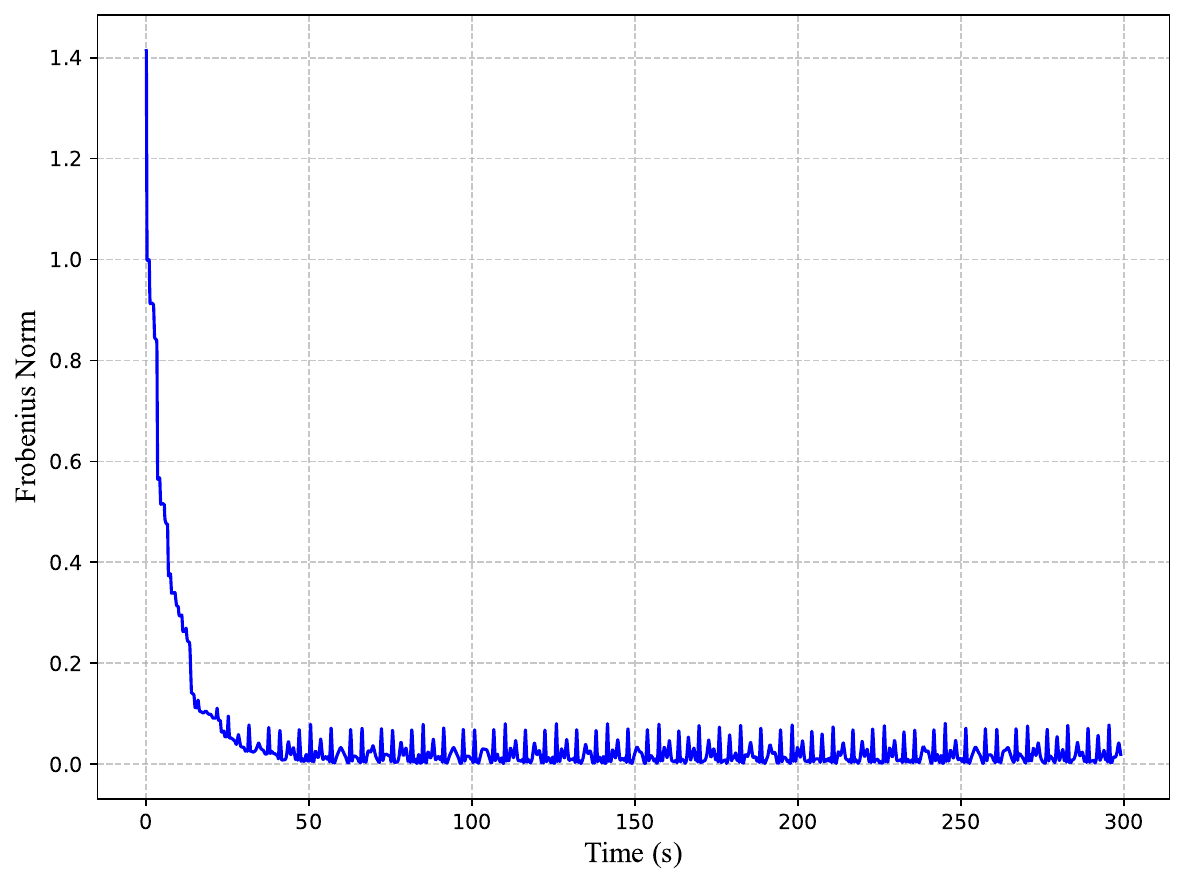}
            \caption{$A_0=0$}
        \end{subfigure}
        \hspace{2em} % Space between figures
        \begin{subfigure}{0.405\textwidth}
            \centering
            \includegraphics[width=\textwidth]{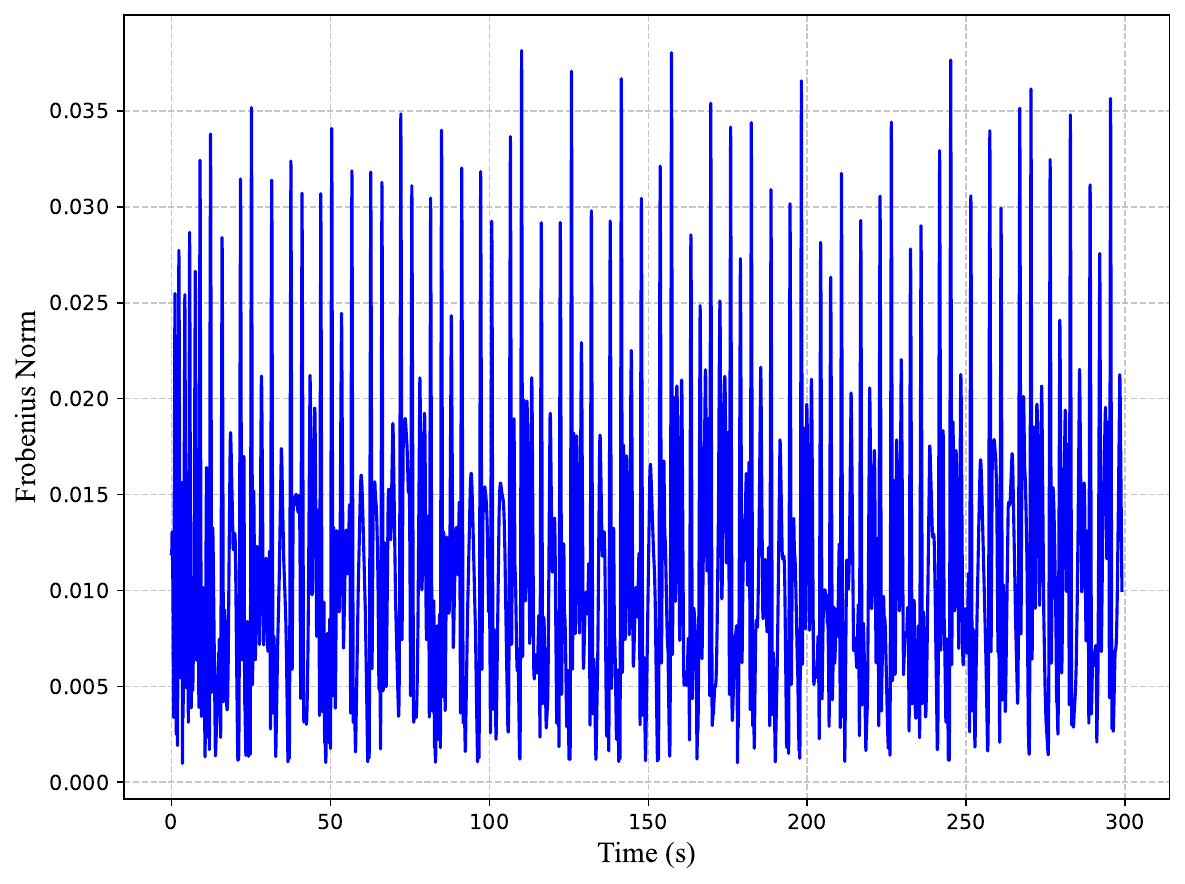}
            \caption{$A_0= J(x_0)$}
        \end{subfigure}
    \end{minipage}
    \caption{The Frobenius norm of the difference between \( A_k \) and \( J(x_k) \)}
    \label{fig1} % Figure label
  \end{figure}

This section investigates the relationship between the system matrix \( A_k \) identified by the APSMC algorithm and the theoretical Jacobian matrix \( J(x_k) \) of the nonlinear system.
 Therefore, to minimize interference, no noise was added to the data, and the full observational data were used for time-varying system identification.
First, consider the case where \( B_k \) is known. In this case, \( B_k \cdot F \cos(\omega t_k) \) can be directly treated as a known term in the algorithm for 
solving the system. Figure~\ref{fig1} compares the Frobenius norm of \( A_k \) and \( J(x_k) \) at each time step.

  In Figure \ref{fig1}(a), the matrix $A_0$ is initialized as a zero matrix and is iteratively updated based on the data. As a result, 
the error is initially large, but over time, with the gradual improvement of the system matrix \( A_k \), it eventually converges to the true Jacobian 
matrix. In Figure \ref{fig1}(b), the initial matrix \( A_0 \) is set to the Jacobian matrix \( J(x_0) \) at the initial time step, making the 
initial error zero. Subsequently, due to the nature of the online algorithm, the error fluctuates within a small range, but \( A_k \) consistently follows the changes in \( J(x_0) \).

Next, consider the case where $u_k$ is known but $B_k$ is unknown. In this scenario, the objective is to estimate the optimal values of $A_k$ and $B_k$ from 
the data. It is important to note that when solving the convex optimization problem \eqref{eqc_{13}}, insufficient excitation—such as single-frequency 
forcing $F \cos(\omega t)$—may lead to column correlation in $\Omega_k$, resulting in a non-unique solution.

\begin{figure}[!ht]
    \centering
    \vspace{1em} % Space between the two rows of figures
    \begin{minipage}{\textwidth}
        \centering
        \begin{subfigure}{0.405\textwidth}
            \centering
            \includegraphics[width=\textwidth]{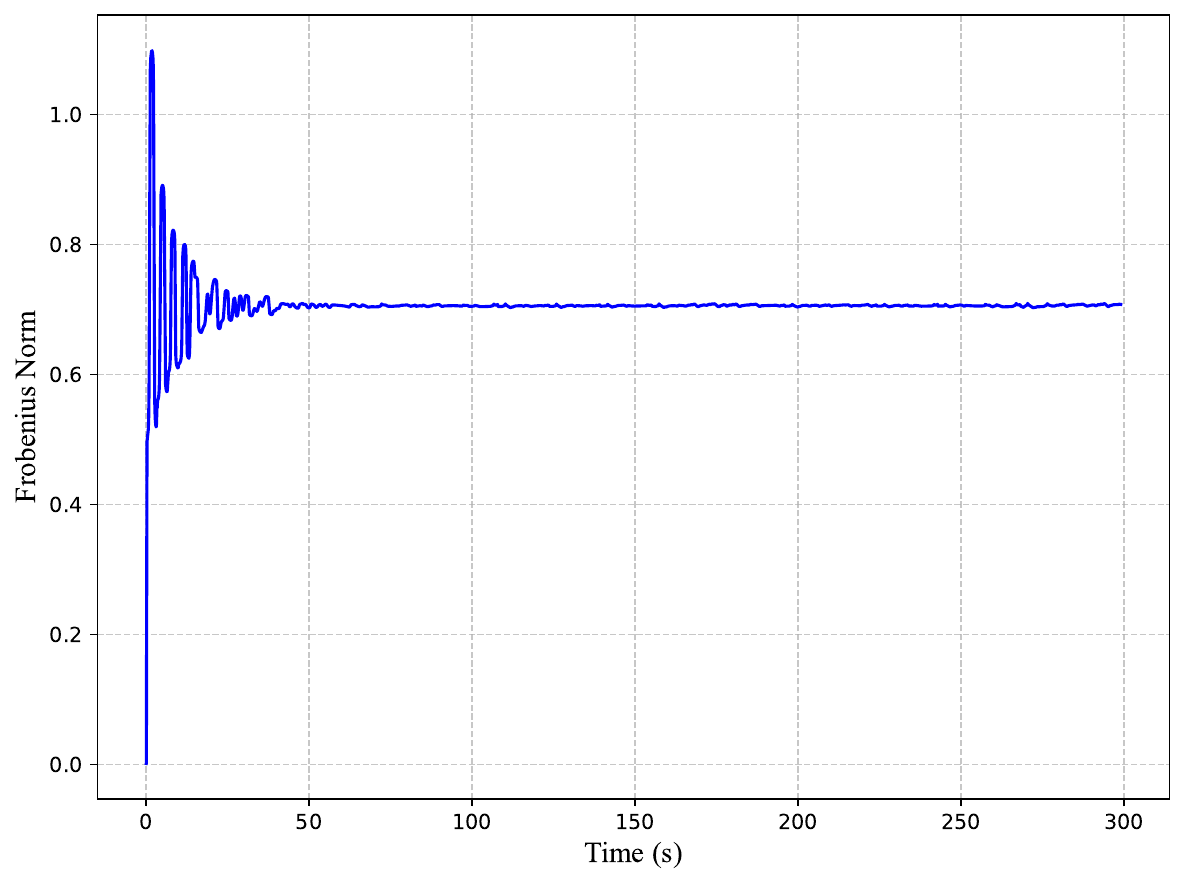}
            \caption{$A_0=I$}
        \end{subfigure}
        \hspace{2em} % Space between figures
        \begin{subfigure}{0.405\textwidth}
            \centering
            \includegraphics[width=\textwidth]{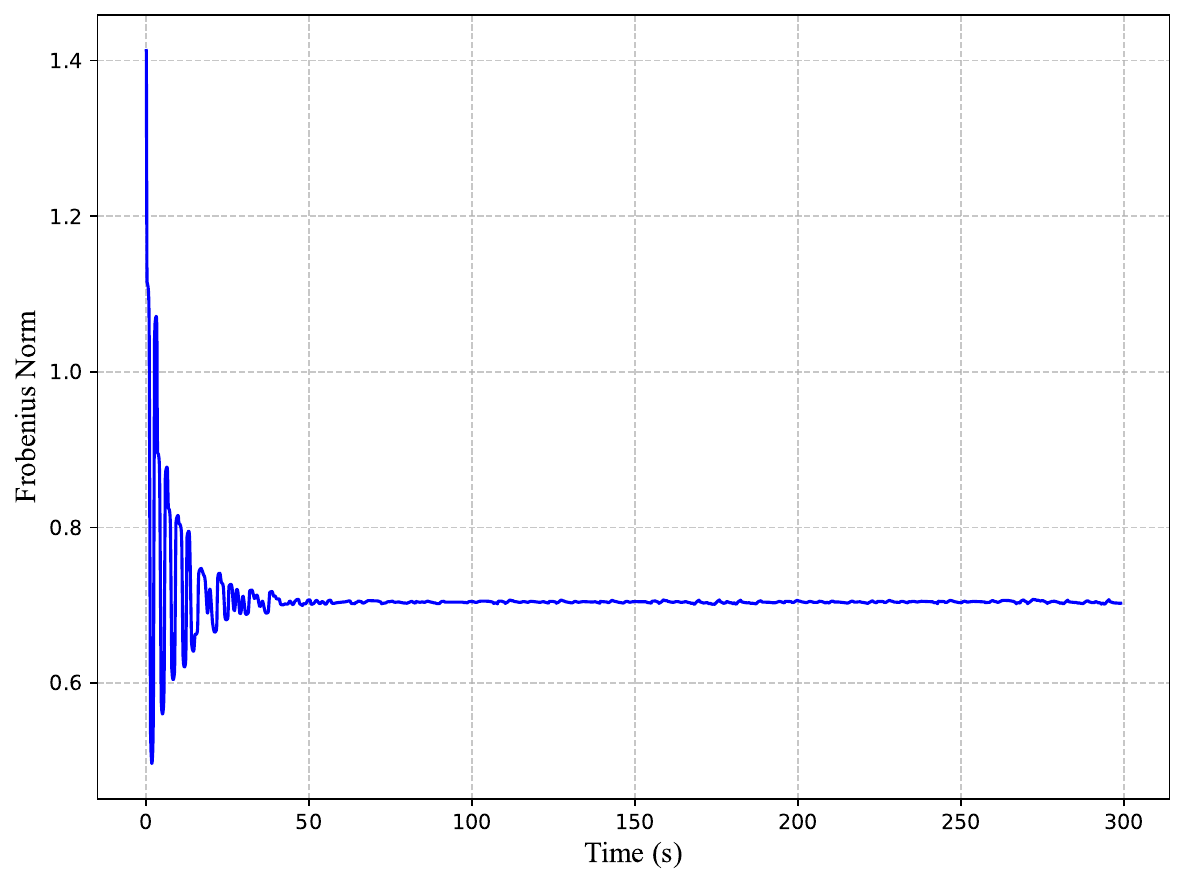}
            \caption{$B_0= I$}
        \end{subfigure}
    \end{minipage}

    \vspace{1em}% Space between the two rows of figures

    \begin{minipage}{\textwidth}
        \centering
        \begin{subfigure}{0.405\textwidth}
            \centering
            \includegraphics[width=\textwidth]{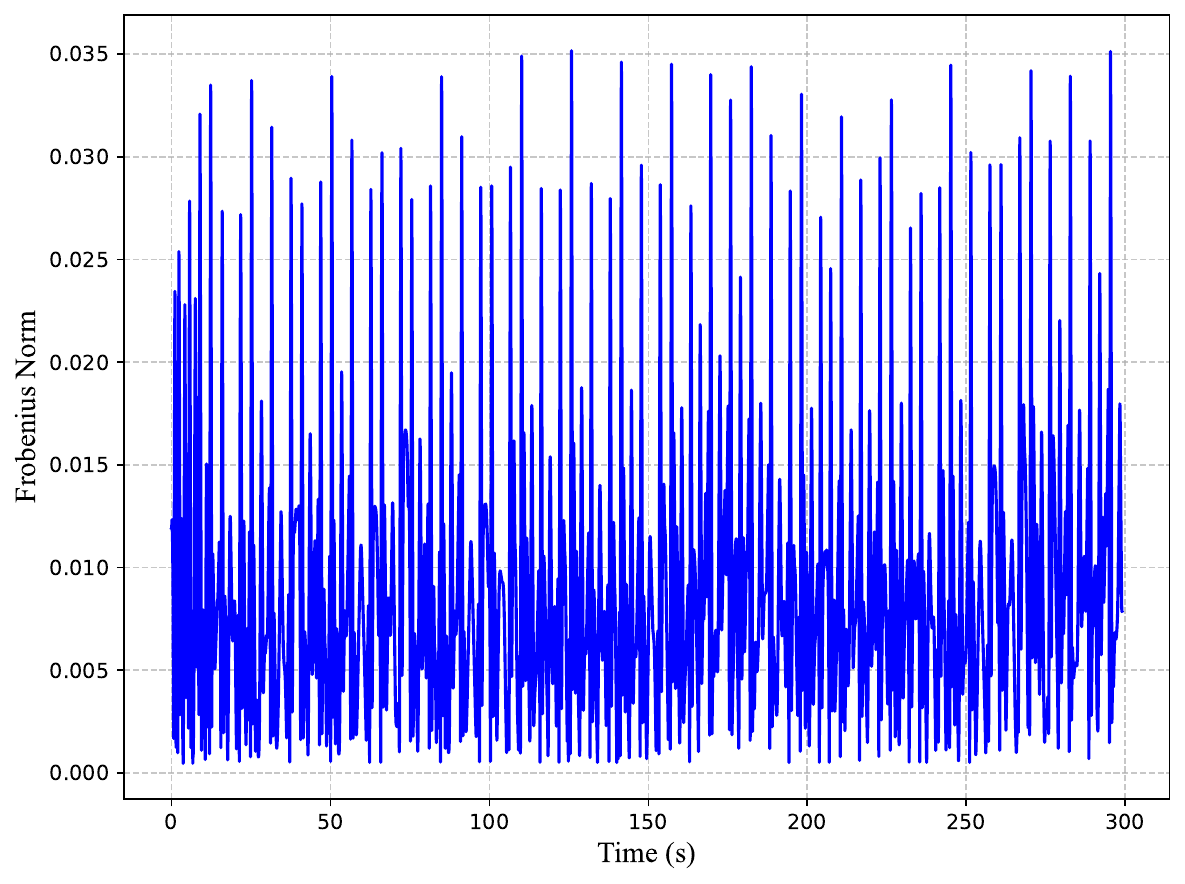}
            \caption{$A_0= J(x_0)$}
        \end{subfigure}
        \hspace{2em} % Space between figures
        \begin{subfigure}{0.405\textwidth}
            \centering
            \includegraphics[width=\textwidth]{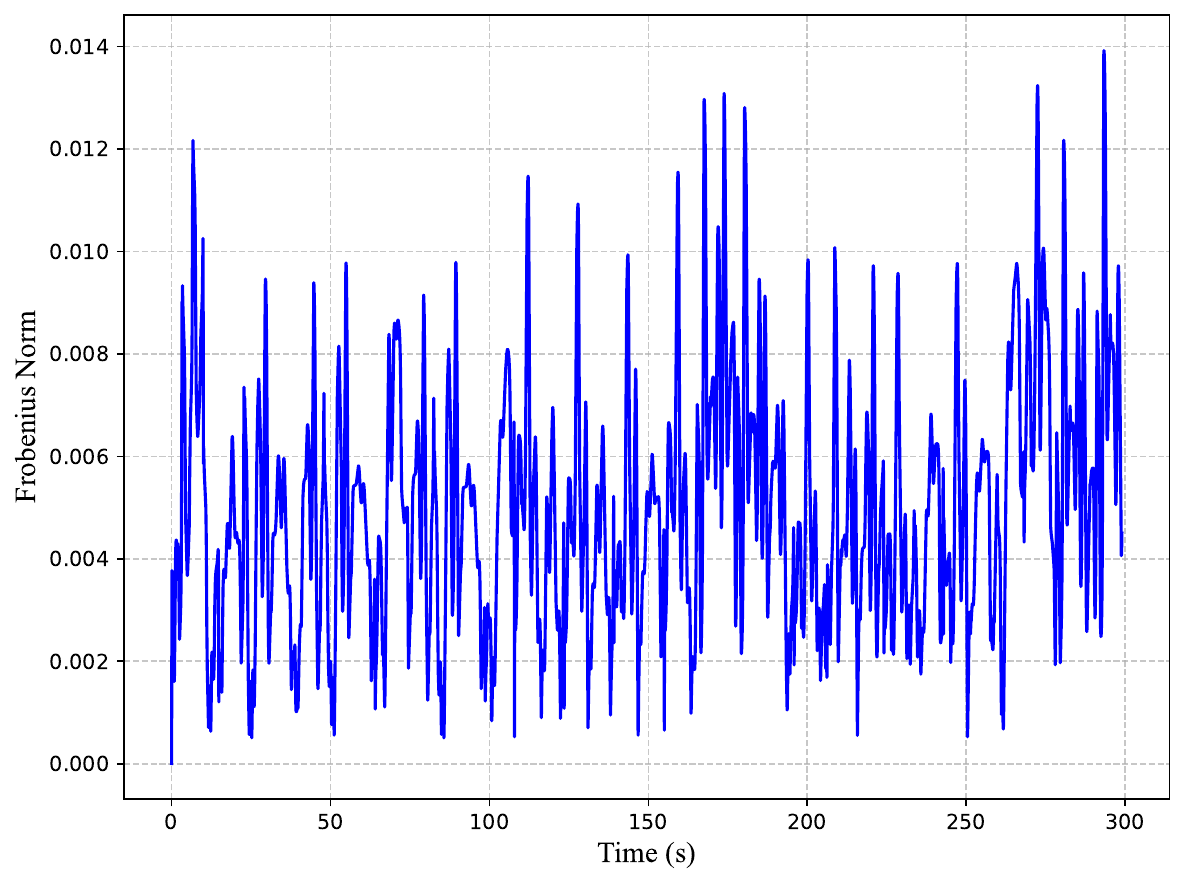}
            \caption{$B_0$ is the theoretical $B_0$ value}
        \end{subfigure}
    \end{minipage}

    \caption{Comparison of the Frobenius norm of \( A_k \) and \( B_k \) under different initial values}
    \label{fig3} % Figure label
  \end{figure}

As a result, purely data-driven identification of \( A_k \) may fail to converge to the true Jacobian \( J(x_k) \), and the solution can be sensitive to the 
choice of initial matrix \( A_0 \). Nevertheless, 
the residual between the estimated and measured states \( \hat{x}_k \) and \( x_k \) may still remain small. 
Figure \ref{fig3} shows the comparison of the Frobenius norm of \( A_k \) and \( B_k \) with the true values.

  For comparison, the initial values are set either as zero matrices or as the true matrices. The resulting prediction errors, quantified by the normalized 
  mean squared error (NMSE) between the estimated state \( \hat{x}_k \) and the measured state \( x_k \), are \(6.05 \times 10^{-5}\) and \(8.35 \times 10^{-8}\), respectively.
  The NMSE is employed as the evaluation metric and is defined as:
  \begin{equation}
      \text{NMSE} = \frac{ \sum_{i=1}^{n} (y_i - \hat{y}_i)^2}{ \sum_{i=1}^{n} (y_i - \bar{y})^2}, \quad   \bar{y} = \frac{1}{n} \sum_{i=1}^{n} y_i
  \end{equation}
  where \(y_i\) denotes the true value, \(\hat{y}_i\) represents the predicted value, \(\bar{y}\) is the mean of the true values, and \(n\) is the total number of samples.
  These results demonstrate that the APSMC algorithm achieves low prediction error even when initialized with zero matrices, despite exhibiting a larger deviation in 
  the Frobenius norm between \(A_k\) and the true Jacobian.

\begin{figure}[!ht]
    \centering
    \vspace{1em} % Space between the two rows of figures
    \begin{minipage}{\textwidth}
        \centering
        \begin{subfigure}{0.405\textwidth}
            \centering
            \includegraphics[width=\textwidth]{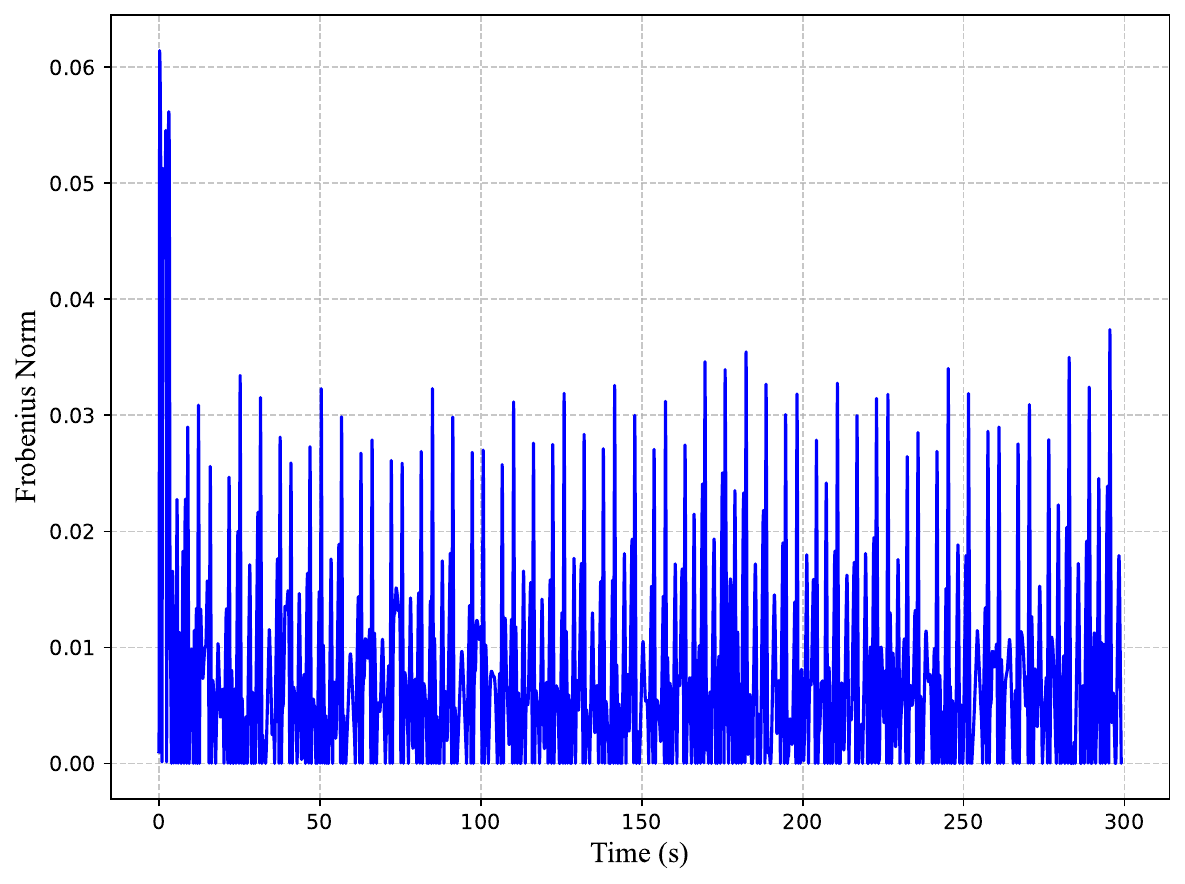}
            \caption{$A_0=I$}
        \end{subfigure}
        \hspace{2em} % Space between figures
        \begin{subfigure}{0.405\textwidth}
            \centering
            \includegraphics[width=\textwidth]{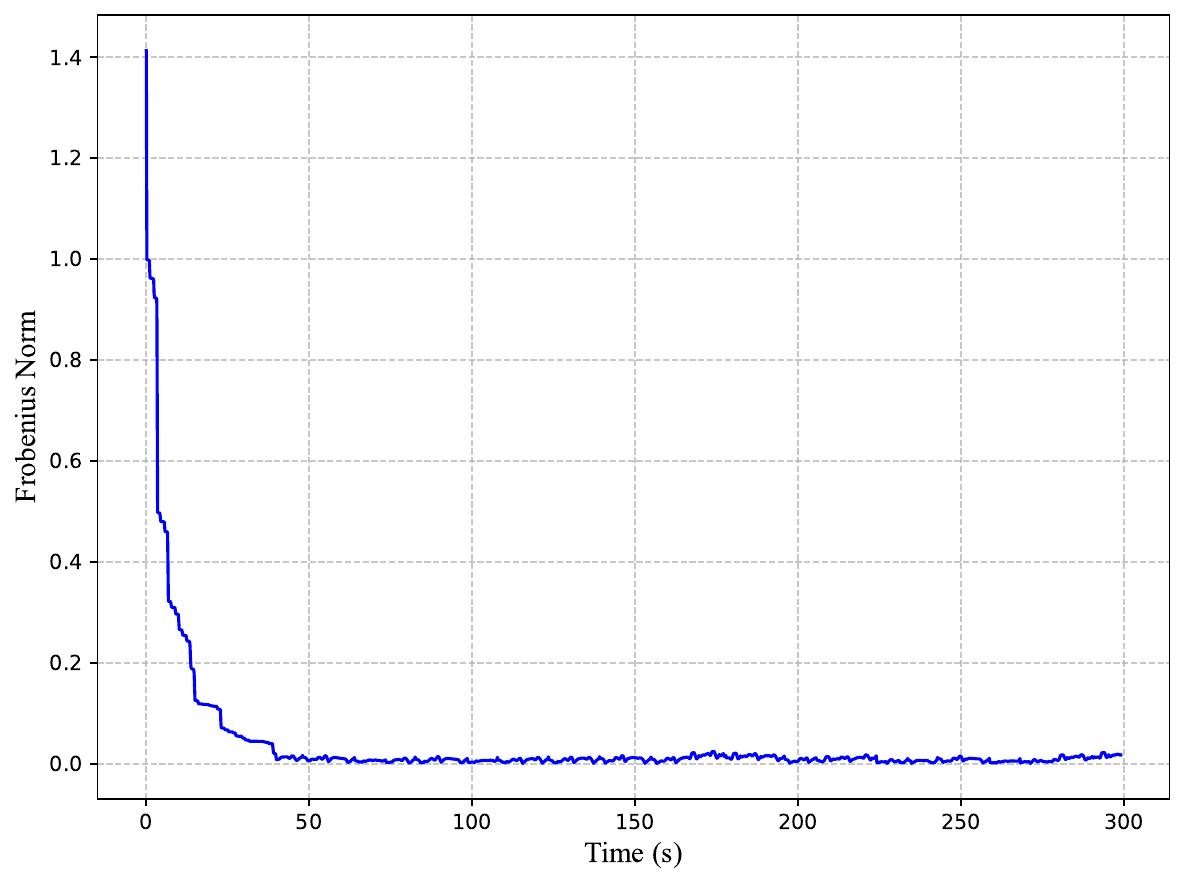}
            \caption{$B_0= I$}
        \end{subfigure}
    \end{minipage}
    \caption{Frobenius norm errors of \( A_k \) and \( B_k \) relative to ground truth under physical constraints}
    \label{fig4} % Figure label
  \end{figure}

  While purely data-driven methods may struggle to recover an \( A_k \) consistent with the true Jacobian \( J(x_k) \), incorporating physical 
  constraints can guide the solution toward one that aligns with it. Assuming the structure of equation~\eqref{eqc_{15}} is known, the form of the 
  Jacobian in equation~\eqref{eqc_{16}} implies that the continuous-time matrix \( A_c^k \), transformed from \( A_k \), should satisfy:
\begin{equation}\label{eqc_{161}}%\eqref{eqc_{161}} 
    A_c^{k} \in \begin{bmatrix} 
        0 & 1 \\ 
        \text{Unknown} & -0.1 
        \end{bmatrix}
\end{equation}
Here, \( A_c^{k} \) represents the continuous matrix corresponding to \( A_k \) at the \( k \)-th sampling time. In this study, we adopt the bilinear transformation 
form to achieve the following:
\begin{equation}
    A_c^{k} = \frac{2}{\Delta t}  \cdot \left( I + A_k \right)^{-1} \cdot \left( A_k - I \right)
\end{equation}
We still start with an initial zero matrix and run the APSMC algorithm. In this case, the changes in the Frobenius norm of \( A_k \) and \( B_k \) relative to the 
true values over time are shown in Figure \ref{fig4}.

  With the incorporation of physical constraints, the error between \( A_k \) and \( J(x_k) \) is significantly reduced. Since 
  the constraints are applied only to \( A_k \), the initial error in \( B_k \) remains large but gradually decreases and converges 
  with minor fluctuations. The NMSE between \( \hat{x}_k \) and \( x_k \) also improves, 
  decreasing to \( 1.22 \times 10^{-5} \), compared to the unconstrained case.

\subsection{Seismic Response Analysis of a Frame Structure}

\subsubsection{Structural Information and Seismic Response Analysis}

\begin{figure}[h]
    \centering
    \includegraphics[width=0.4\textwidth]{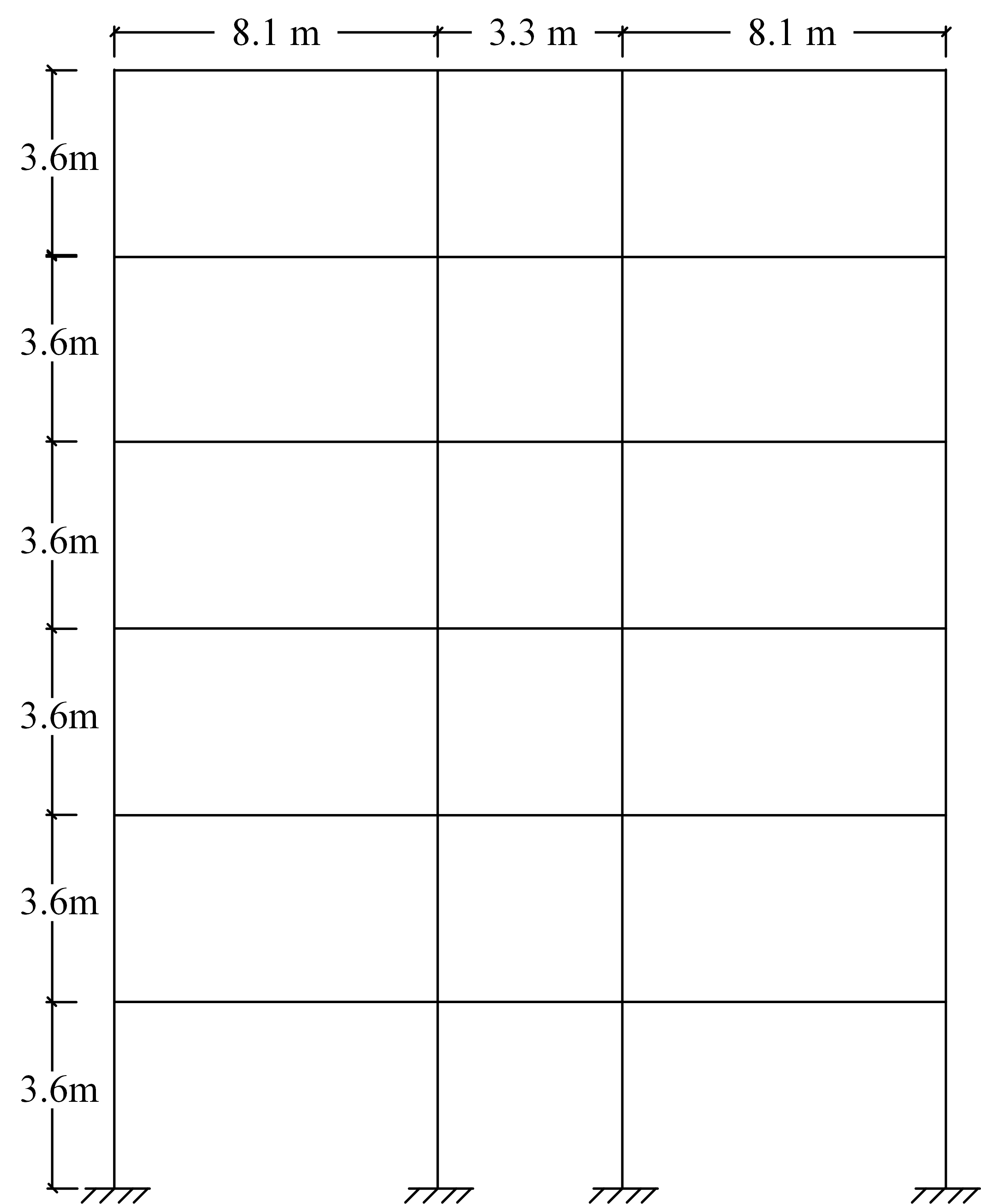}
    \caption{Two-dimensional computational diagram}\label{fig:ssk}
\end{figure}

The example structure, built in 2010 for office use, has a seismic design intensity of 7 degrees, corresponding to a PGA of 0.10g for a 10\% exceedance probability 
in 50 years. The building has six floors, each 3.6 meters high. A two-dimensional schematic of the structural model is shown in Figure~\ref{fig:ssk}. It consists 
of three spans in the transverse  direction: 3.3 meters for the middle span and 8.1 meters for each end span. Columns are 650 mm $\times$  650 mm, and beams are 250 mm $\times$  800 mm 
in cross-section. The materials used include C35 concrete and HRB400 reinforcement. Additional details can be found in~\cite{wen2022rapid}.

\begin{figure}[!ht]
    \centering
    \vspace{1em} % Space between the two rows of figures
    \begin{minipage}{\textwidth}
        \centering
        \begin{subfigure}{0.405\textwidth}
            \centering
            \includegraphics[width=\textwidth]{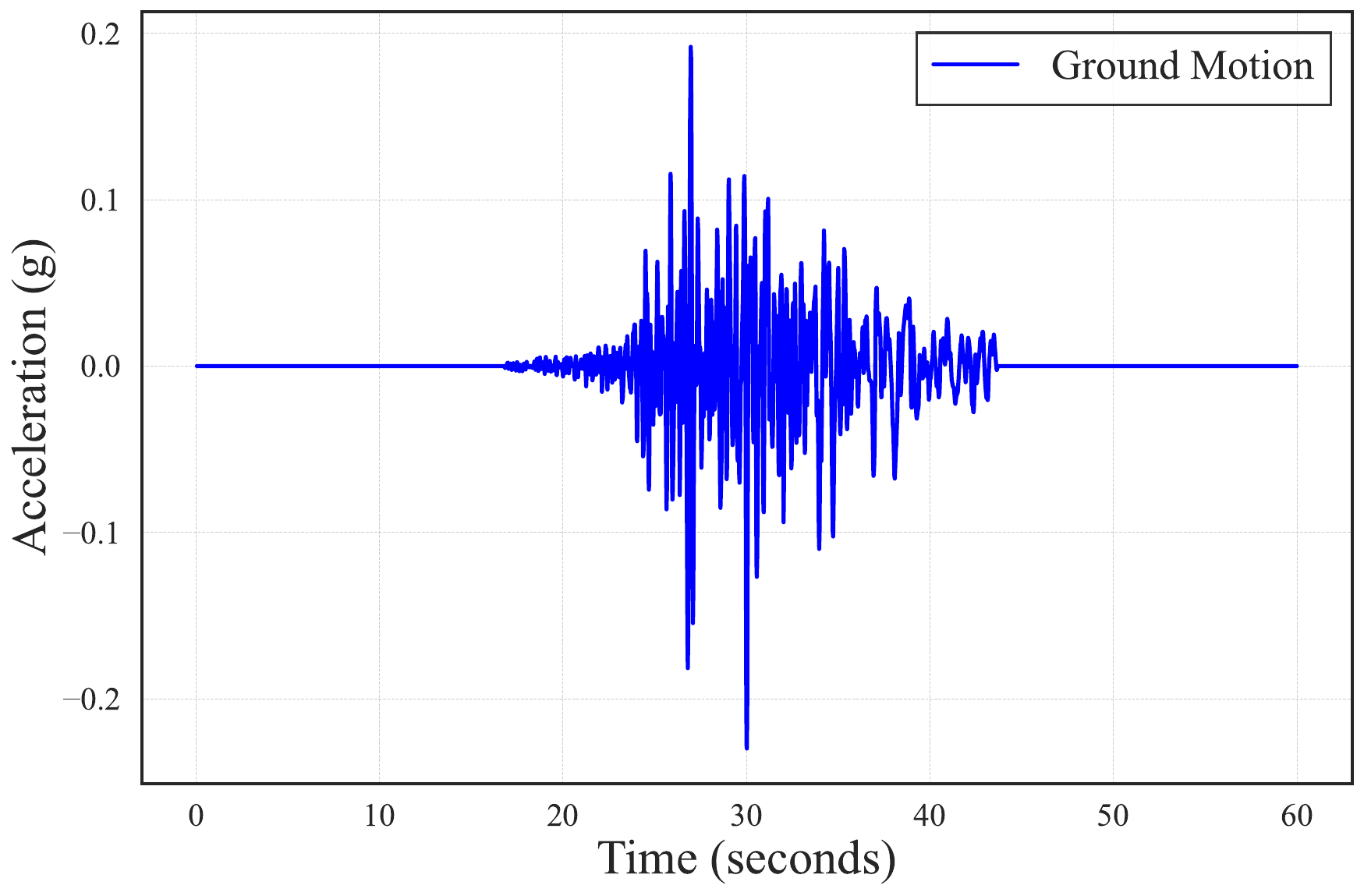}
            \caption{Time history}
        \end{subfigure}
        \hspace{2em} % Space between figures
        \begin{subfigure}{0.405\textwidth}
            \centering
            \includegraphics[width=\textwidth]{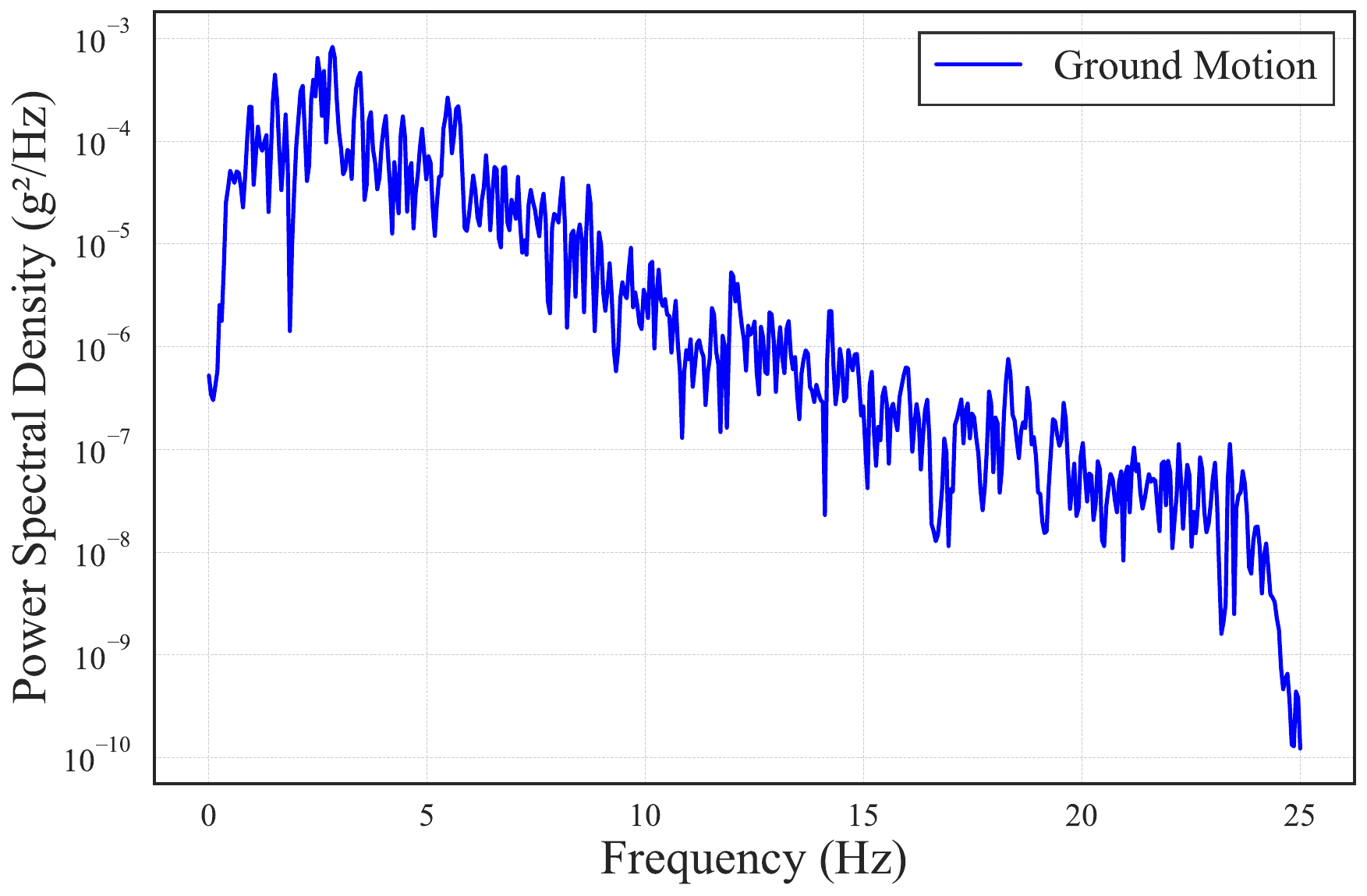}
            \caption{Power spectral density}
        \end{subfigure}
    \end{minipage}
    \caption{Time history and power spectral density of the ground motion.}
    \label{fig10} % Figure label
\end{figure}

The ground motion record from the 1994 Northridge earthquake, with Record Sequence Number 968, was selected from the PEER NGA WEST-2 
database \cite{baker2006pacific}. The recording station is 46.74 km from the rupture, and both the site and structure correspond to 
Chinese Site Class II. The acceleration time history and its power spectral density are shown in Figure~\ref{fig10}.

Finite element modeling is performed in OpenSees, with concrete and reinforcement simulated using the Concrete02 and Steel02 material models, respectively. Beam and column cross-sections are defined using fiber sections. All elements are modeled as nonlinear 
force-based beam-column elements, with three integration points per beam and five per column. Only columns include P–Delta effects; beams adopt linear transformation.

\begin{figure}[!ht]
    \centering
    \begin{subfigure}{0.31\textwidth}
        \centering
        \includegraphics[width=\textwidth]{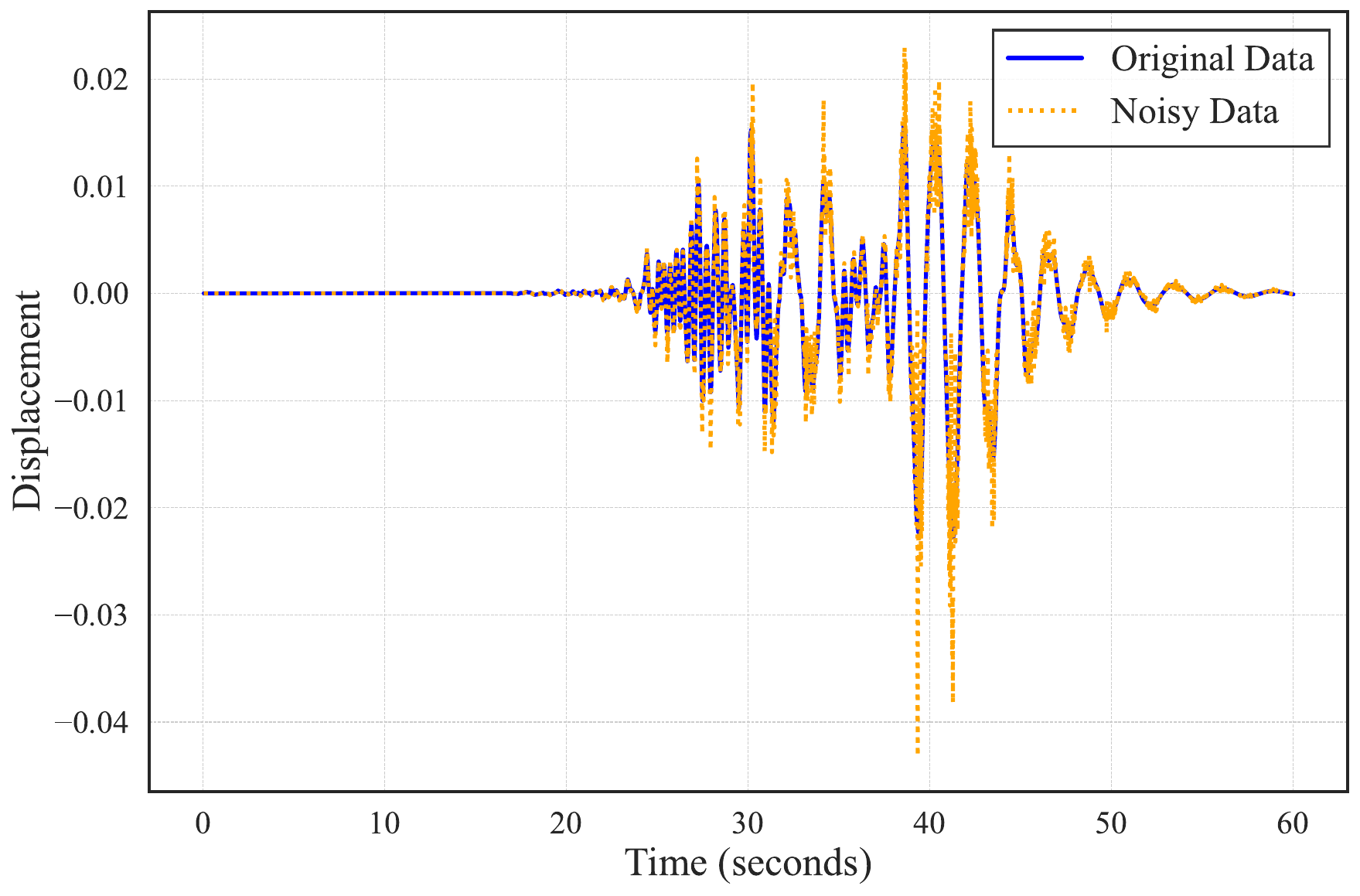}
        \caption{Displacement \( x_2(t) \)}
    \end{subfigure}
    \hspace{0.5em}% Space between figures
    \begin{subfigure}{0.31\textwidth}
        \centering
        \includegraphics[width=\textwidth]{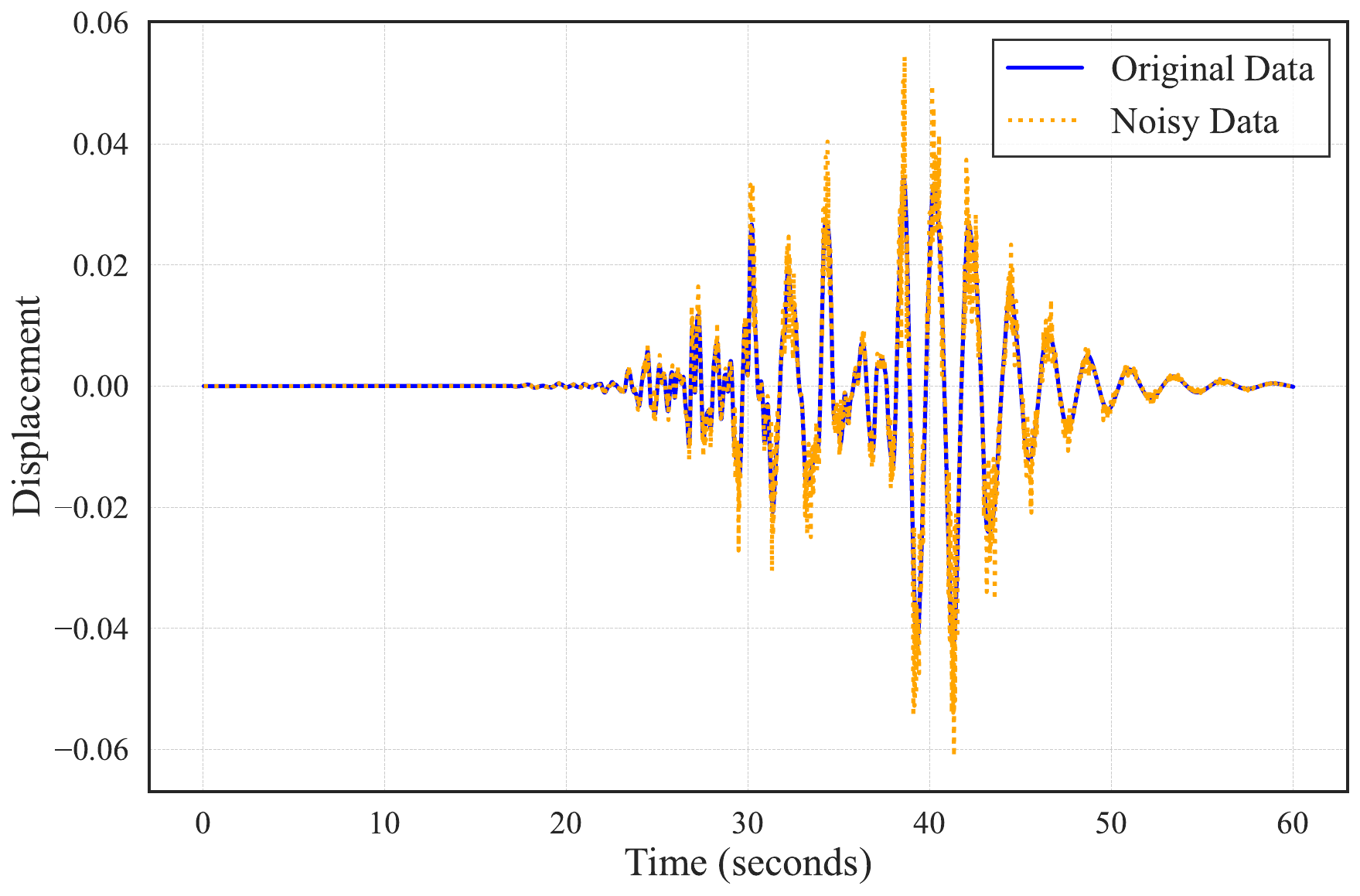}
        \caption{Displacement \( x_4(t) \)}
    \end{subfigure}
    \hspace{0.5em}% Space between figures
    \begin{subfigure}{0.31\textwidth}
        \centering
        \includegraphics[width=\textwidth]{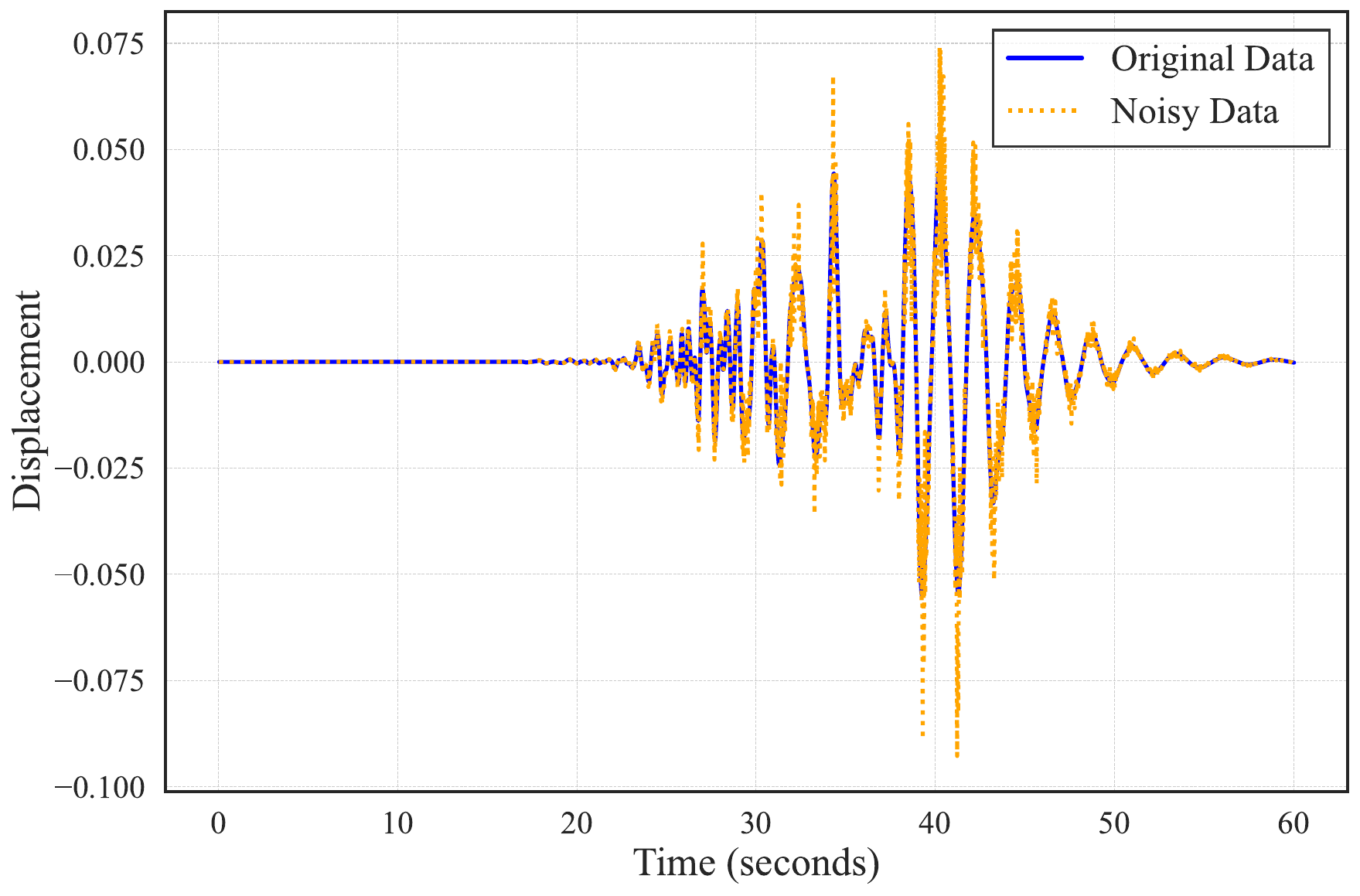}
        \caption{Displacement \( x_6(t) \)}
    \end{subfigure}
  
    \vspace{1em}% Space between the two rows of figures
    \begin{subfigure}{0.31\textwidth}
        \centering
        \includegraphics[width=\textwidth]{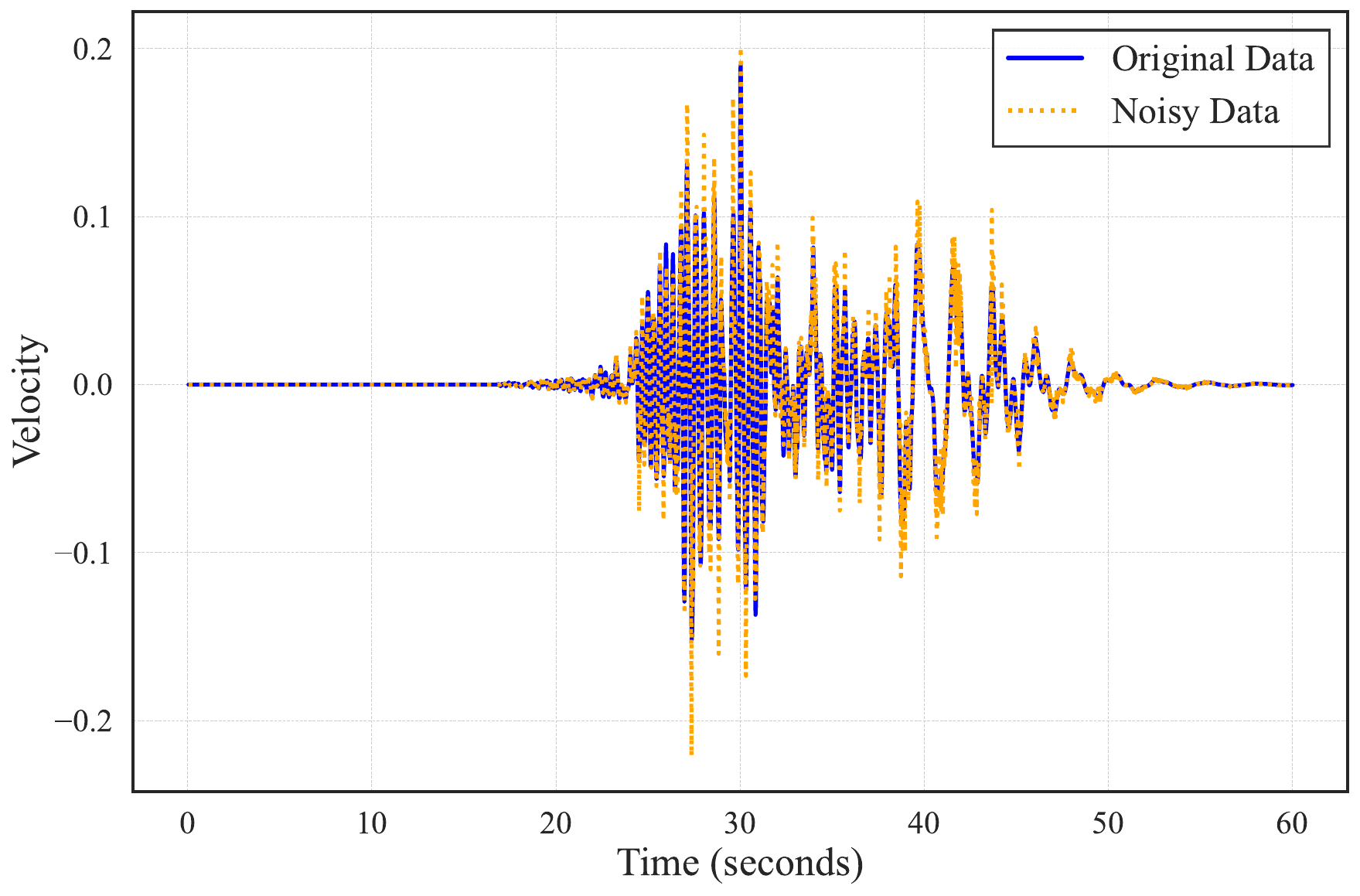}
        \caption{Velocity \( \dot{x}_2(t) \)}
    \end{subfigure}
    \hspace{0.5em}% Space between figures
    \begin{subfigure}{0.31\textwidth}
        \centering
        \includegraphics[width=\textwidth]{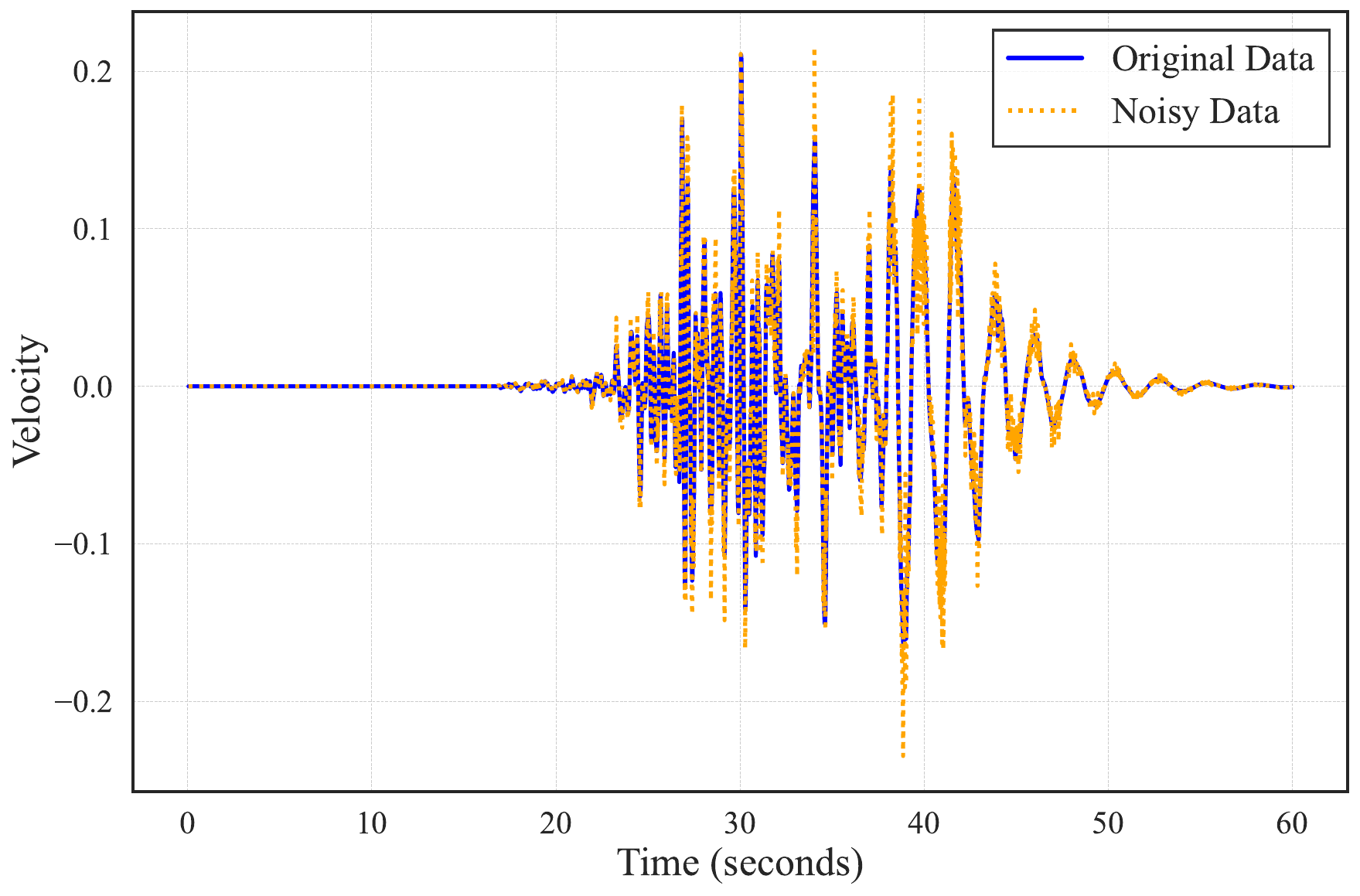}
        \caption{Velocity \( \dot{x}_4(t) \)}
    \end{subfigure}
    \hspace{0.5em}% Space between figures
    \begin{subfigure}{0.31\textwidth}
        \centering
        \includegraphics[width=\textwidth]{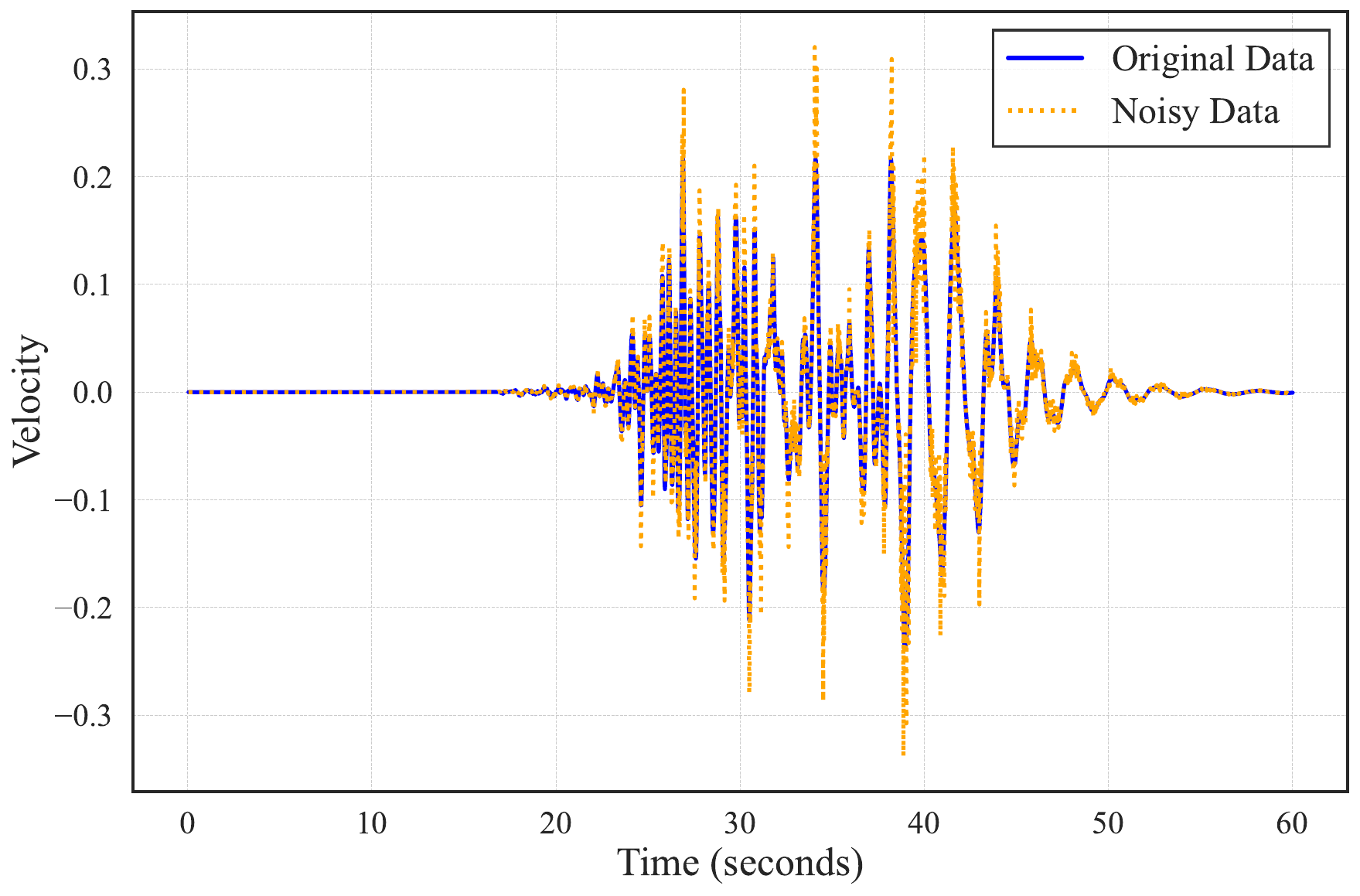}
        \caption{Velocity \( \dot{x}_6(t) \)}
    \end{subfigure}
    \caption{Time response of structural seismic response data with noise.}
    \label{fig:16}
\end{figure}

Rayleigh damping is used with a 5\% damping ratio for the first two modes. Nonlinear time-history analysis is performed using the Newmark 
average acceleration method~\cite{chopra2011dynamics}. To evaluate the robustness of APSMC to noise, structural response data are augmented 
with nonstationary white noise at 30\% of signal amplitude. Figure~\ref{fig:16} shows the 
resulting response distortions on the 2nd, 4th, and 6th floors, with similar effects observed on other floors.

\subsubsection{Impact of Physical Constraints in APSMC}

Considering that the duration of the ground motion is less than 60 seconds and taking into account the characteristics of the structural seismic response, this 
study selects the response data from 16 to 40 seconds for implementing the APSMC algorithm and other comparative methods. Given a time step of 0.02 seconds, this 
corresponds to a total of 1400 data points.

To emphasize the physical interpretability of the identified system matrix and facilitate the application of physical constraints, both displacement and velocity 
data are used as inputs. Under this configuration, the Dynamic Mode Decomposition with Control (DMDc) algorithm~\cite{proctorDynamicModeDecomposition2016} is 
adopted as a baseline. DMDc constructs an optimal linear operator that approximates the temporal evolution of the system, capturing its dominant dynamic behavior.

\begin{figure}[!ht]
    \centering
    \begin{subfigure}{0.295\textwidth}
        \centering
        \includegraphics[width=\textwidth]{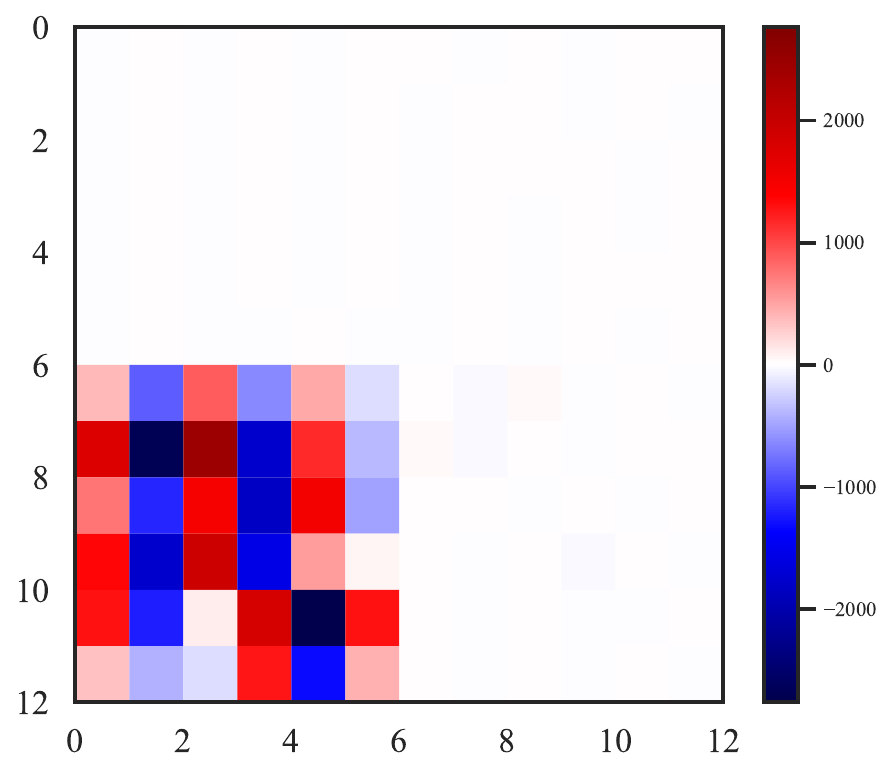}
        \caption{DMDc}
    \end{subfigure}
    \hspace{0.5em}
    \begin{subfigure}{0.29\textwidth}
        \centering
        \includegraphics[width=\textwidth]{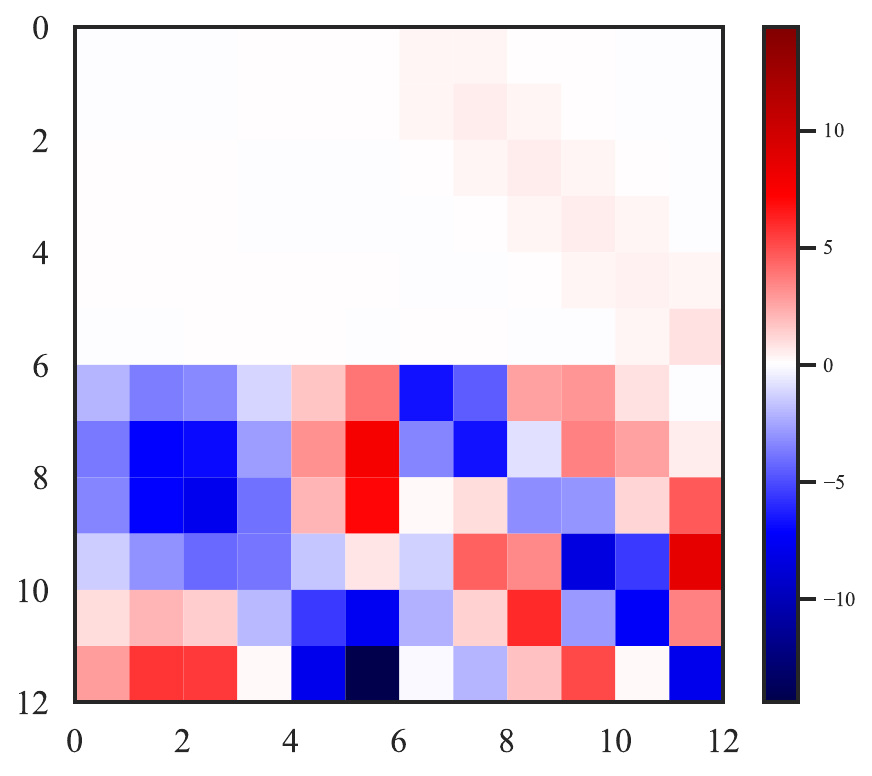}
        \caption{APSMC Unconstrained}
    \end{subfigure}
    \hspace{0.5em}
    \begin{subfigure}{0.29\textwidth}
        \centering
        \includegraphics[width=\textwidth]{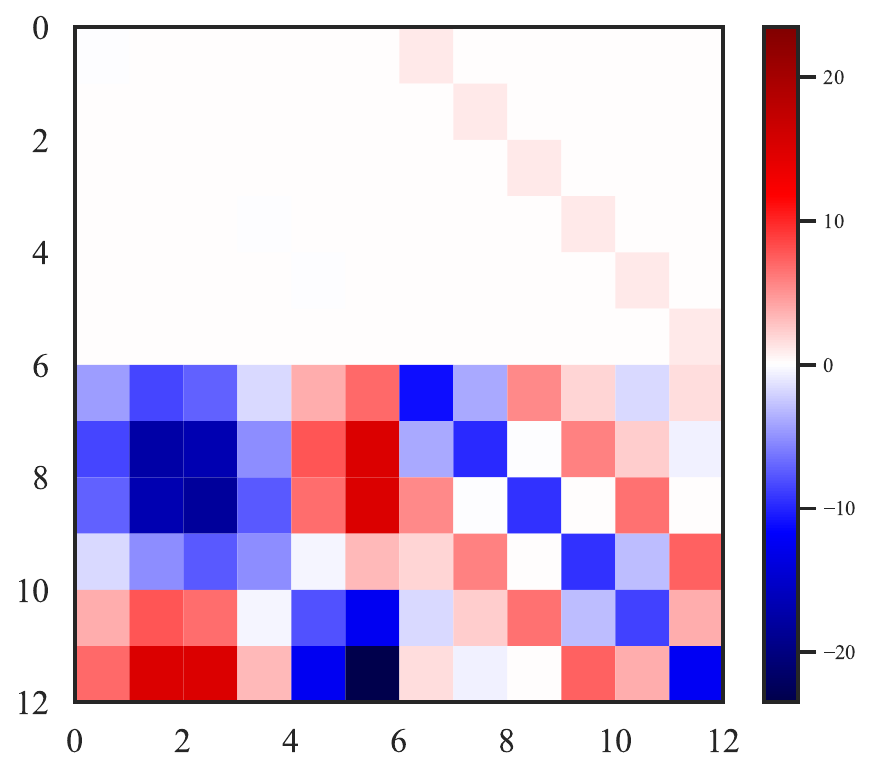}
        \caption{APSMC Constrained}
    \end{subfigure}
    \caption{System matrices identified from noise-free data}
    \label{fig:12}
\end{figure}

\begin{figure}[!ht]
    \centering
    \begin{subfigure}{0.29\textwidth}
        \centering
        \includegraphics[width=\textwidth]{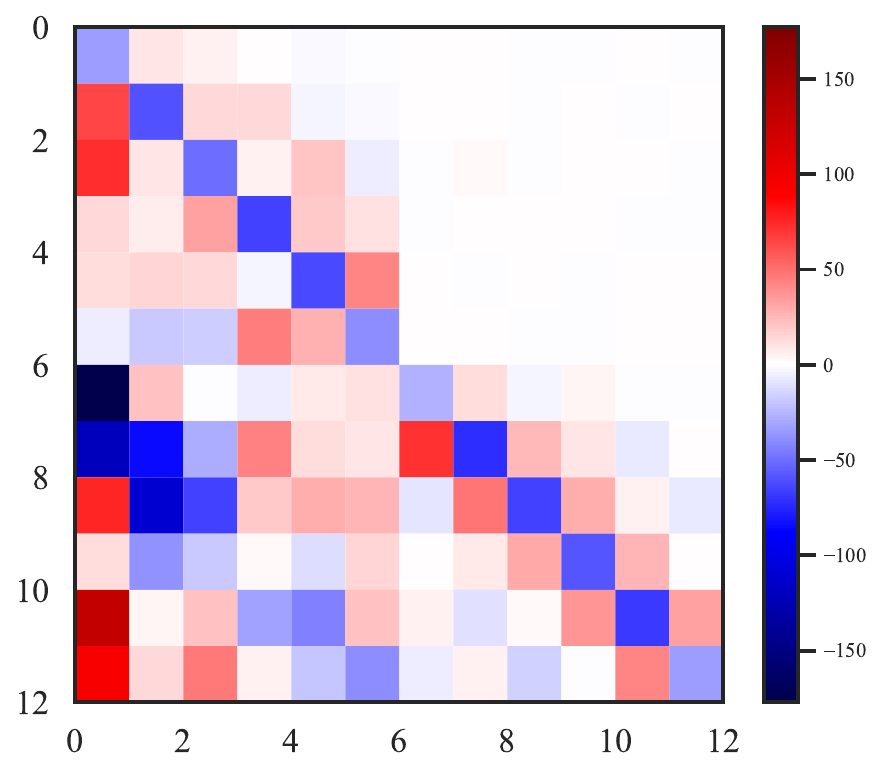}
        \caption{DMDc}
    \end{subfigure}
    \hspace{1em}% Space between figures
    \begin{subfigure}{0.29\textwidth}
        \centering
        \includegraphics[width=\textwidth]{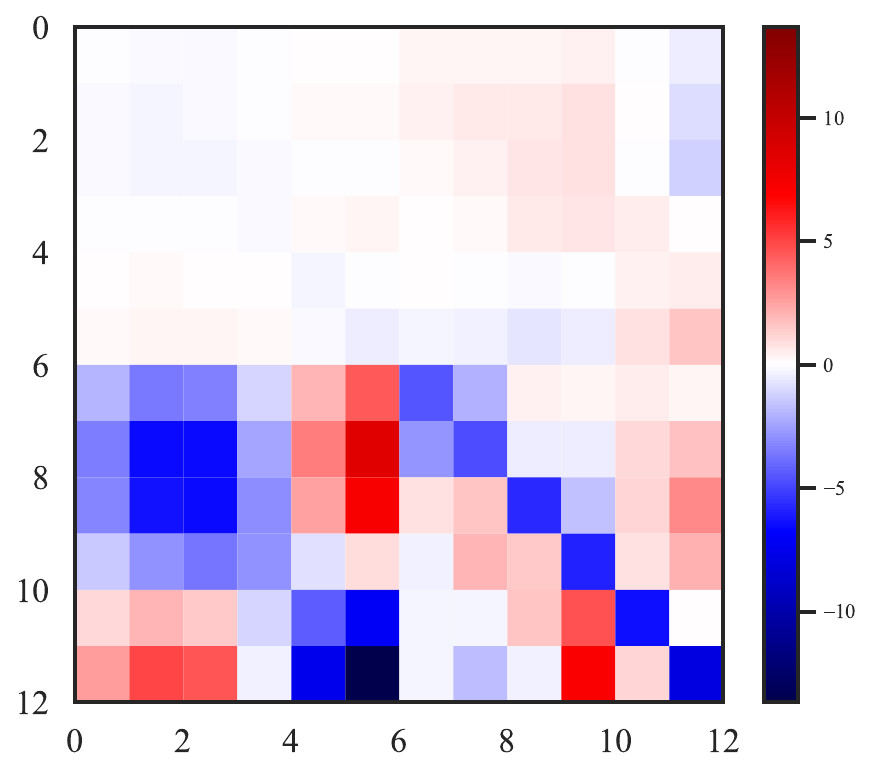}
        \caption{APSMC Unconstrained}
    \end{subfigure}
    \hspace{1em}% Space between figures
    \begin{subfigure}{0.29\textwidth}
        \centering
        \includegraphics[width=\textwidth]{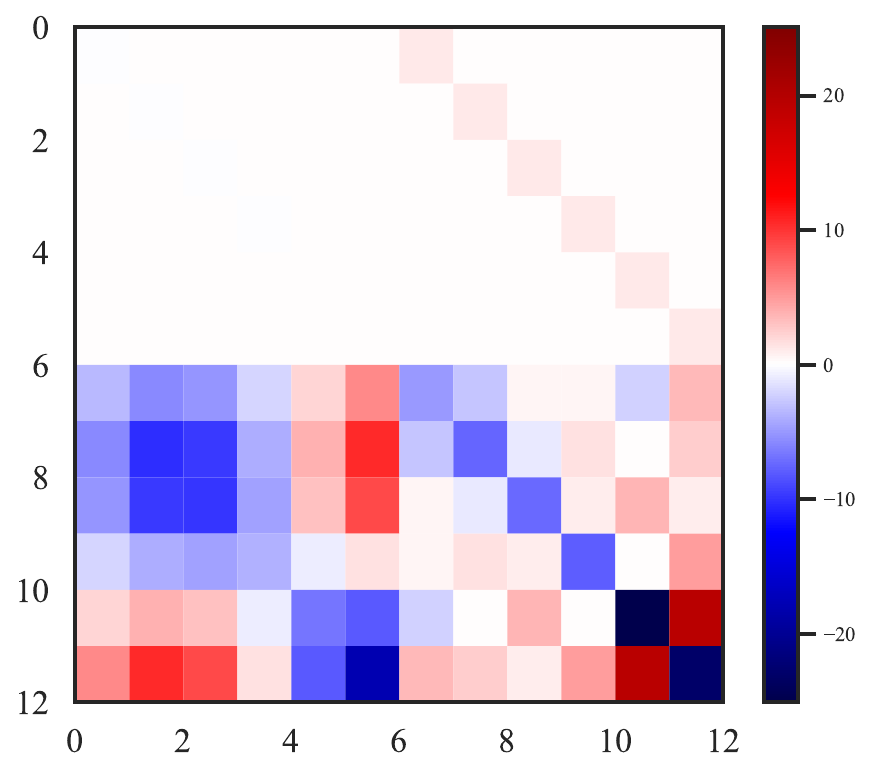}
        \caption{APSMC Constrained}
    \end{subfigure}
    \caption{System matrices identified from noisy data}
    \label{fig:14}
\end{figure}

Figures~\ref{fig:12} and \ref{fig:14} compare the identified system matrices obtained under noise-free and noisy conditions using three 
methods: (a) DMDc, (b) APSMC without physical constraints, and (c) APSMC with physical constraints.

To better interpret the system matrices presented in Figures~\ref{fig:12} and \ref{fig:14}, we first examine the 
theoretical structure they are expected to follow based on structural dynamics.
Under strong seismic excitation, structures exhibit nonlinear behavior due to material inelasticity and enter the elastic–plastic range. 
While this precludes a fully linear description of the global response, the system can still be approximated as linear within sufficiently 
short time intervals. Accordingly, the matrices identified by the APSMC algorithm are expected to conform to the following linear structural dynamic form:
\begin{equation}
    \begin{bmatrix}
      \ddot{q} \\
      \dot{q}
    \end{bmatrix} =
    \begin{bmatrix}
      0 & I \\
      -M^{-1}K & -M^{-1}C_1
    \end{bmatrix}
    \begin{bmatrix}
      \dot{q} \\
      q
    \end{bmatrix} +
    \begin{bmatrix}
      0 \\
      M^{-1}
    \end{bmatrix} F,
\end{equation}
where \(M \in \mathbb{R}^{n \times n}\) is the mass matrix, \(C_1 \in \mathbb{R}^{n \times n}\) is the damping matrix, \(K \in \mathbb{R}^{n \times n}\) is the 
stiffness matrix, \(q \in \mathbb{R}^n\) is the displacement vector, and \(F \in \mathbb{R}^n\) is the external force vector.
However, since the DMDc model seeks a globally optimal linear approximation based on the entire dataset, the resulting system matrix shown in Figure~\ref{fig:12}(a) 
does not exhibit the structural form expected from physical modeling.

In contrast, the APSMC algorithm adopts a time-varying linear model, enabling clearer structural interpretation, 
as shown in Figure~\ref{fig:12}(b). Theoretically, the system matrix should have a block structure: the upper-left \( \mathbb{R}^{6 \times 6} \) block as a 
zero matrix, the upper-right as an identity matrix, the lower-left as a symmetric matrix \( -M^{-1}K \), and the lower-right as a symmetric 
matrix \( -M^{-1}C_1 \). Figure~\ref{fig:12}(b) shows that the left two blocks closely match this theoretical structure.

However, the upper-right block deviates from an ideal identity matrix and instead exhibits a tridiagonal form, while the lower-right block lacks clear symmetry. 
 Theoretically, given the assumption 
of Rayleigh damping, the damping matrix \(C_1\) should be symmetric, and consequently, \(-M^{-1}C_1\) should also be symmetric. The imperfect correspondence in these blocks 
is likely because the contributions of the damping component and the identity matrix to the overall structural response are relatively minor, making them more difficult 
to accurately extract directly from the data.

Based on the above analysis, we impose a structural constraint on the discrete system matrix $A_k$ at each update step of the APSMC algorithm. Specifically, 
after applying the bilinear transformation, the resulting continuous-time system matrix $A_k^c$ is required to satisfy the following form:
\begin{equation}
    A_k^c \in \begin{bmatrix}
        0 & I \\
        \text{Symmetric} & \text{Symmetric}
    \end{bmatrix},\quad A_k^c = \frac{2}{\Delta t} \cdot \left( I + A_k \right)^{-1} \cdot \left( A_k - I \right)
\end{equation}
The constrained continuous-time matrix is then transformed back to discrete form, and the APSMC algorithm proceeds with proximal gradient updates. The resulting matrices 
are presented in Figures~\ref{fig:12}(c) and~\ref{fig:14}(c).
Unlike the previously discussed Duffing oscillator case, the nonlinear time-history analysis performed using OpenSees does not yield accurate system
 matrices at each time step that can serve as ground truth for validation.

Therefore, to assess the accuracy of the system matrices identified by APSMC and other comparative methods, prediction validation is performed using 
response data from 35 consecutive time steps following the 40-second mark, which are not used during training. 
Table~\ref{tab:nmse} summarizes the prediction performance of the three methods under both noise-free and noisy conditions using the NMSE metric. 
\begin{table}[htbp]
    \centering
    \caption{Prediction NMSE under Noise-Free and Noisy Conditions}
    \begin{tabular}{lcc}
    \toprule
    \textbf{Method} & \textbf{Noise-Free} & \textbf{Noisy} \\
    \midrule
    APSMC Constrained & 3.23\% & 4.35\% \\
    APSMC  Unconstrained  & 6.00\% & 12.56\% \\
    DMDc                & 26.46\% & 34.54\% \\
    \bottomrule
    \end{tabular}
    \label{tab:nmse}
\end{table}
    
\begin{figure}[!ht]
    \centering
    \begin{subfigure}{0.31\textwidth}
        \centering
        \includegraphics[width=\textwidth]{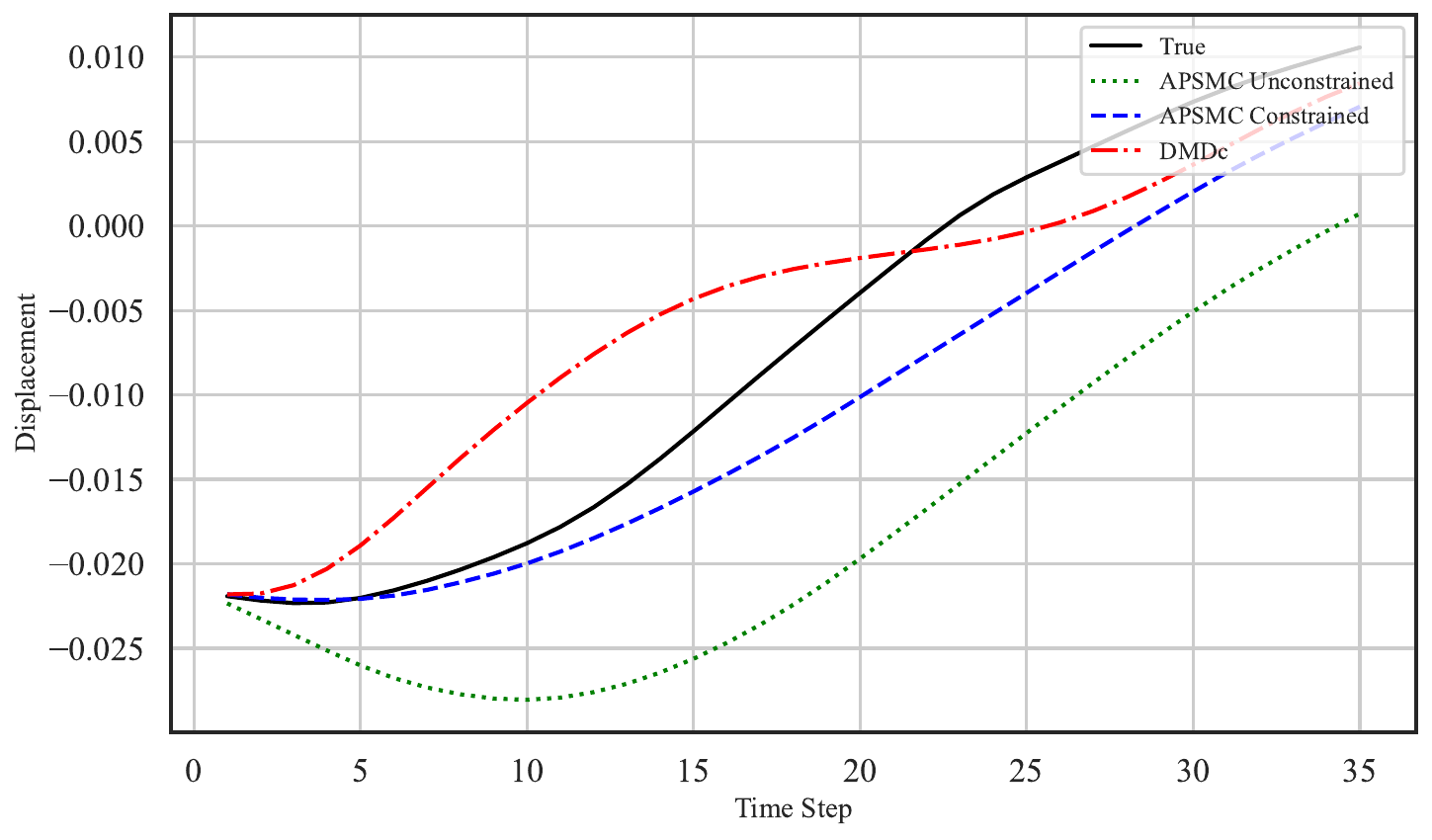}
        \caption{Displacement \( x_2(t) \)}
    \end{subfigure}
    \hspace{0.5em}% Space between figures
    \begin{subfigure}{0.31\textwidth}
        \centering
        \includegraphics[width=\textwidth]{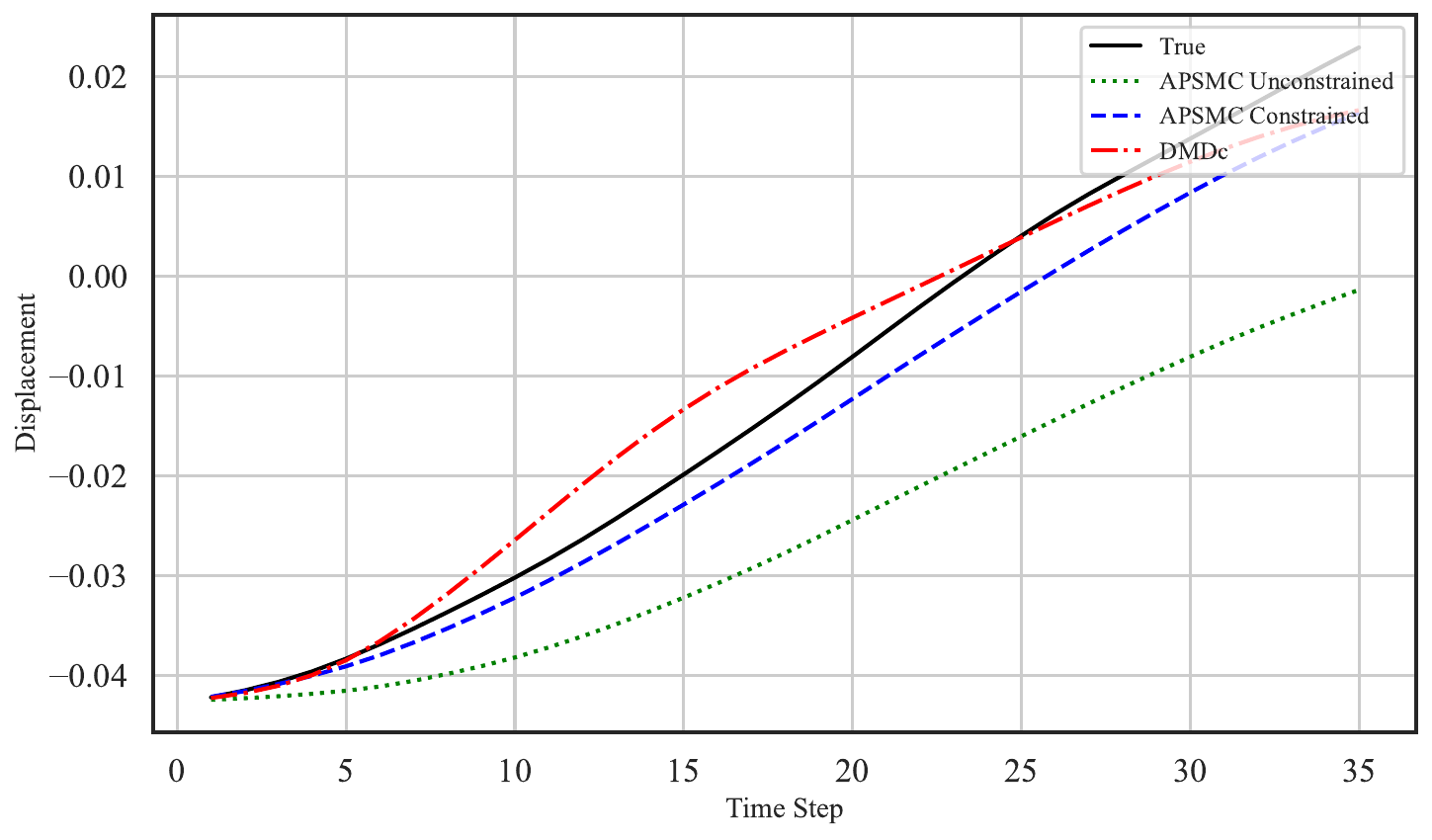}
        \caption{Displacement \( x_4(t) \)}
    \end{subfigure}
    \hspace{0.5em}% Space between figures
    \begin{subfigure}{0.31\textwidth}
        \centering
        \includegraphics[width=\textwidth]{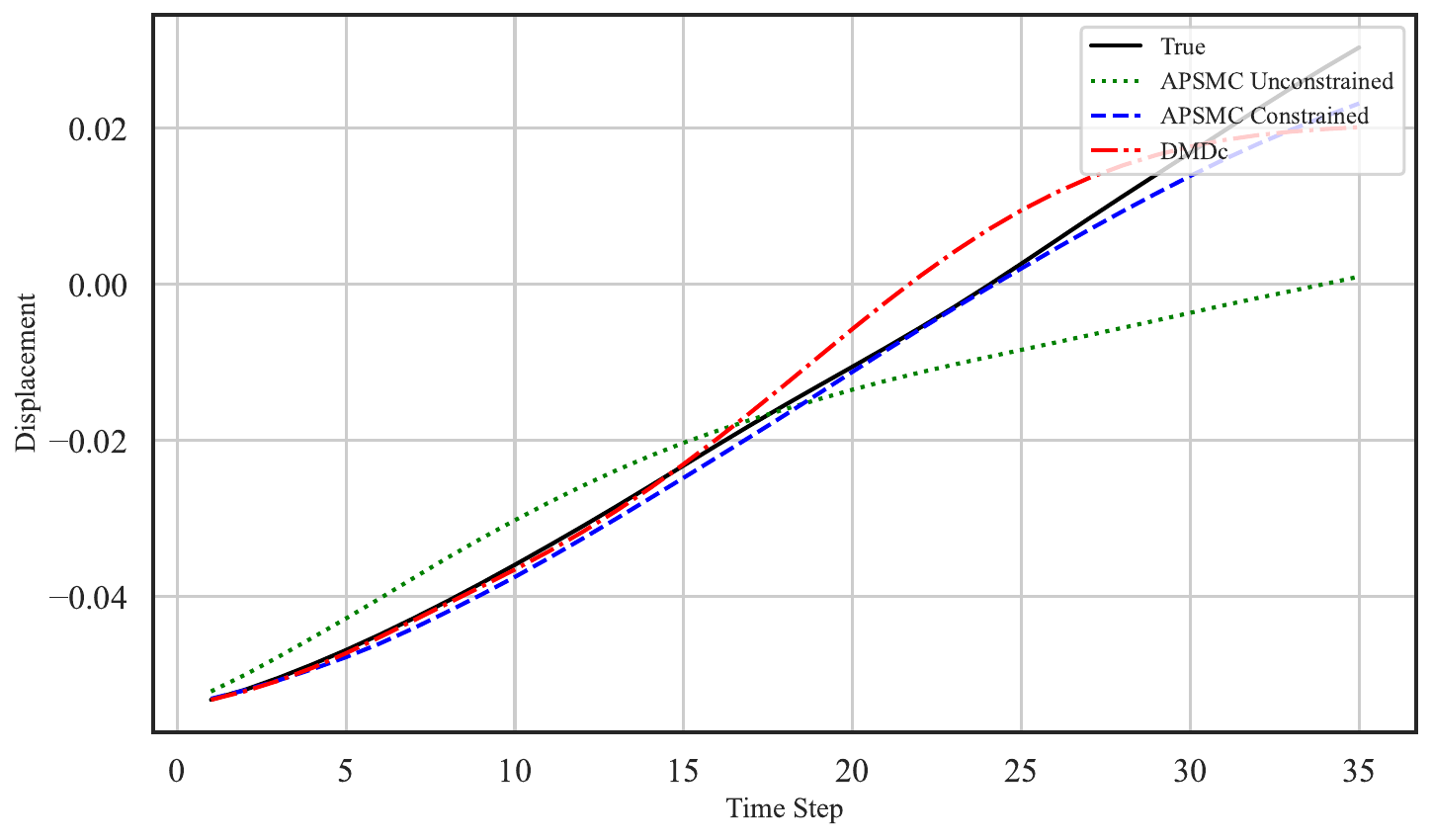}
        \caption{Displacement \( x_6(t) \)}
    \end{subfigure}
  
    \vspace{1em}% Space between the two rows of figures
    \begin{subfigure}{0.31\textwidth}
        \centering
        \includegraphics[width=\textwidth]{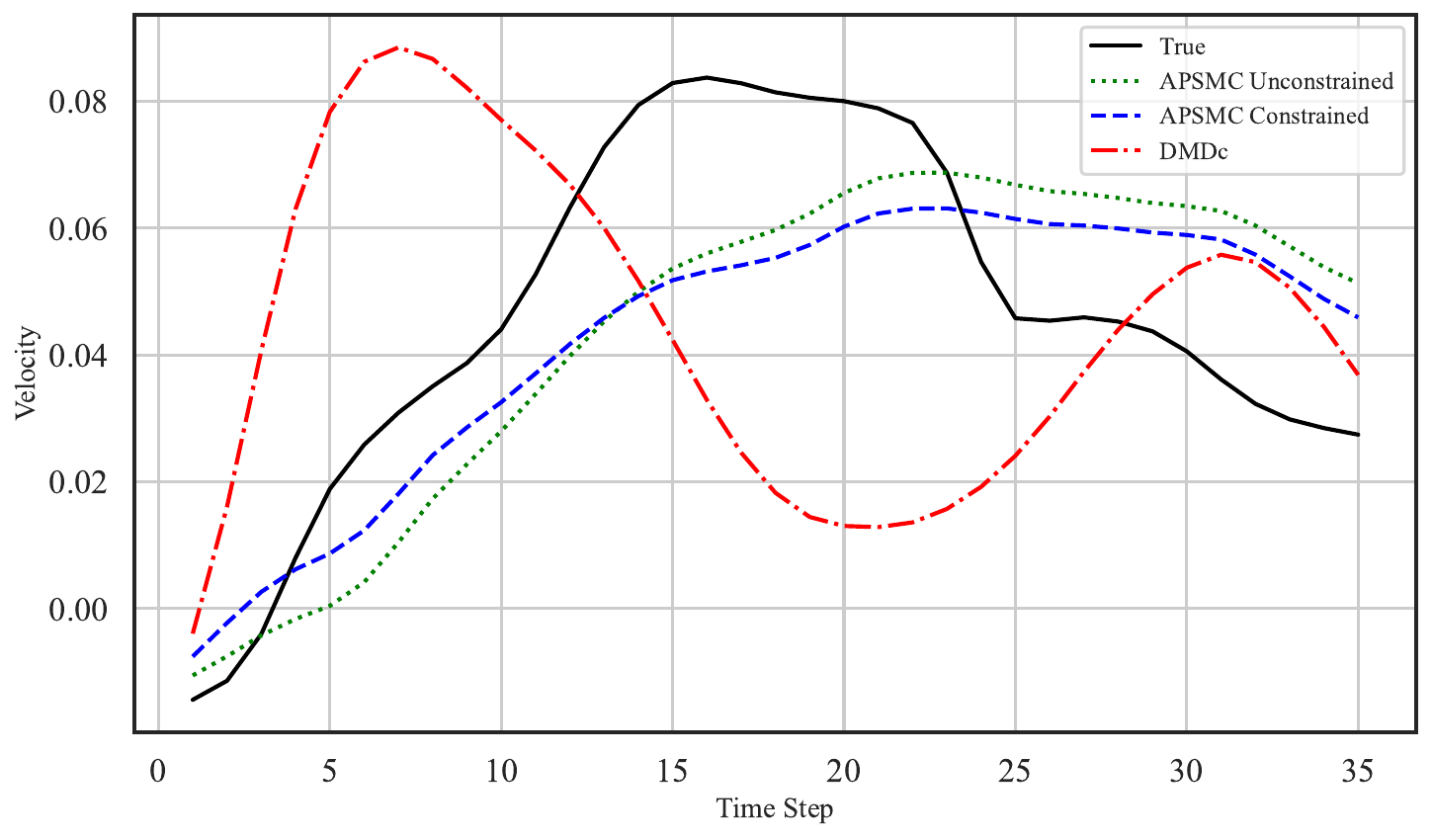}
        \caption{Velocity \( \dot{x}_2(t) \)}
    \end{subfigure}
    \hspace{0.5em}% Space between figures
    \begin{subfigure}{0.31\textwidth}
        \centering
        \includegraphics[width=\textwidth]{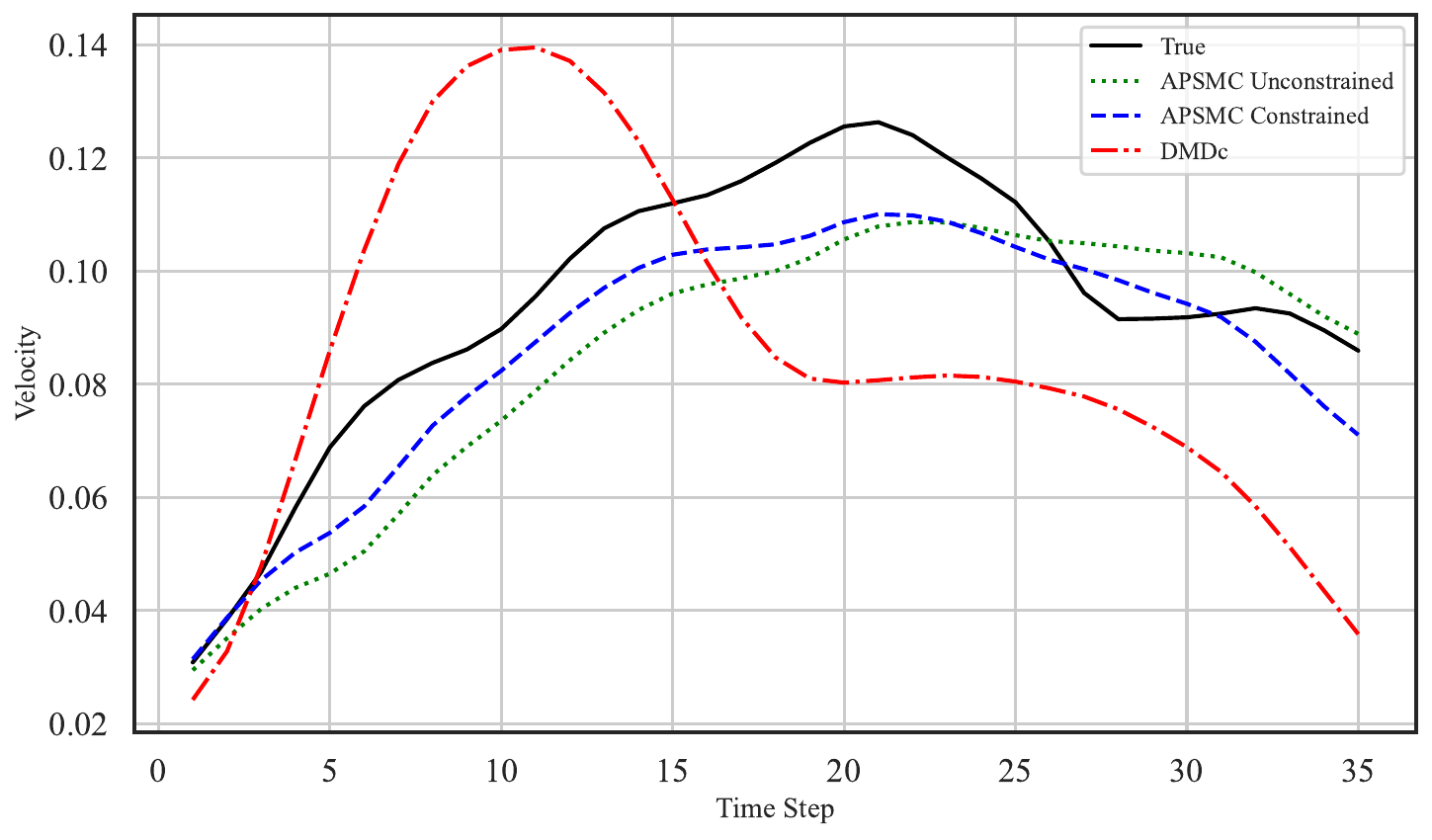}
        \caption{Velocity \( \dot{x}_4(t) \)}
    \end{subfigure}
    \hspace{0.5em}% Space between figures
    \begin{subfigure}{0.31\textwidth}
        \centering
        \includegraphics[width=\textwidth]{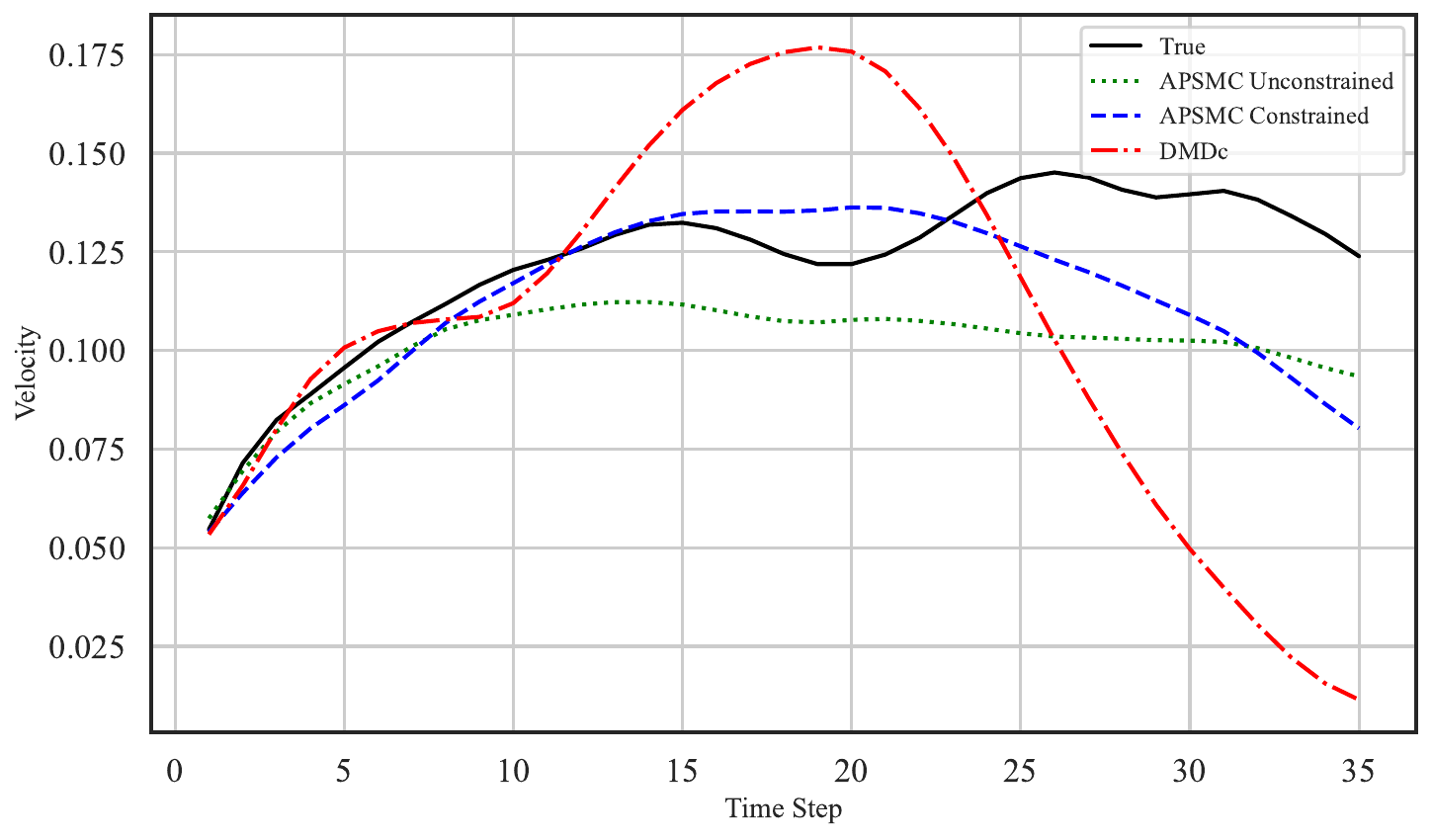}
        \caption{Velocity \( \dot{x}_6(t) \)}
    \end{subfigure}
    \caption{Comparative analysis between true data and predicted results.}
    \label{fig:13}
\end{figure}

The results show that APSMC with physical constraints consistently achieves the lowest NMSE under both noise-free and noisy conditions, demonstrating superior accuracy 
and robustness. While the unconstrained APSMC performs similarly in the absence of noise, its accuracy declines significantly when noise is introduced. In contrast, DMDc, 
which fits a globally optimal linear model, fails to capture the system’s time-varying nonlinear behavior, leading to poor short-term predictions in both cases. For visual 
comparison, Figure~\ref{fig:13} shows the predicted responses of all three models trained on noisy data alongside the ground truth.

\section{Experimental Evaluation of the Proposed Framework}\label{sec:Case}

The preceding numerical simulations demonstrated the capability of APSMC in modeling nonlinear dynamics and highlighted the role of physical constraints 
in overcoming challenges typically faced by purely data-driven methods. This section focuses on validating the framework using laboratory-scale experimental data, 
with an emphasis on denoising performance and generalization ability compared to classical time-domain algorithms.

To showcase the generality of APSMC, initial system models are constructed using the ERA \cite{juang1985eigensystem} and the 
Observer/Kalman Filter Identification combined with ERA (OKID+ERA) \cite{juangIdentificationObserverKalman}. Under noisy conditions, APSMC is shown to outperform
 both methods in estimation accuracy.

Due to the limited quantity of data obtained from single-excitation laboratory experiments, the observed improvement in generalization is modest. A more 
comprehensive evaluation of APSMC’s denoising capability using long-term monitoring data from a real bridge is available in \cite{chen5097818adaptive}.

\subsection{Scaled Model Description and Experimental Procedures}

This study employs a 1:10 scaled laboratory model of a three-span curved continuous girder bridge, based on the third span of Bridge No.~1 at a 
Shenzhen interchange. Details of the structural model and vibration testing setup can be found in~\cite{chen2025principal}.

\begin{figure}[htb]
    \centering
    \includegraphics[width=0.9\linewidth]{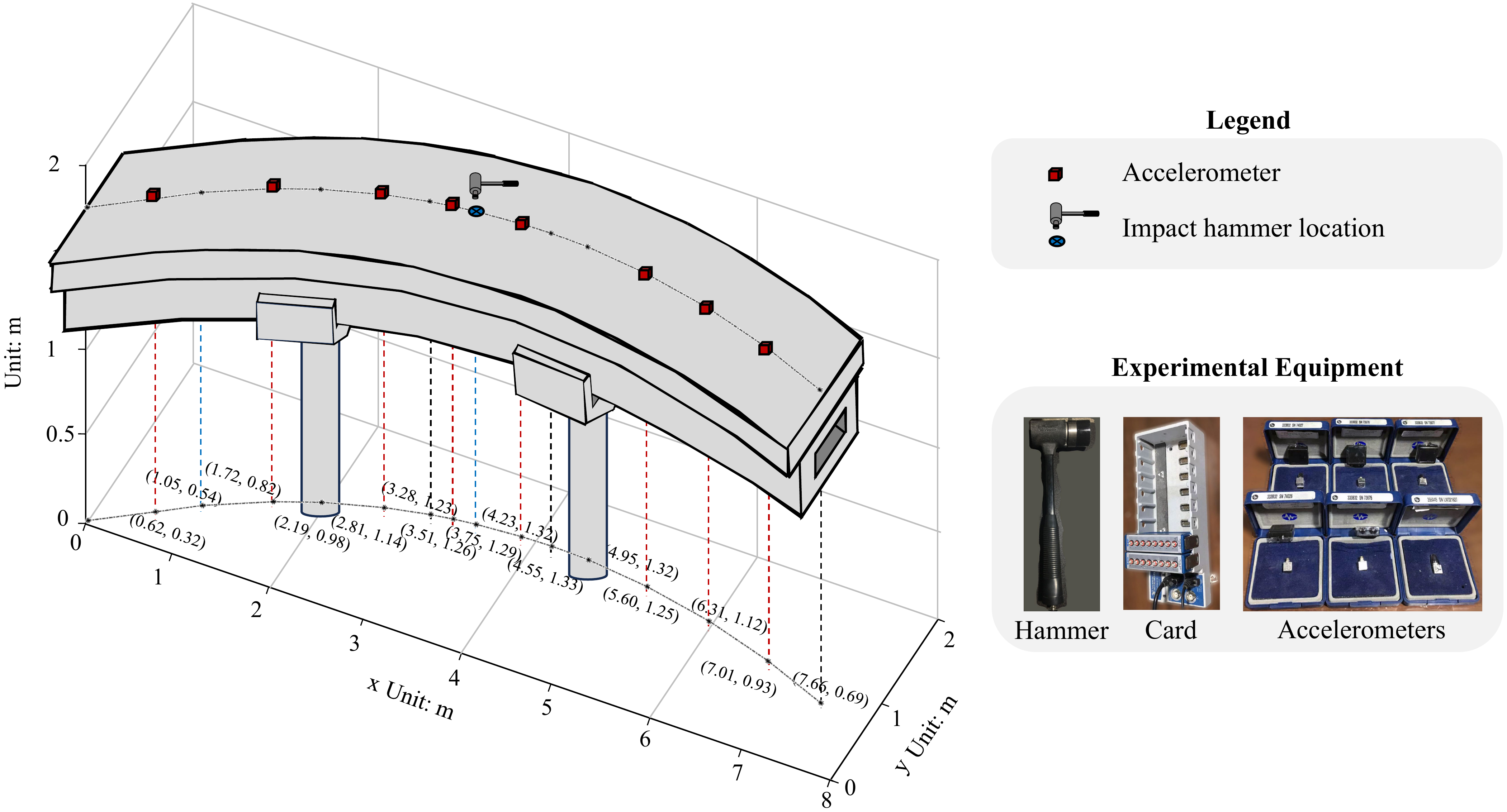}
    \caption{Schematic of the scaled bridge model and test equipment.}
    \label{fig111}
\end{figure}

Figure~\ref{fig111} shows the schematic of the experimental setup. The main girders, cap beams, and piers were cast using C40 concrete. Girders and 
cap beams were reinforced with stirrups, while piers included longitudinal bars and spiral stirrups. The central piers were bolted to the girders.

The model was excited using an impact hammer along the negative \( z \)-axis, and responses were recorded by eight accelerometers. Except for the tri-axial 
sensors at both ends of the bridge—whose positive directions align with the \( z \)-, \( x \)-, and \( y \)-axes as shown in Figure~\ref{fig111}—the others 
were uniaxial sensors measuring vertical accelerations along the positive \( z \)-axis.
Data were collected at a 3200 Hz sampling rate over 10 seconds per test. To assess the generalization capability of the APSMC algorithm and ensure 
repeatability, 20 independent impact tests were conducted.

\subsection{Experimental Validation of the APSMC Algorithm}
 The analysis is based on 20 
unfiltered 10-second data segments collected from eight accelerometers, capturing raw structural responses under impact excitation.

\begin{figure}[!ht]
  \centering
  \begin{subfigure}{0.4\textwidth}
      \centering
      \includegraphics[width=\textwidth]{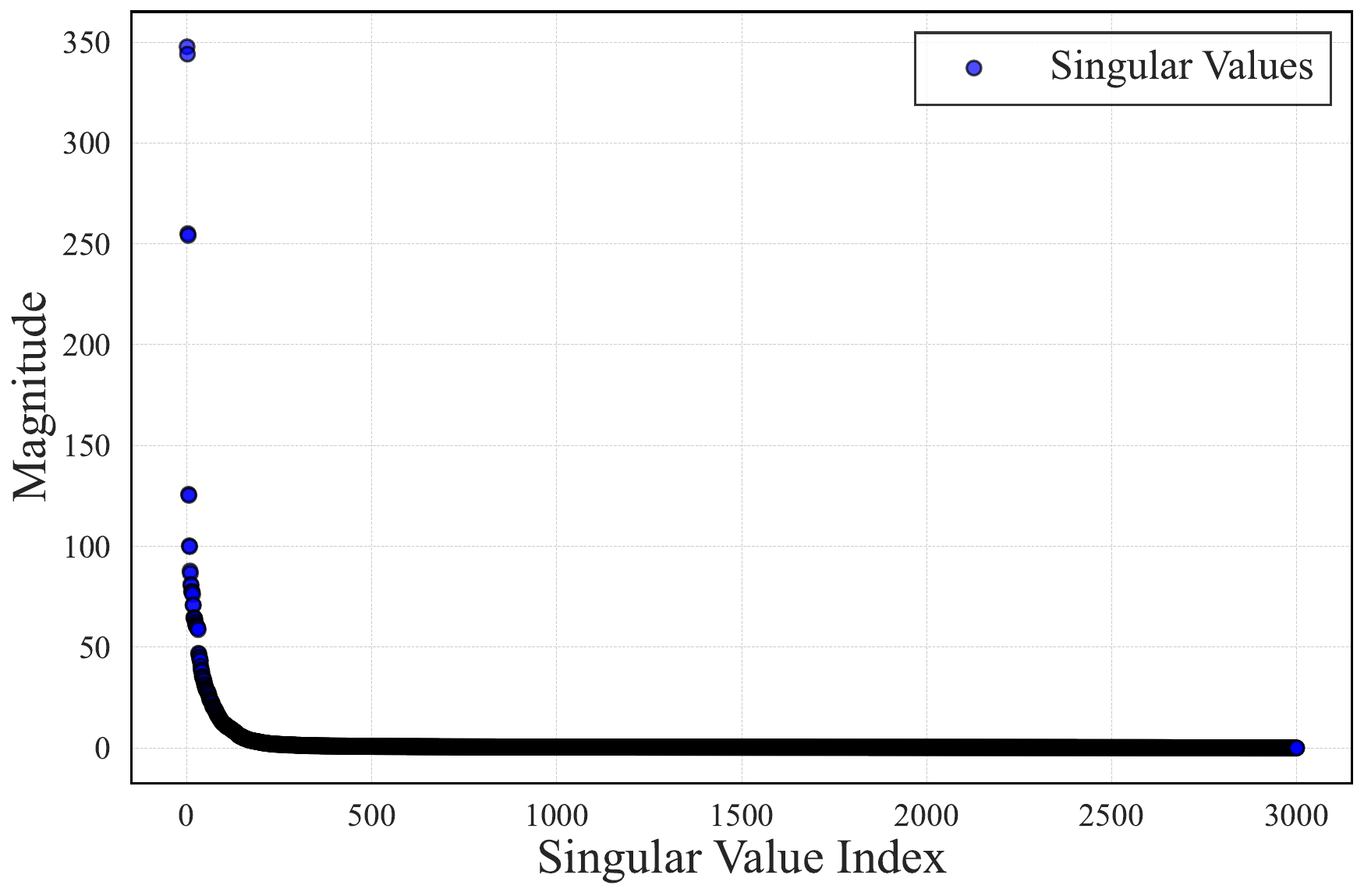}
      \caption{Singular values of \(H\)}\label{fig2}%\ref{fig2}
  \end{subfigure}
  \hspace{2em}% Space between figures
  \begin{subfigure}{0.4\textwidth}
      \centering
      \includegraphics[width=\textwidth]{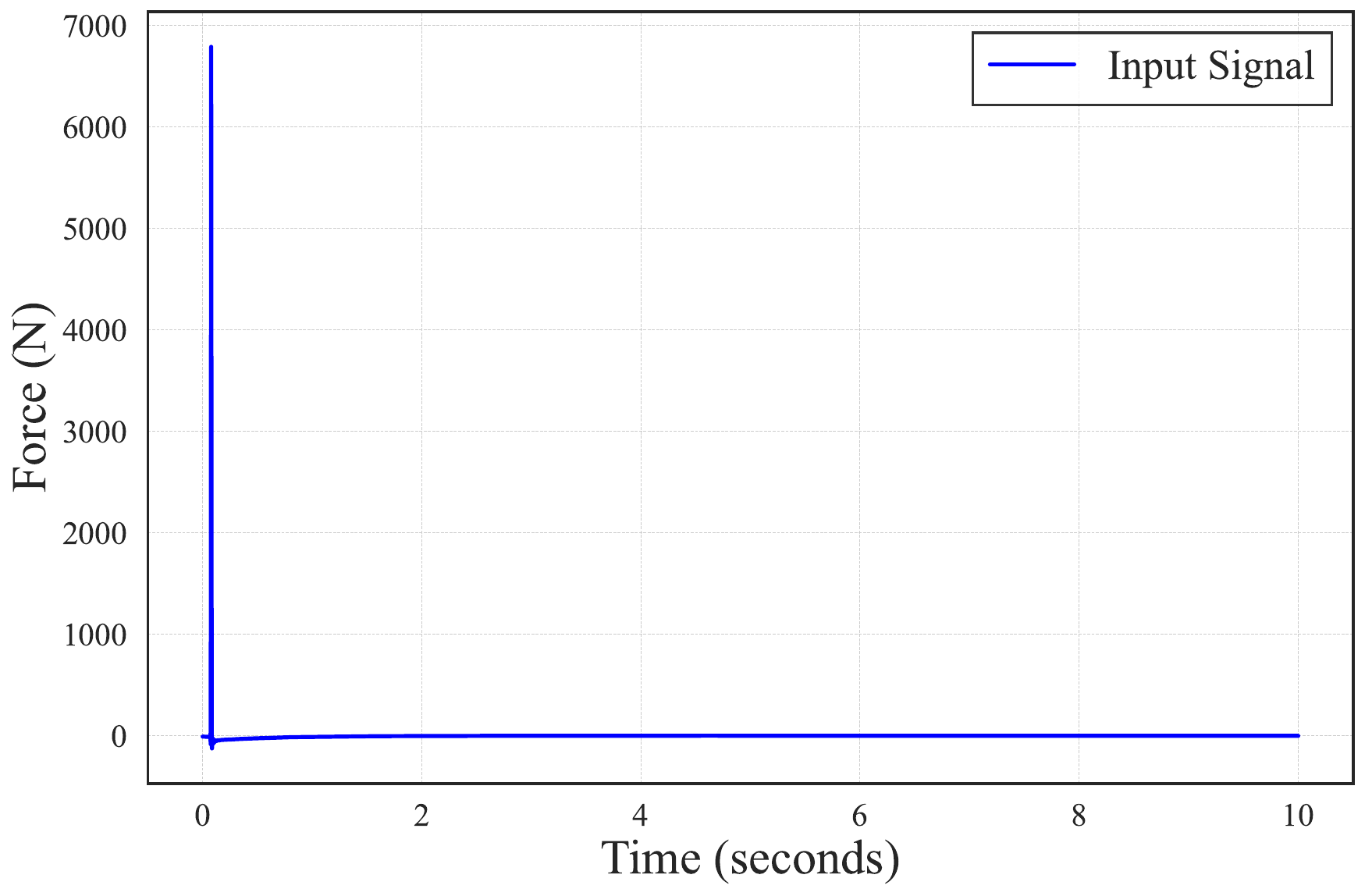}
      \caption{Input signal \(F\)}\label{fig8}%\ref{fig8}
  \end{subfigure}
  \caption{Singular values employed for determining model order and the corresponding input signal}
\end{figure}

 Given that 
 the monitored data in the scaled experiment corresponds to impulse responses (see Figure \ref{fig8} for the input load diagram), the ERA algorithm 
 was used to fit the data and obtain the initial model \( A \), \( B \), and \( C \). To achieve this, a Hankel matrix \( H \) of dimensions \( 12m \times n \) is 
 constructed as follows:
 \begin{equation}
     H = \begin{bmatrix}
         y_0 & y_1  & \cdots & y_{n-1} \\
         y_1 & y_2  & \cdots & y_n \\
         y_2 & y_3  & \cdots & y_{n+1} \\
         \vdots  & \vdots & \ddots & \vdots \\
         y_{m-1} & y_m  & \cdots & y_{m+n-2}
     \end{bmatrix}
 \end{equation}
 where \( y_j \in \mathbb{R}^{12} \) represents the sensor data at time step \( j \). In fact, selecting different values for \( m \) and \( n \) 
for analysis can significantly impact the model prediction results \cite{panStructureTimedelayEmbedding2020,bruntonChaosIntermittentlyForced,arbabiErgodicTheoryDynamic2017}. 
After extensive trial calculations, this study chose \( m \) and \( n \) both to be 3000 for subsequent analysis. The singular values of \( H \) at this
 point are shown in Figure \ref{fig2}, indicating that the energy of the singular values is predominantly concentrated within the first 300 values.

  Moreover, the first 577 singular values preserve 95\% of the total energy. It is necessary to determine the signal subspace and noise subspace based
   on the distribution of singular values. Although there is existing literature on determining model order in the presence of white noise \cite{gavish2014optimal} 
   or sparse noise \cite{candesRobustPrincipalComponent2009} in observational data, from the perspective of validating the method's effectiveness, this study 
   considers using the first 50 to 300 singular values and the first 577 orders for subsequent prediction comparisons.

\begin{figure}[!ht]
    \centering
    \begin{subfigure}{0.327\textwidth}
        \centering
        \includegraphics[width=\textwidth]{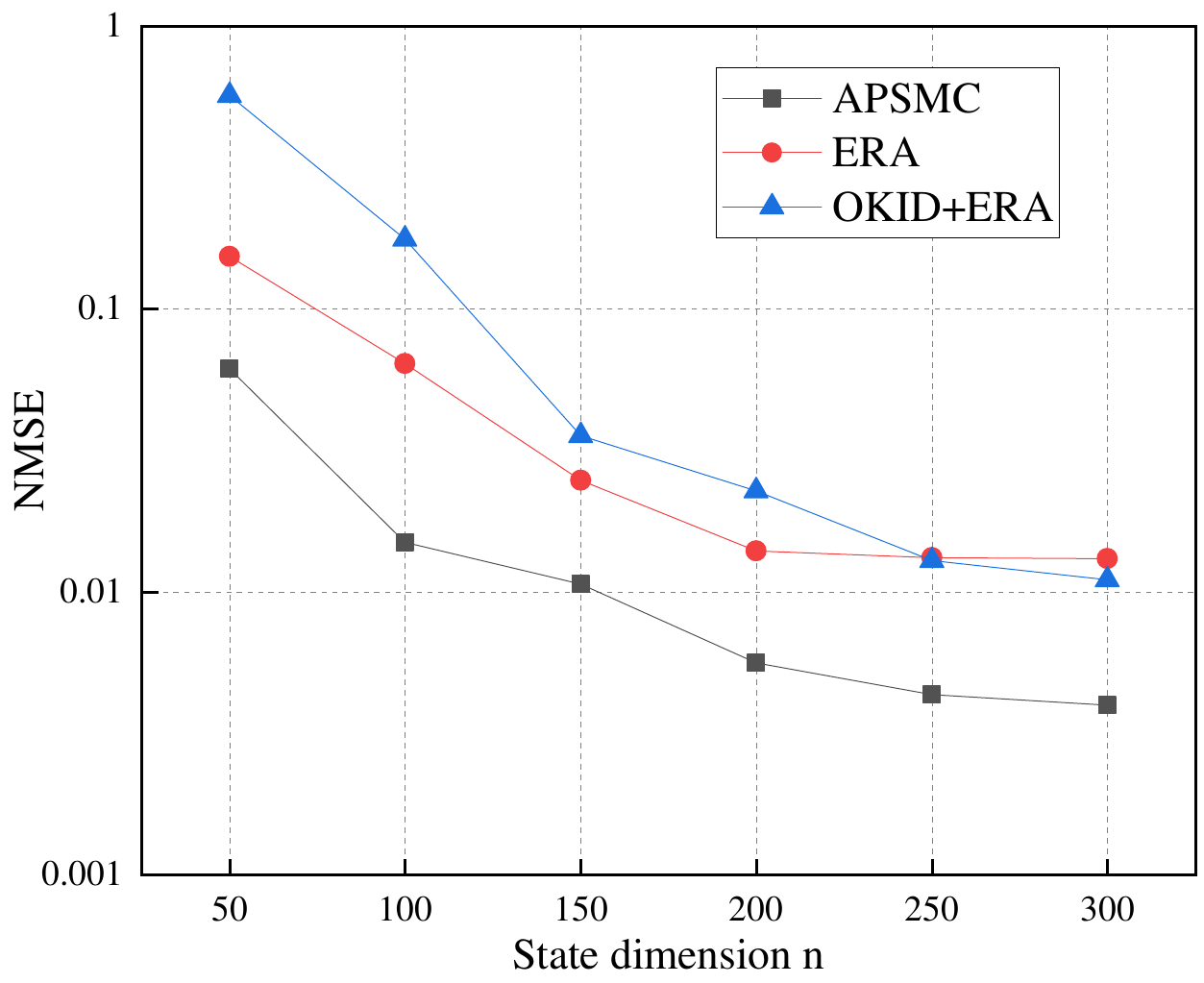}
        \caption{Prediction Accuracy of \(\hat{y}_{k|k-1}\)}\label{fig7}%\ref{fig7}
    \end{subfigure}
    \hspace{2em}% Space between figures
    \begin{subfigure}{0.4\textwidth}
        \centering
        \includegraphics[width=\textwidth]{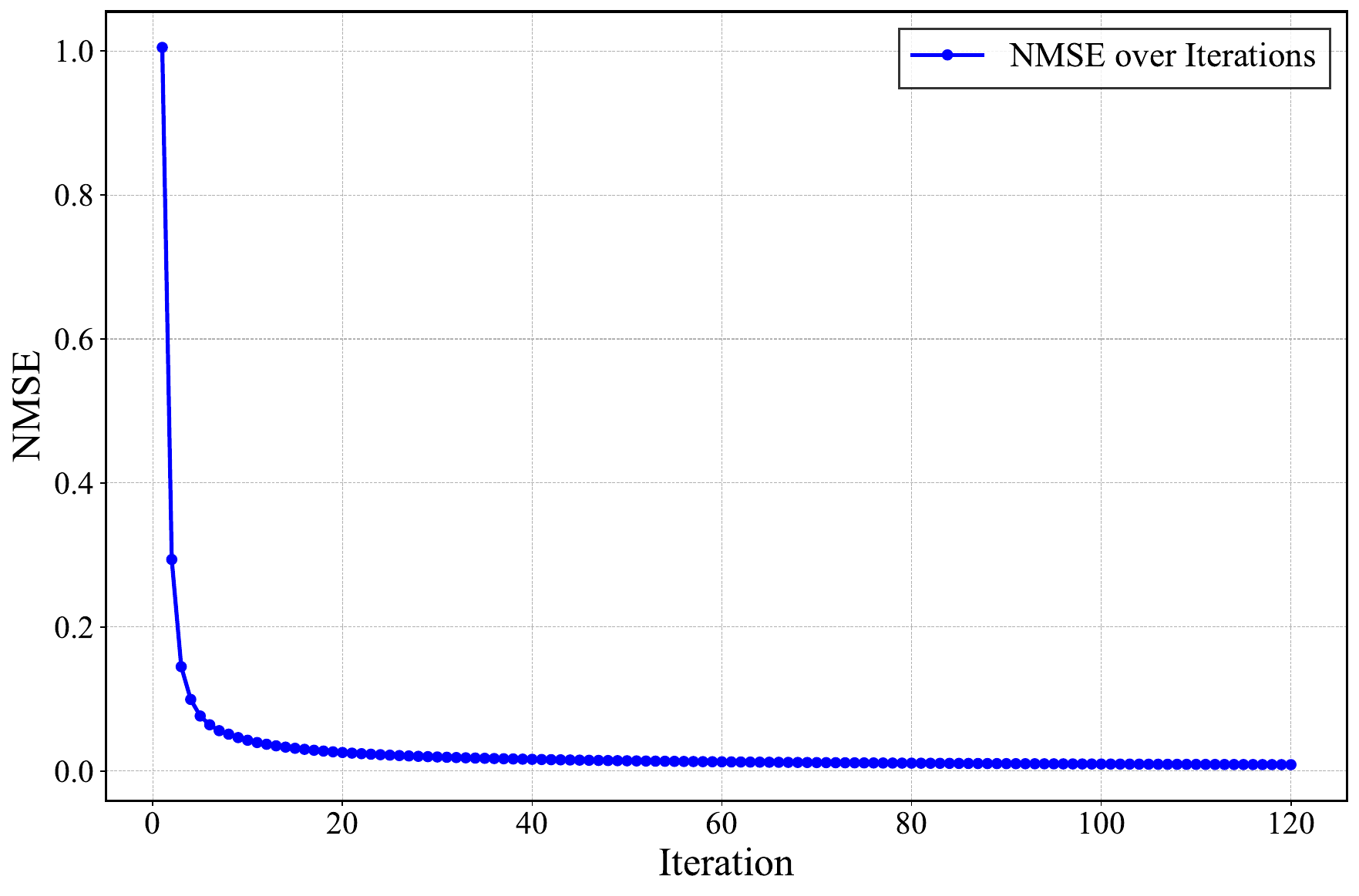}
        \caption{Convergence Curve for \(n=300\)}\label{fig6}%\ref{fig6}
    \end{subfigure}
    \caption{Prediction Accuracy and Convergence Analysis}
    \label{fig:combined}
\end{figure}

To construct the initial model, the determined model order and selected Hankel matrix dimensions were input into the ERA algorithm, resulting in a state-space 
model with matrices \( A \), \( B \), and \( C \). Following the acquisition of the initial model, the APSMC algorithm was employed for prediction analysis. 
Figure \ref{fig7} presents the NMSE values for \(\hat{y}_{k|k-1}\) compared to the monitored \( y_k \) under different model orders. It is evident that the 
prediction accuracy of the APSMC algorithm is improved by nearly an order of magnitude.

In fact, once the model order \( n \) of the ERA model reached 200 (with an NMSE of 1.325\%), its prediction accuracy 
remained largely stagnant despite further increases in model order. Even at \( n = 577 \), the NMSE of the ERA model was 1.315\%, while the OKID+ERA model 
achieved an NMSE of 0.905\%. This discrepancy is likely because both the initial
 model construction and prediction using the OKID+ERA model incorporated the force \( F_k \), whereas the APSM and ERA models did not utilize the 
 force \( F_k \) during either the prediction or model construction processes.

 In addition, to further demonstrate the effectiveness of the method and the convergence behavior of the APSMC algorithm, a randomly generated \( C \) matrix was used 
 during runtime for the case with a model order of \( n = 300 \), instead of the \( C \) matrix obtained from ERA. Figure \ref{fig6} illustrates the convergence 
 curve for this case, where each iteration corresponds to an update of \( A_k \) using 10 seconds of data, over a total runtime of 9.09 minutes. The runtime was 
 measured using Python on a PC equipped with a 3.4 GHz Intel Core i7-14700K processor. 
 The final NMSE achieved is 0.894\%. The corresponding data covers a total duration of 20 minutes, indicating that the computational efficiency of the algorithm is 
 sufficient for real-time updates.

To further verify whether the APSM-updated model accurately simulates the actual structure, the model achieving an NMSE of 0.894\% was 
used as the initial model for subsequent tests with 19 additional experimental datasets. The results are summarized in Table \ref{tab:impact_data}. It can be observed that 
the time-domain fitting errors are generally around 3\%, with occasional values near 1\% or 17\%, likely due to inherent experimental randomness.

Due to possible discrepancies between some repeated impact experiments and the first impact test, cases with large errors across all three methods are highlighted in red in Table~\ref{tab:impact_data}.
The stability and convergence speed of the algorithm depend on the choice of learning rate: an excessively high learning rate may cause instability, while a rate that 
is too low may result in slow convergence. Table \ref{tab:impact_data} also lists the corresponding learning rates used for each dataset.

\begin{table}[t]
    \centering
    \caption{Performance Comparison of Different Methods and Learning Rates}
    \begin{tabular}{lcccc}
    \toprule
    \textbf{Sequence} & \textbf{APSMC} & \textbf{ERA} & \textbf{OKID+ERA} & \textbf{Learning Rate} \\
    \midrule
    2  & 2.474\% & 2.907\% & 6.550\% & 0.04 \\
    3  & 2.229\% & 2.667\% & 7.141\% & 0.04 \\
    4  & 2.272\% & 2.552\% & 3.769\% & 0.04 \\
    5  & 1.446\% & 1.942\% & 2.641\% & 0.04 \\
    6  & 2.173\% & 2.539\% & 22.209\% & 100 \\
    7  & 3.297\% & 3.072\% & 5.675\% & 0.5 \\
    8  & 1.974\% & 2.172\% & 3.029\% & 0.04 \\
    9  & 2.782\% & 2.881\% & 3.972\% & 0.04 \\
    10 & 3.528\% & 3.498\% & 5.249\% & 0.4 \\
    11 & 0.767\% & 0.927\% & 16.954\% & 80 \\
    12 & 3.493\% & 3.981\% & 9.208\% & 0.007 \\
    13 & 0.703\% & 0.904\% & 16.351\% & 0.04 \\
    14 & 3.078\% & 4.128\% & 10.896\% & 0.004 \\
    15 & \textcolor{red}{\textbf{18.847\%}} & \textcolor{red}{\textbf{10.065\%}} & \textcolor{red}{\textbf{36.490\%}} & 0.08 \\
    16 & 7.042\% & 2.444\% & 38.328\% & 0.2 \\
    17 & 0.837\% & 0.920\% & 16.242\% & 0.04 \\
    18 & 3.302\% & 8.234\% & 9.755\% & 0.005 \\
    19 & \textcolor{red}{\textbf{17.028\%}} & \textcolor{red}{\textbf{11.164\%}} & \textcolor{red}{\textbf{39.132\%}} & 0.09 \\
    20 & 0.765\% & 0.797\% & 15.803\% & 0.04 \\
    \bottomrule
    \end{tabular}
    \label{tab:impact_data}
\end{table}

\begin{figure}[!ht]
  \centering
  \begin{subfigure}{0.29\textwidth}
      \centering
      \includegraphics[width=\textwidth]{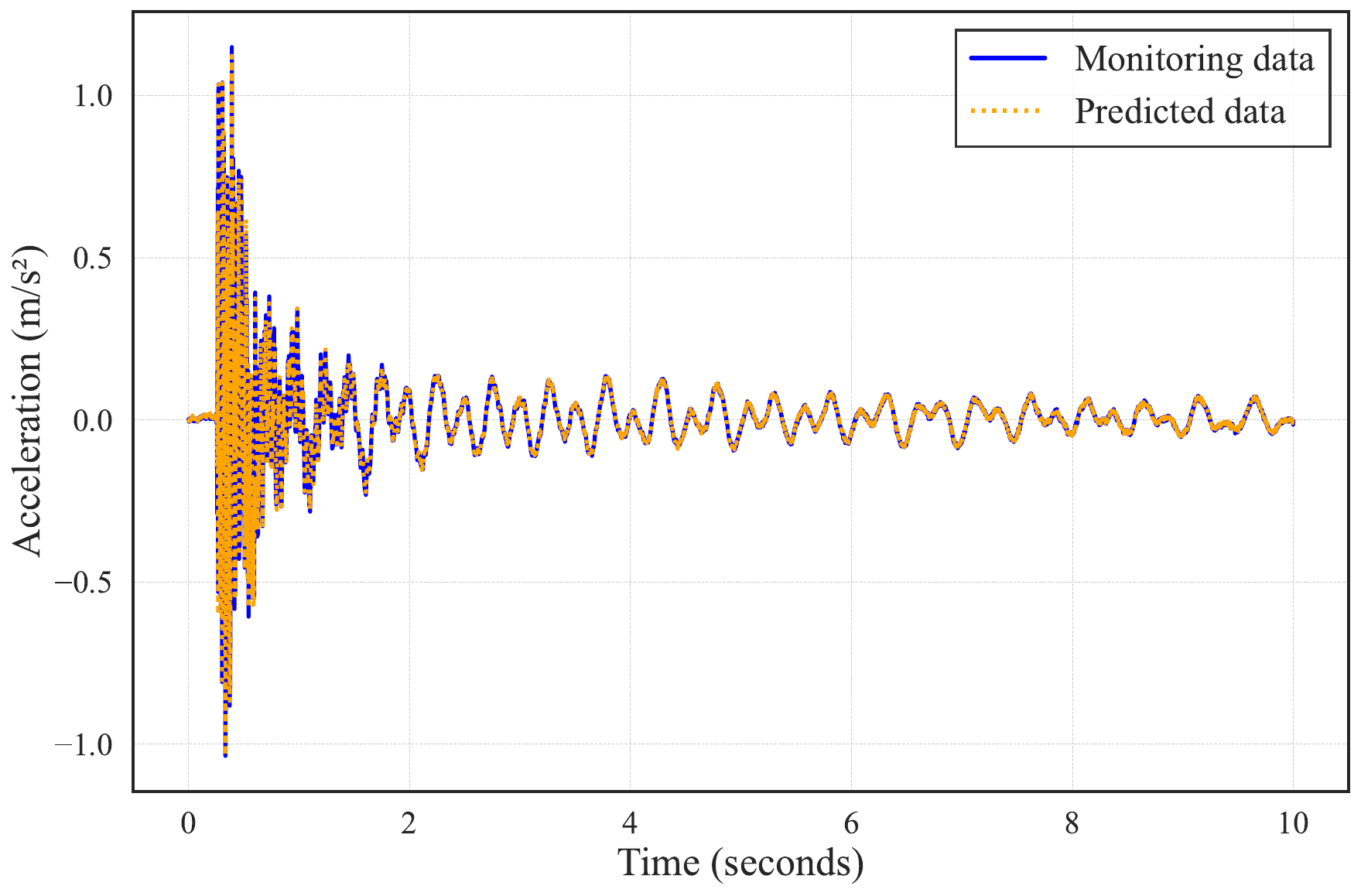}
      \caption{Monitoring Data \( y_1(t) \)}
  \end{subfigure}
  \hspace{2em}% Space between figures
  \begin{subfigure}{0.29\textwidth}
      \centering
      \includegraphics[width=\textwidth]{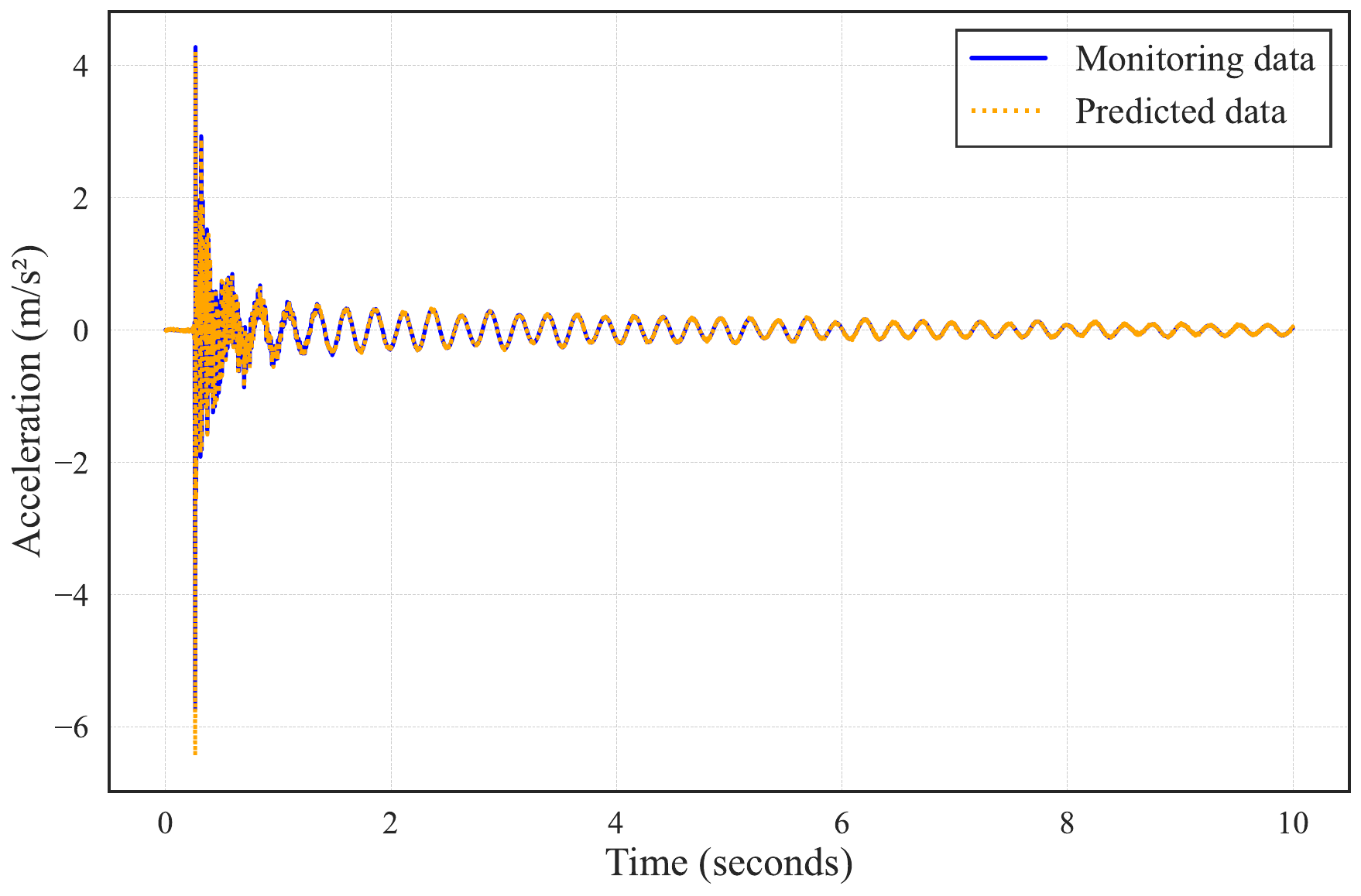}
      \caption{Monitoring Data \( y_2(t) \)}
  \end{subfigure}
  \hspace{2em}% Space between figures
  \begin{subfigure}{0.29\textwidth}
      \centering
      \includegraphics[width=\textwidth]{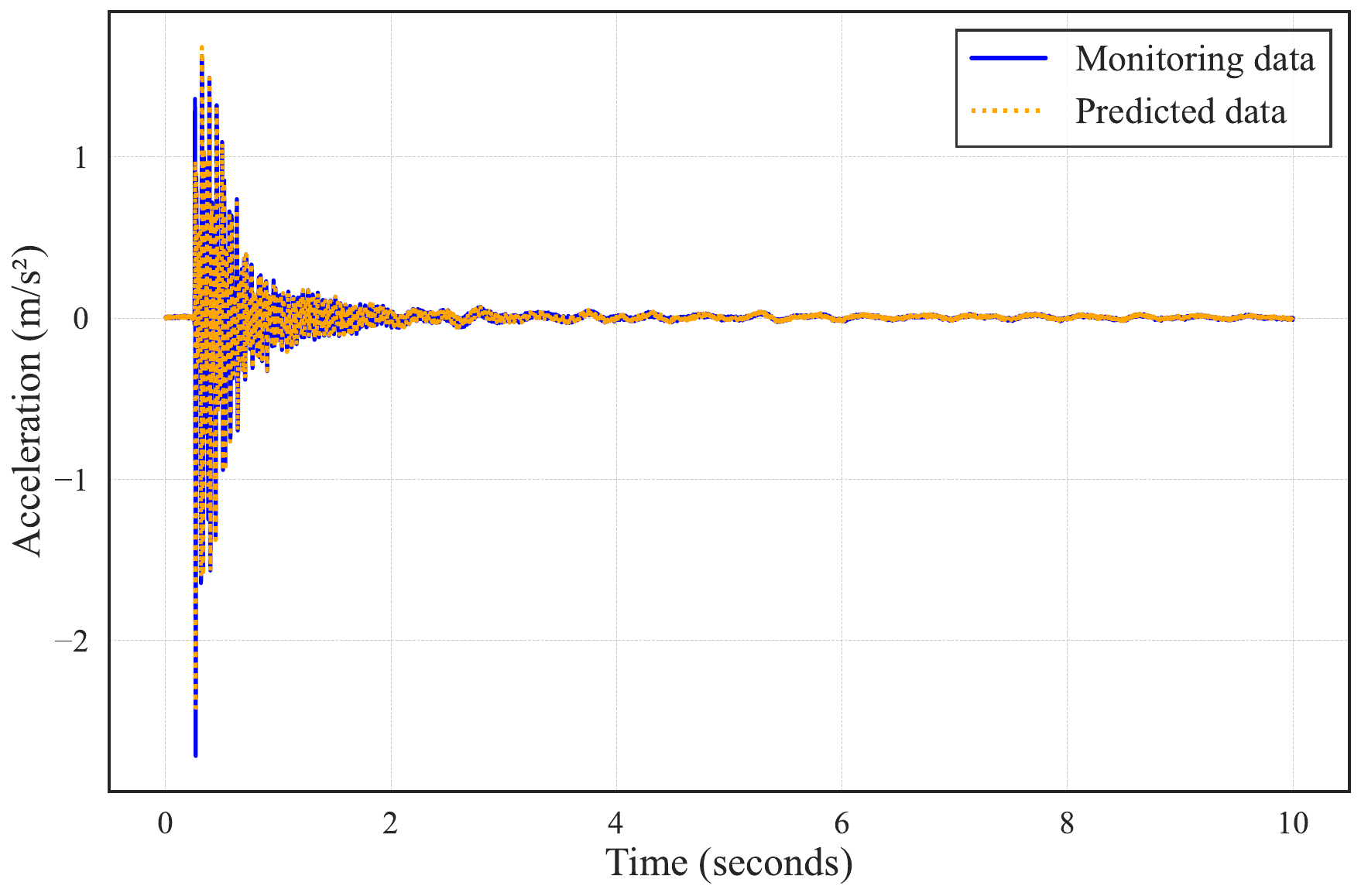}
      \caption{Monitoring Data \( y_3(t) \)}
  \end{subfigure}

  \vspace{1em}% Space between the two rows of figures
  \begin{subfigure}{0.29\textwidth}
      \centering
      \includegraphics[width=\textwidth]{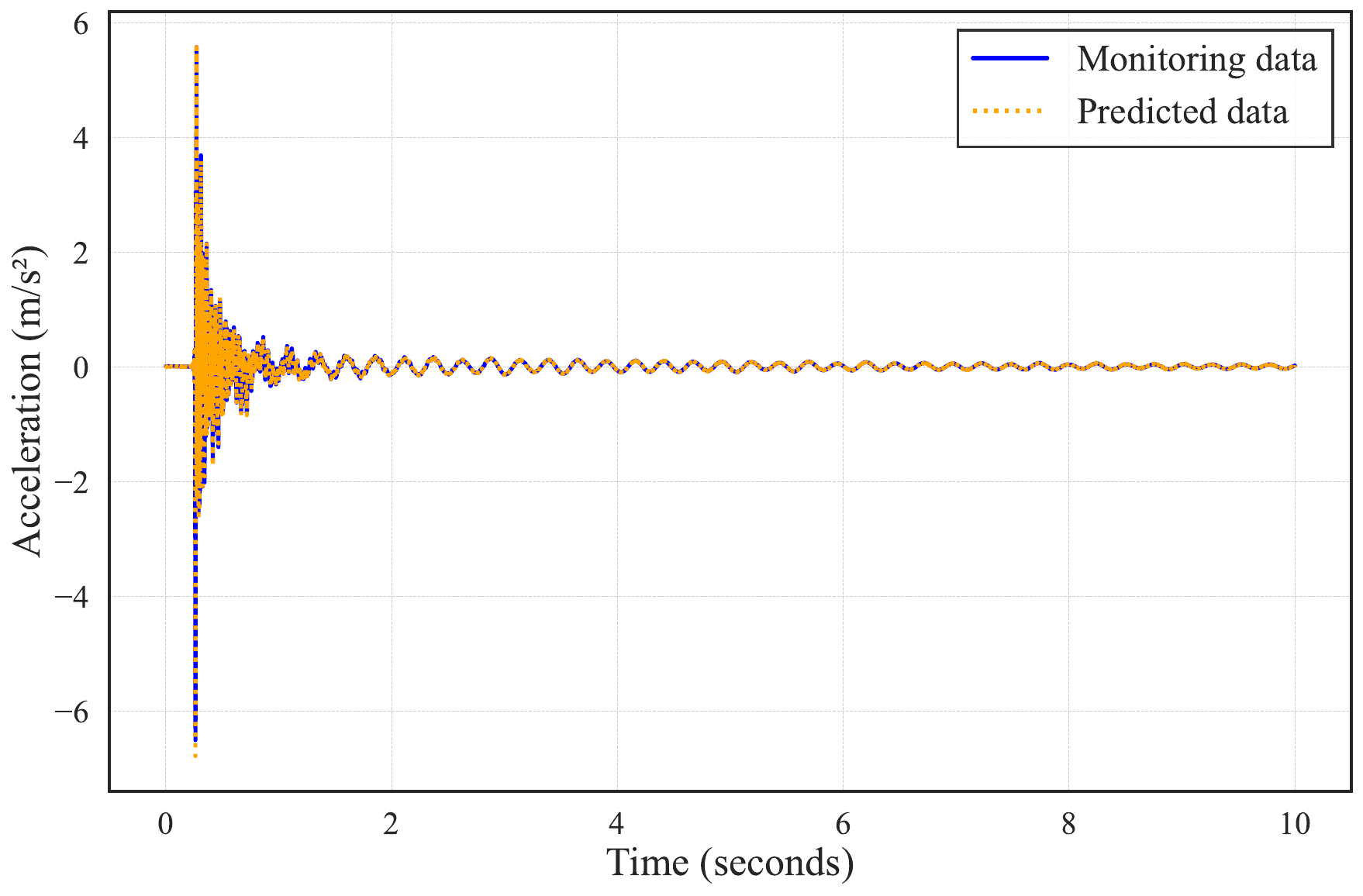}
      \caption{Monitoring Data \( y_4(t) \)}
  \end{subfigure}
  \hspace{2em}% Space between figures
  \begin{subfigure}{0.29\textwidth}
      \centering
      \includegraphics[width=\textwidth]{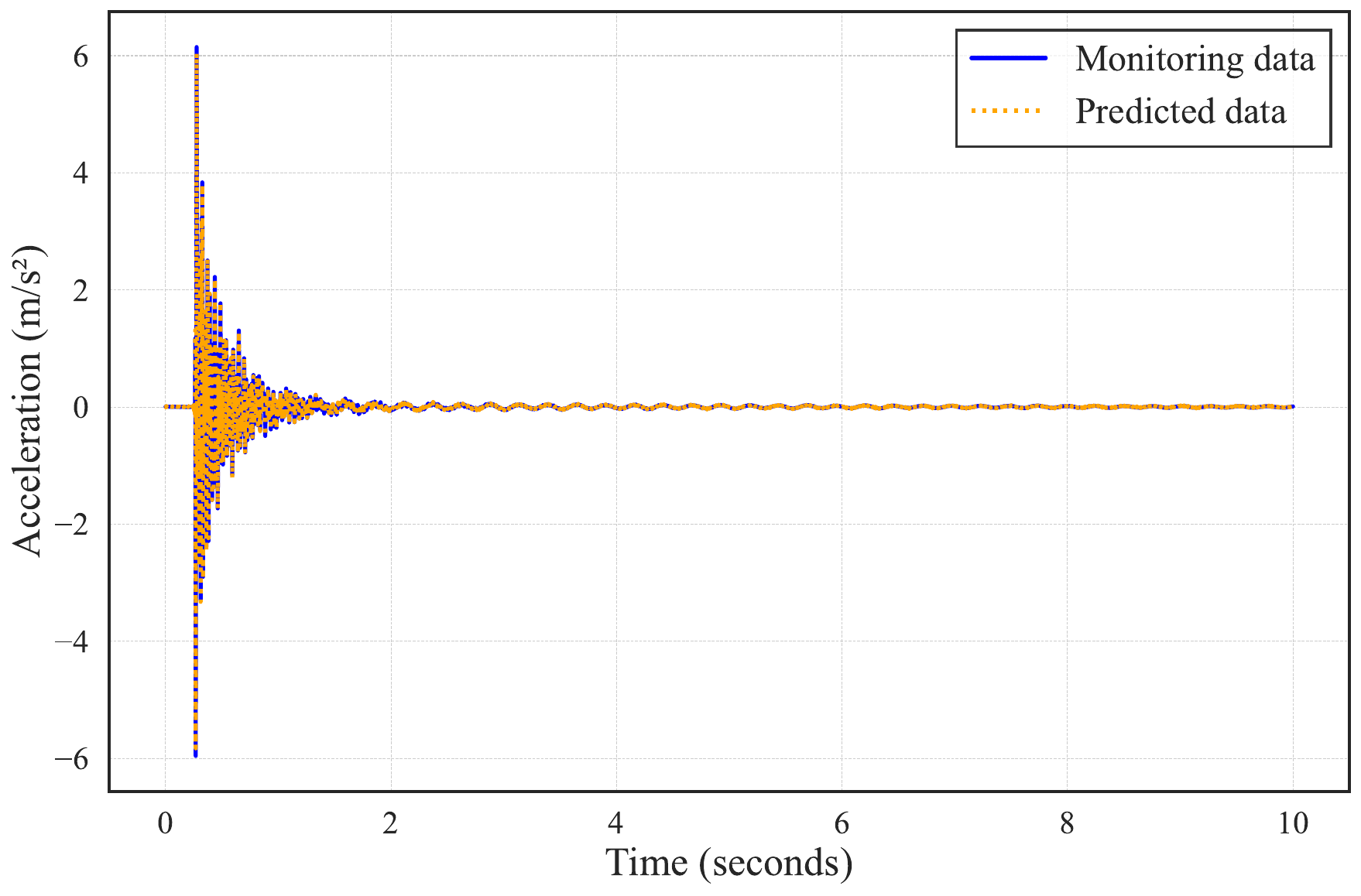}
      \caption{Monitoring Data \( y_5(t) \)}
  \end{subfigure}
  \hspace{2em}% Space between figures
  \begin{subfigure}{0.29\textwidth}
      \centering
      \includegraphics[width=\textwidth]{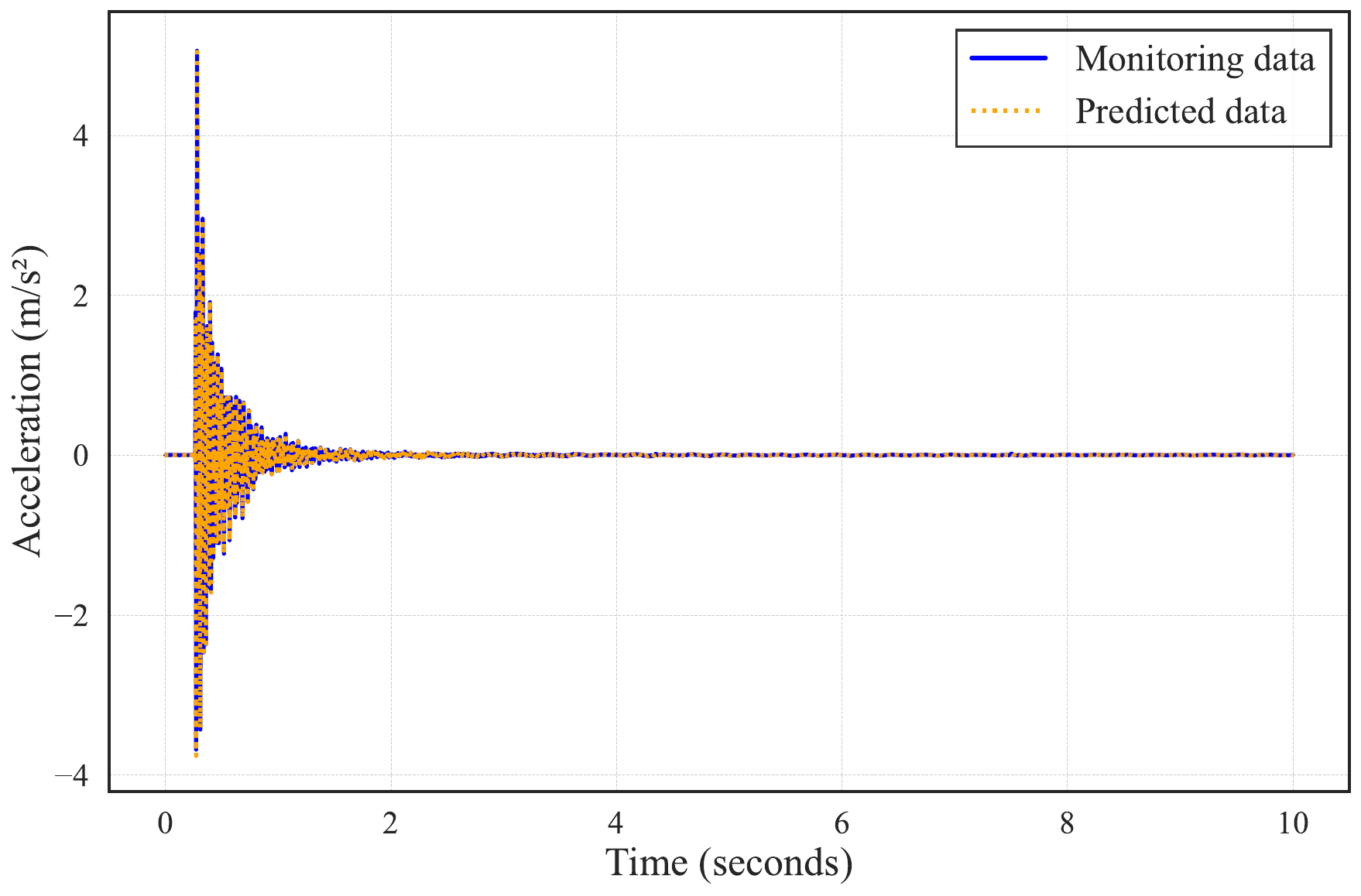}
      \caption{Monitoring Data \( y_6(t) \)}
  \end{subfigure}
  \caption{Comparative Analysis of Monitoring and Predicted Data for Various \( y_1 \sim y_{6} \)}
  \label{fig:15}%\ref{fig:15}
\end{figure}

\begin{figure}[!ht]
  \centering
  \begin{subfigure}{0.29\textwidth}
      \centering
      \includegraphics[width=\textwidth]{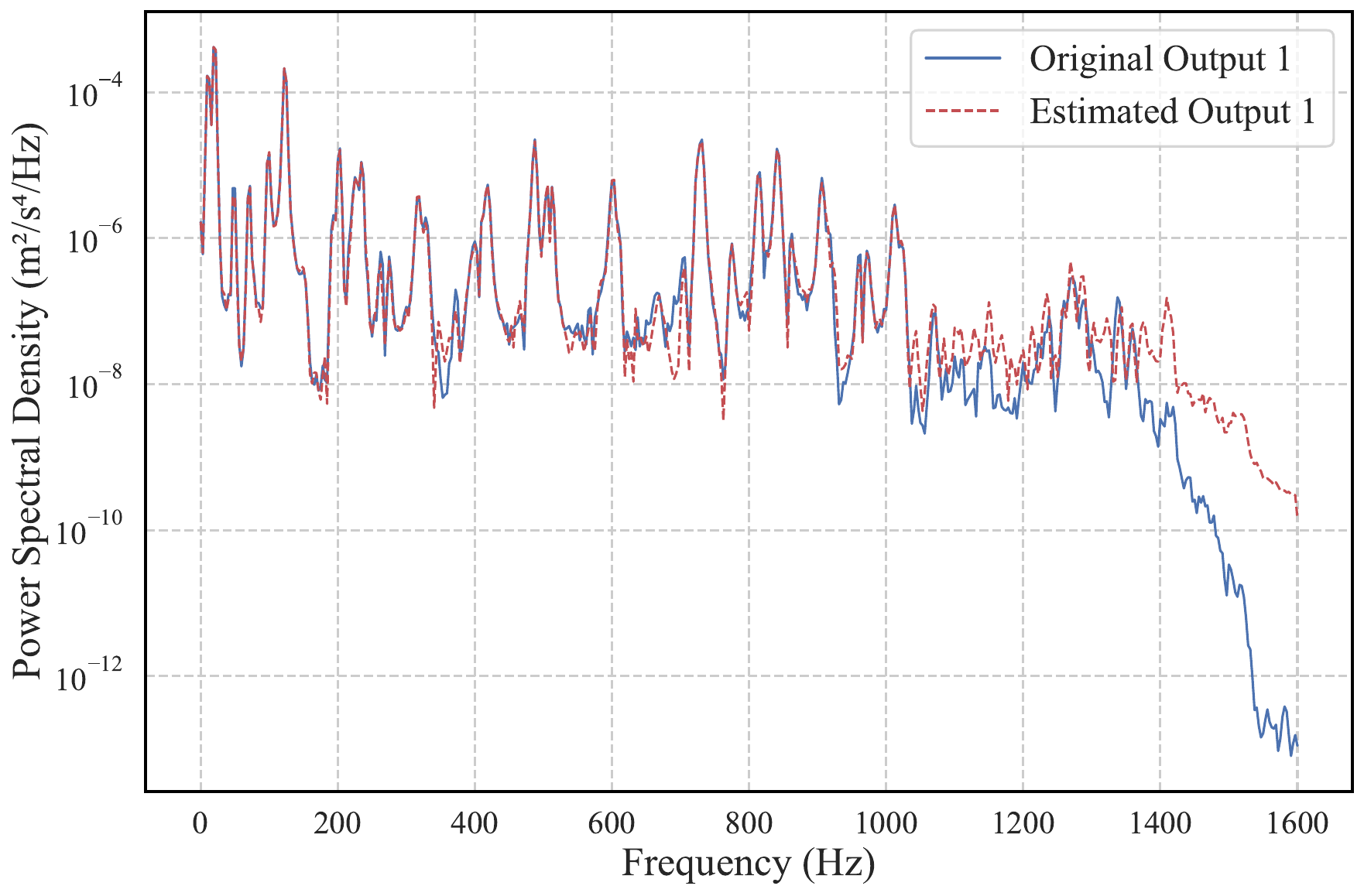}
      \caption{PSD of Monitoring Data \( y_1(t) \)}
  \end{subfigure}
  \hspace{2em}% Space between figures
  \begin{subfigure}{0.29\textwidth}
      \centering
      \includegraphics[width=\textwidth]{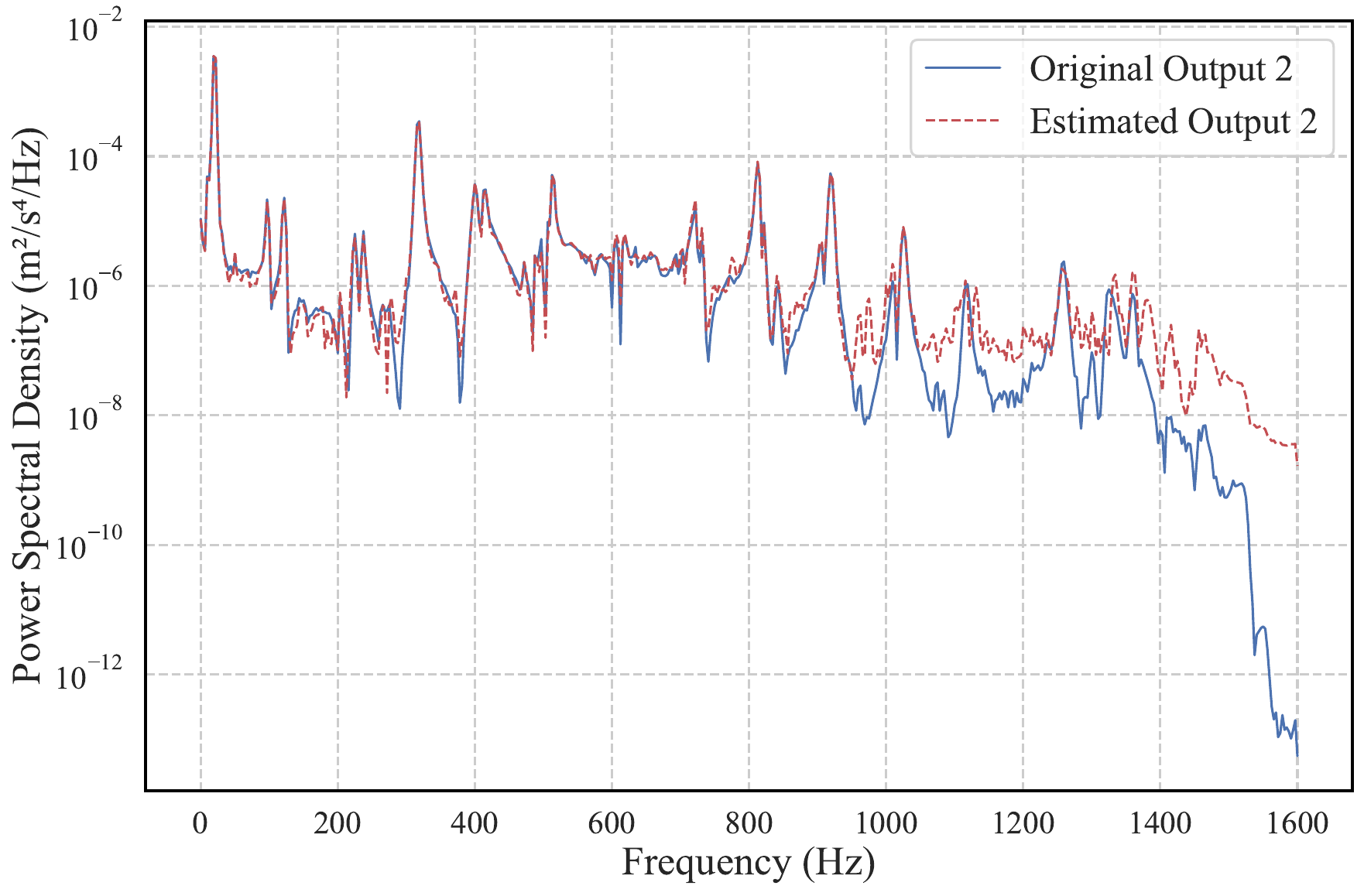}
      \caption{PSD of Monitoring Data \( y_2(t) \)}
  \end{subfigure}
  \hspace{2em}% Space between figures
  \begin{subfigure}{0.29\textwidth}
      \centering
      \includegraphics[width=\textwidth]{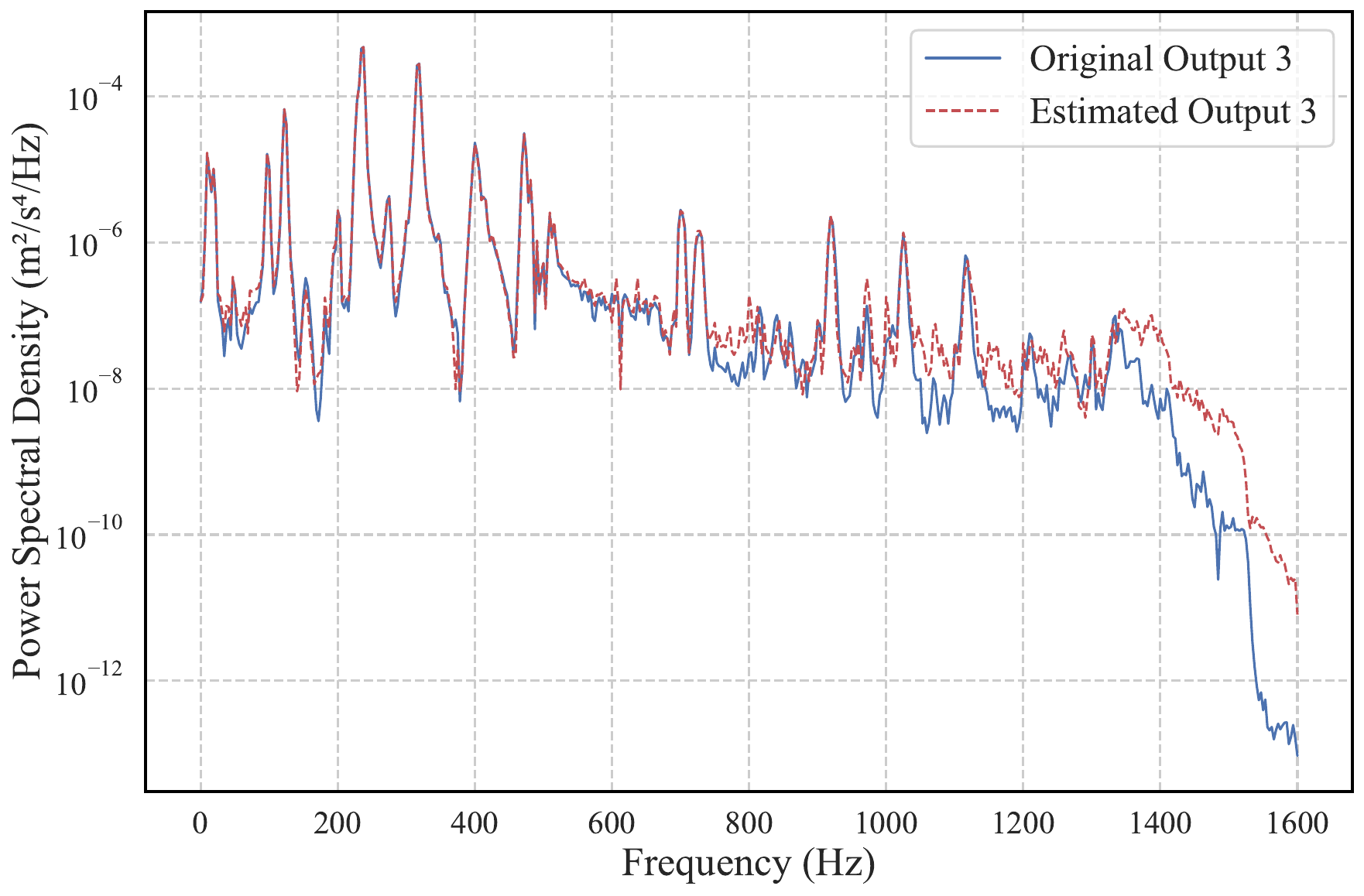}
      \caption{PSD of Monitoring Data \( y_3(t) \)}
  \end{subfigure}

  \vspace{1em}% Space between the two rows of figures
  \begin{subfigure}{0.29\textwidth}
      \centering
      \includegraphics[width=\textwidth]{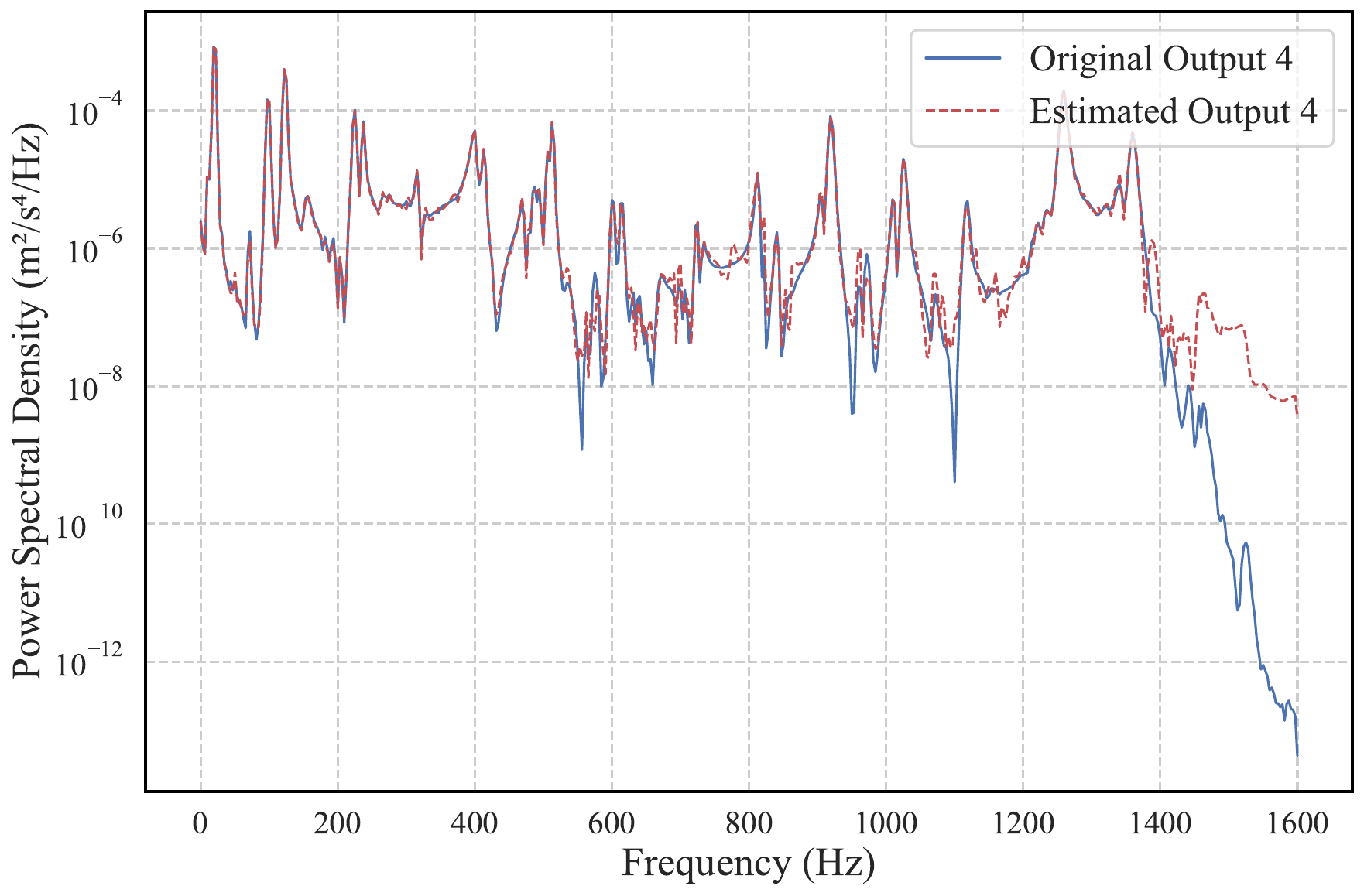}
      \caption{PSD of Monitoring Data \( y_4(t) \)}
  \end{subfigure}
  \hspace{2em}% Space between figures
  \begin{subfigure}{0.29\textwidth}
      \centering
      \includegraphics[width=\textwidth]{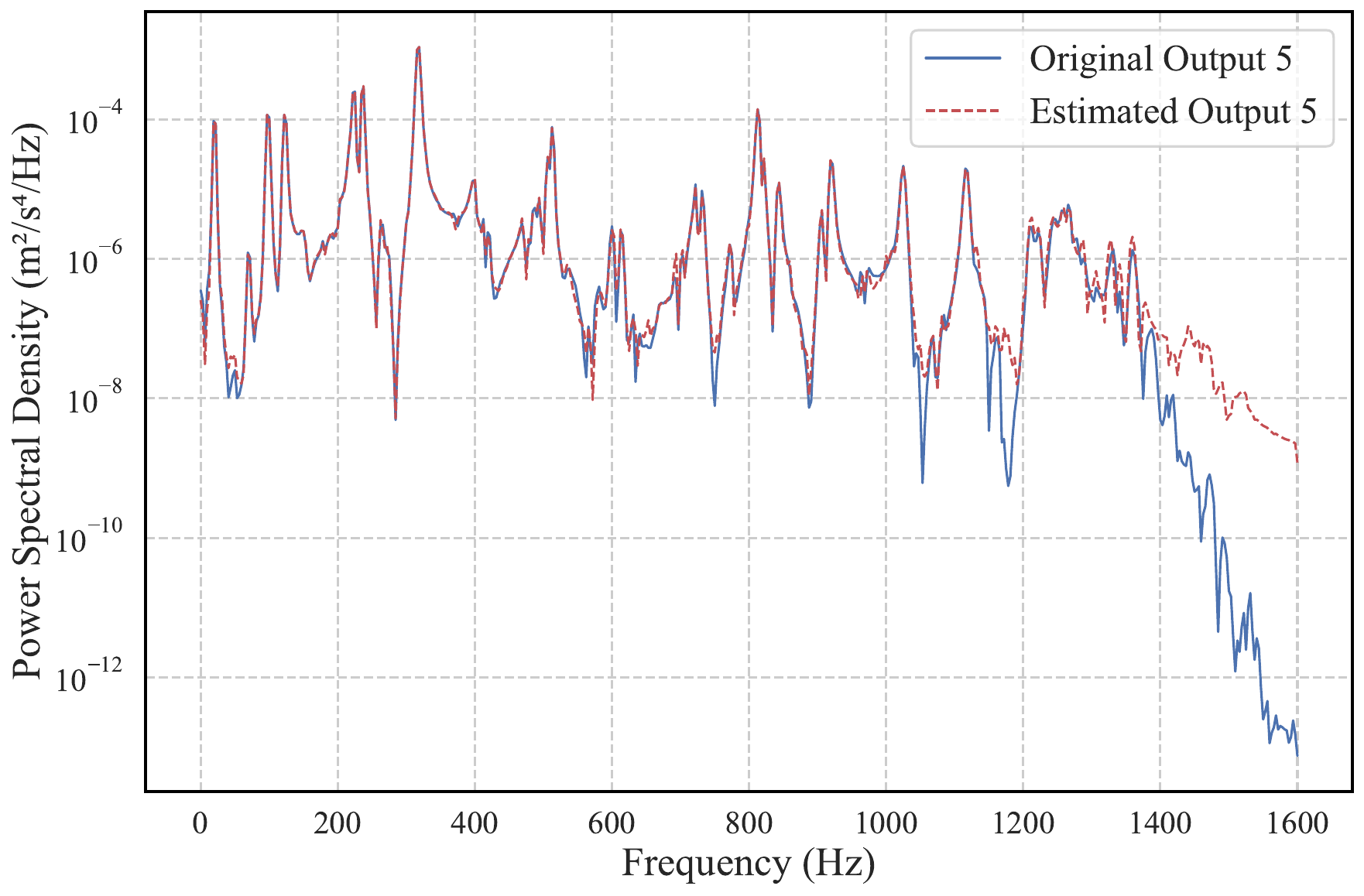}
      \caption{PSD of Monitoring Data \( y_5(t) \)}
  \end{subfigure}
  \hspace{2em}% Space between figures
  \begin{subfigure}{0.29\textwidth}
      \centering
      \includegraphics[width=\textwidth]{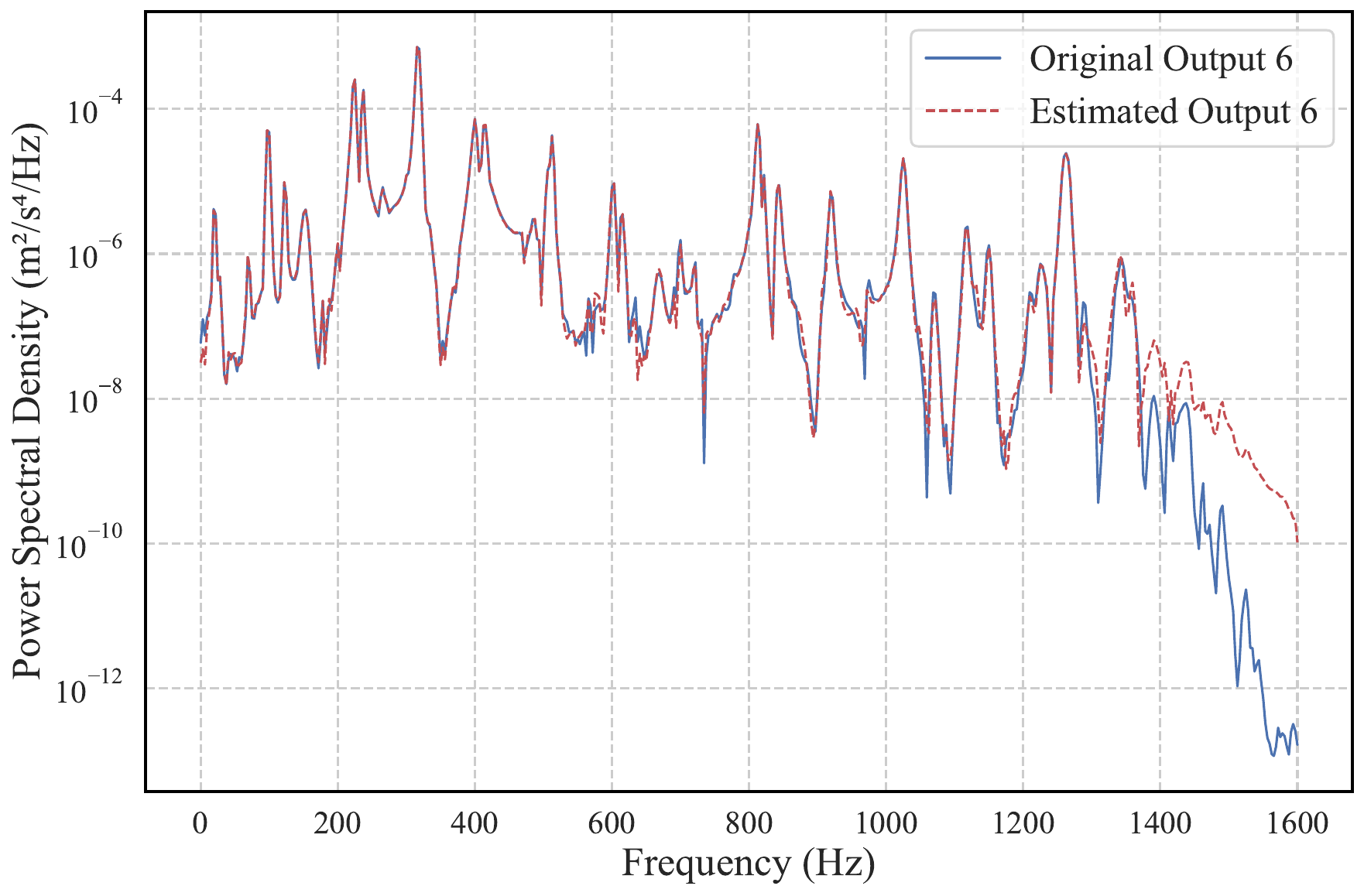}
      \caption{PSD of Monitoring Data \( y_6(t) \)}
  \end{subfigure}

  \caption{Comparative Analysis of Power Spectral Density for Various Monitoring Data \( y_1 \sim y_{6} \)}
  \label{fig:9}%\ref{fig:9}
\end{figure}

\begin{figure}[!ht]
  \centering
  \begin{subfigure}{0.29\textwidth}
      \centering
      \includegraphics[width=\textwidth]{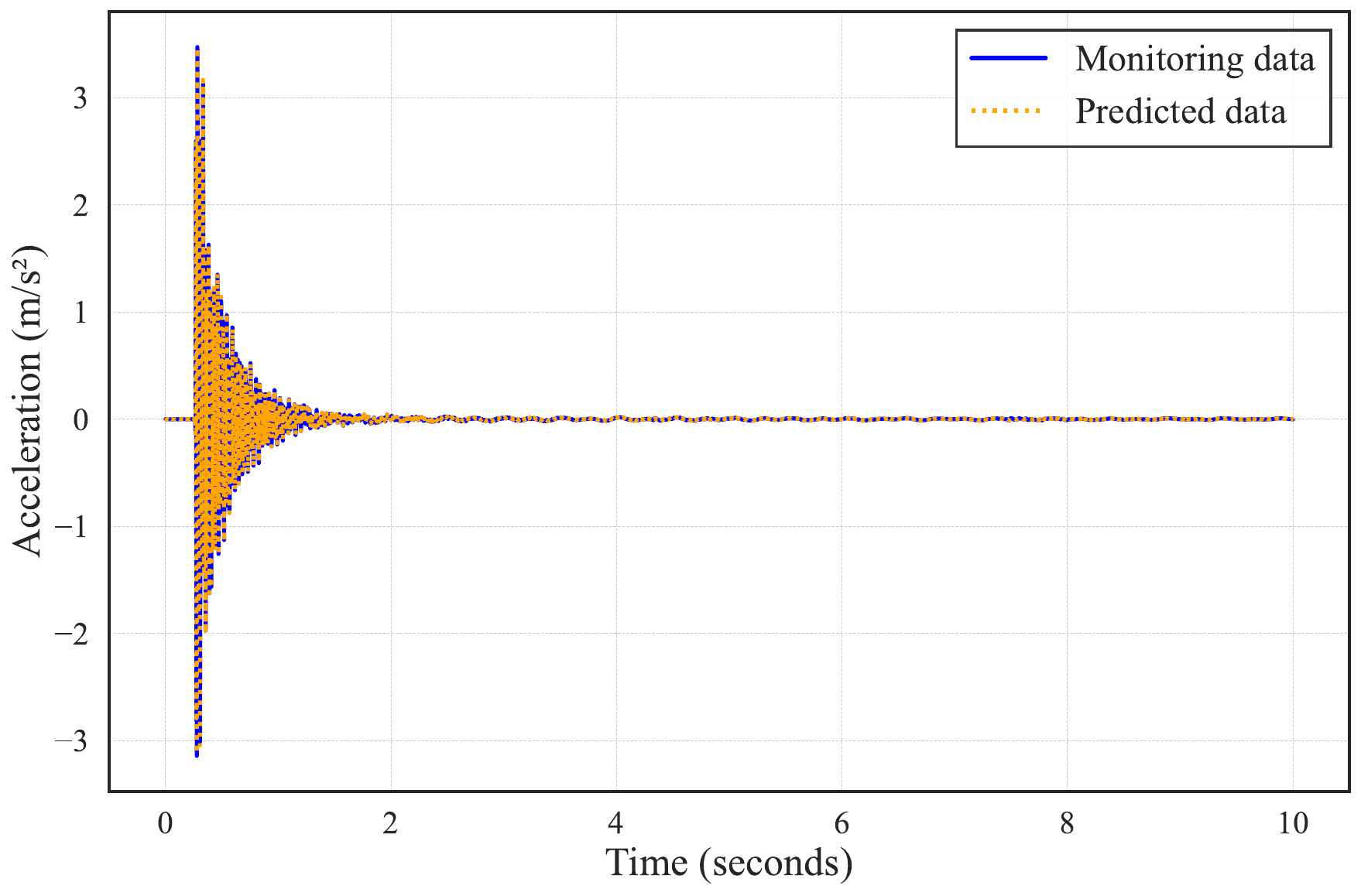}
      \caption{Monitoring Data \( y_7(t) \)}
  \end{subfigure}
  \hspace{2em}% Space between figures
  \begin{subfigure}{0.29\textwidth}
      \centering
      \includegraphics[width=\textwidth]{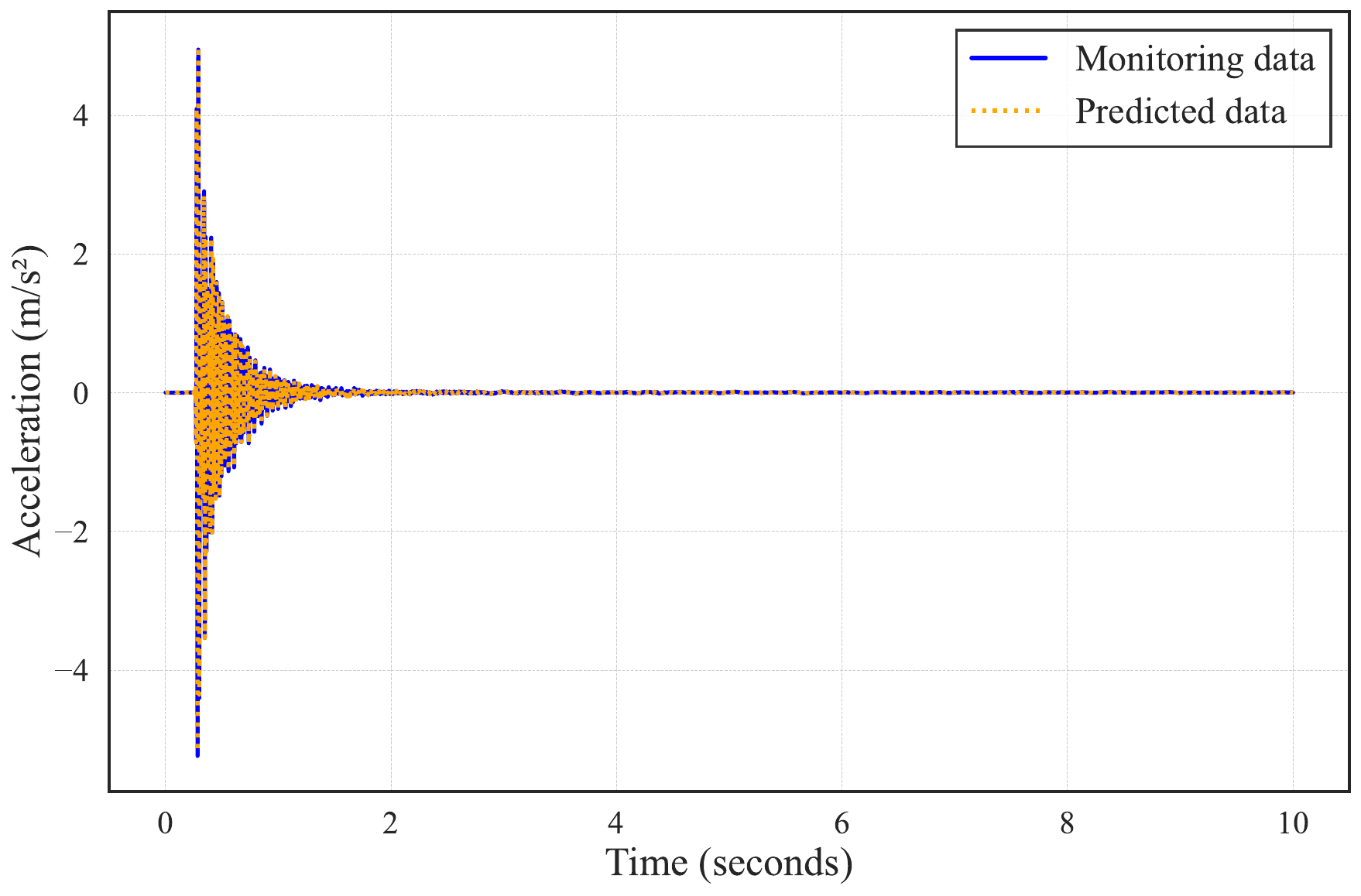}
      \caption{Monitoring Data \( y_8(t) \)}
  \end{subfigure}
  \hspace{2em}% Space between figures
  \begin{subfigure}{0.29\textwidth}
      \centering
      \includegraphics[width=\textwidth]{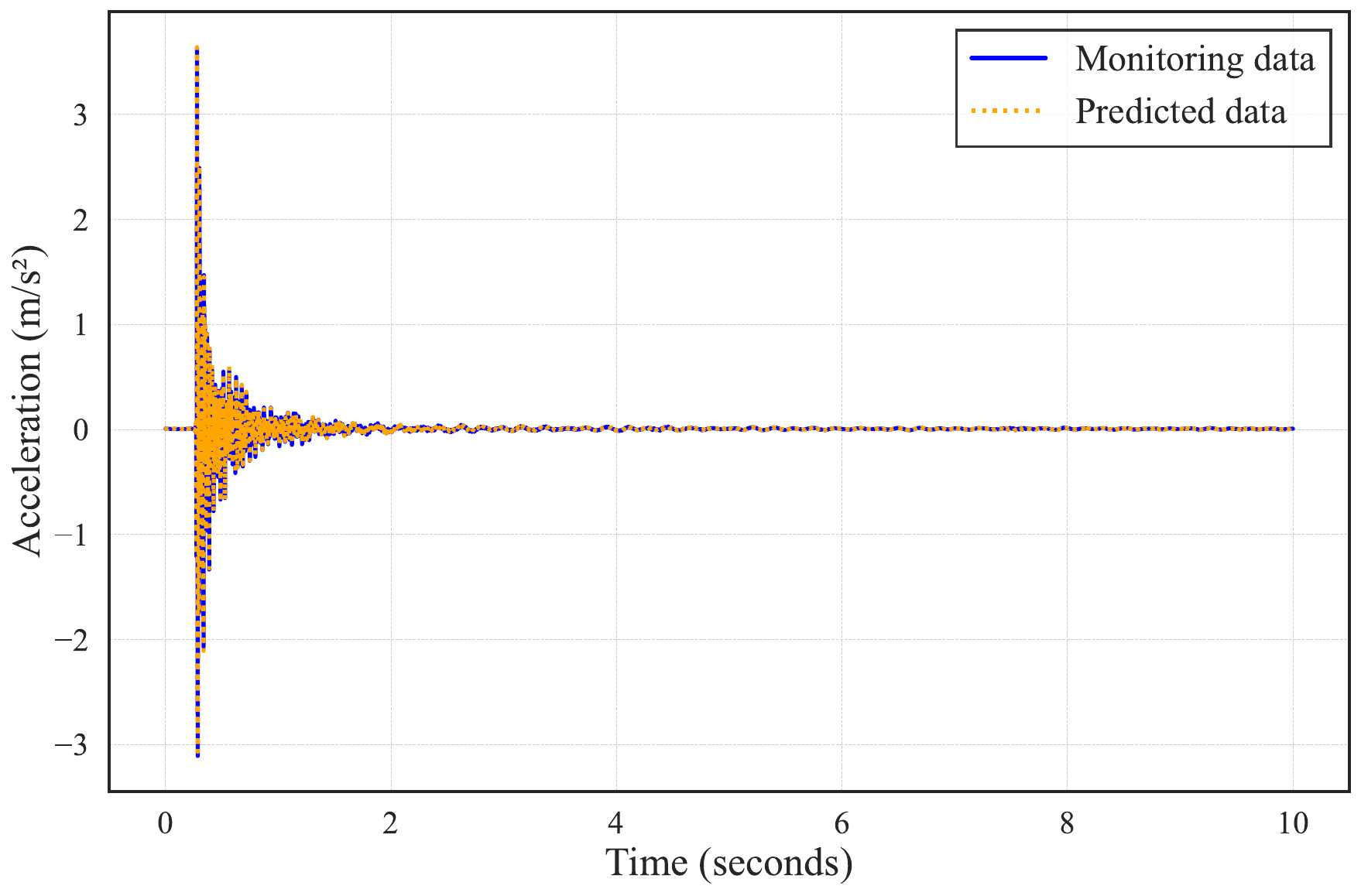}
      \caption{Monitoring Data \( y_9(t) \)}
  \end{subfigure}

  \vspace{1em}% Space between the two rows of figures
  \begin{subfigure}{0.29\textwidth}
      \centering
      \includegraphics[width=\textwidth]{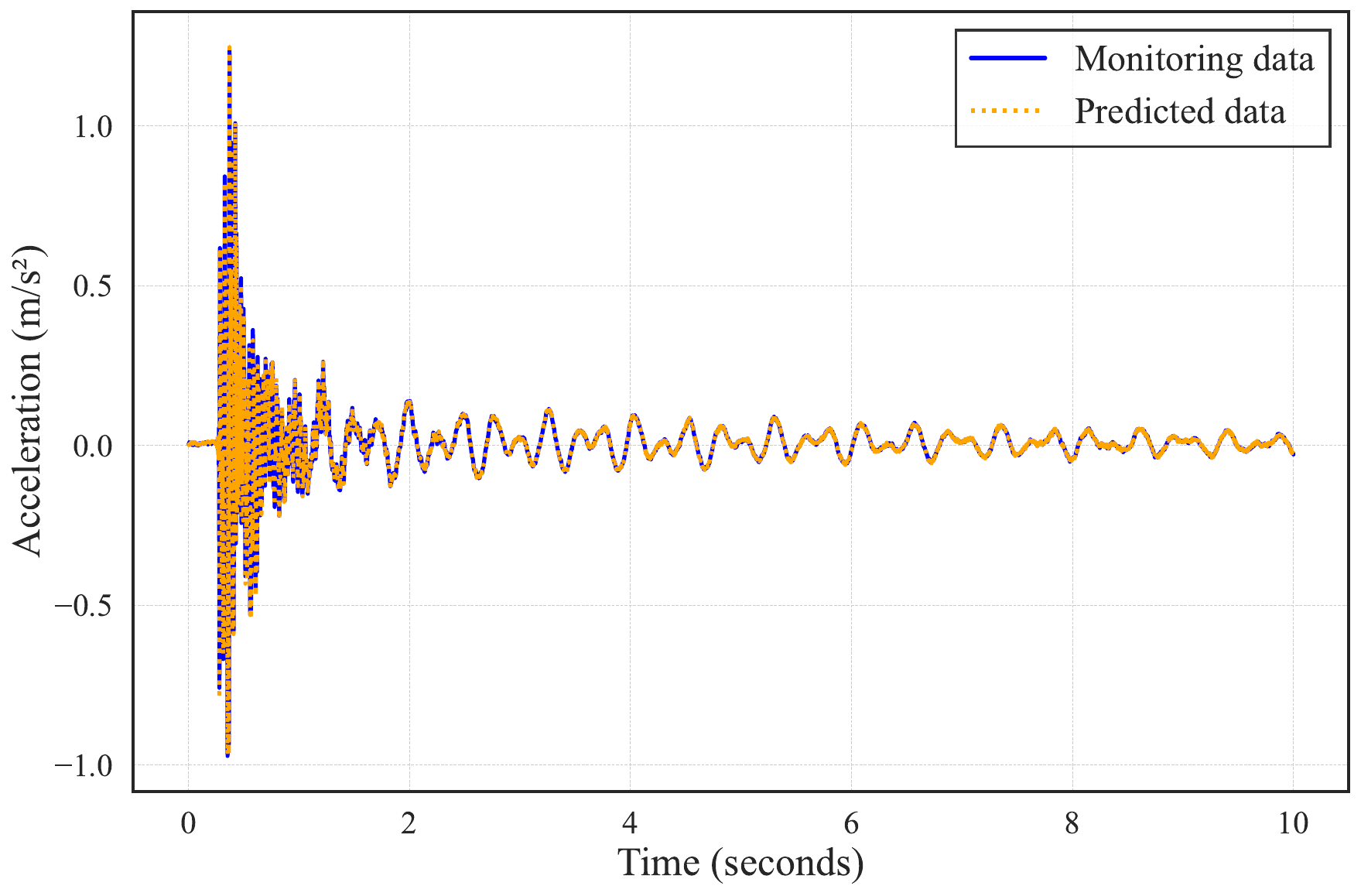}
      \caption{Monitoring Data \( y_{10}(t) \)}
  \end{subfigure}
  \hspace{2em}% Space between figures
  \begin{subfigure}{0.29\textwidth}
      \centering
      \includegraphics[width=\textwidth]{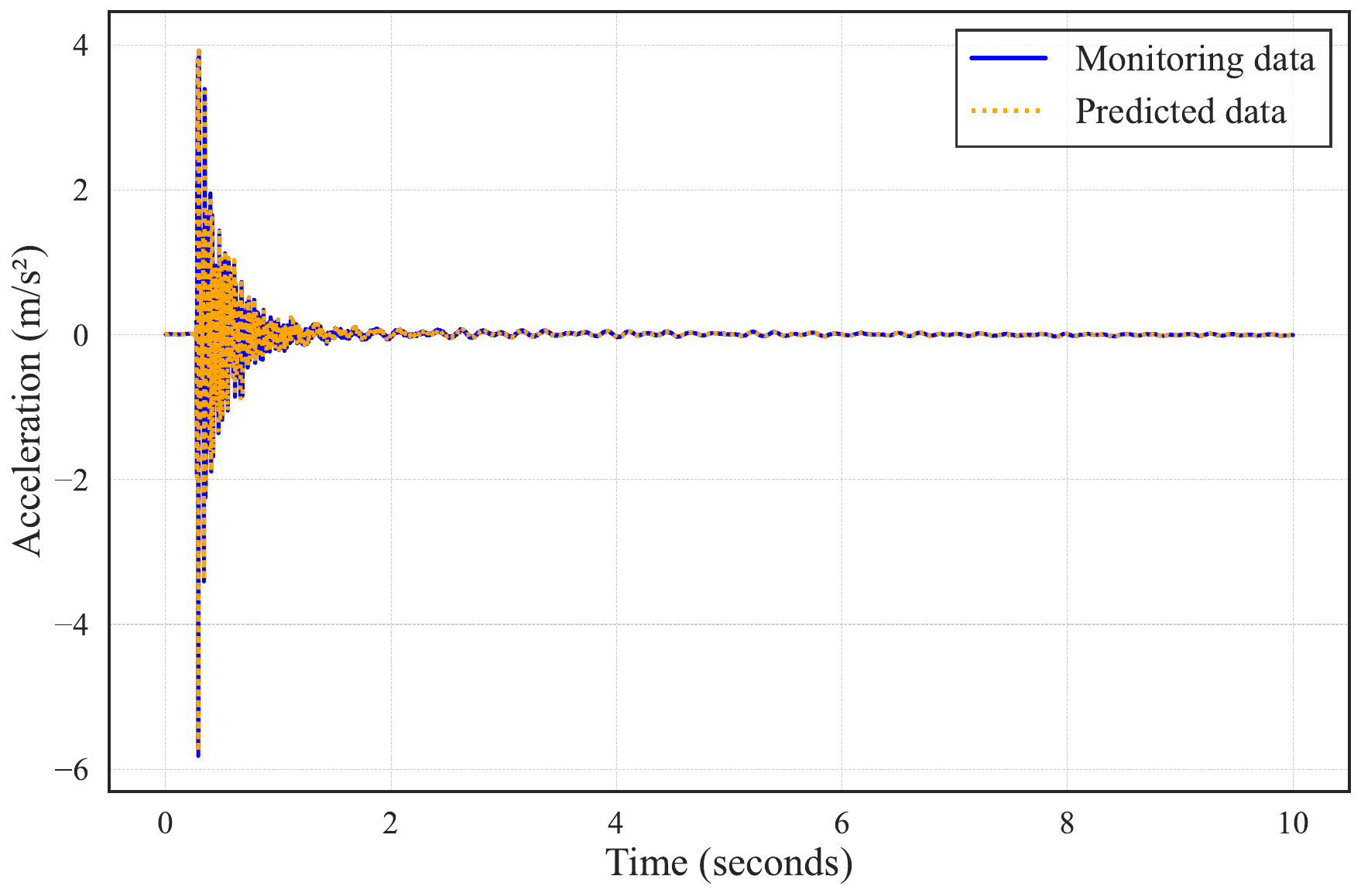}
      \caption{Monitoring Data \( y_{11}(t) \)}
  \end{subfigure}
  \hspace{2em}% Space between figures
  \begin{subfigure}{0.29\textwidth}
      \centering
      \includegraphics[width=\textwidth]{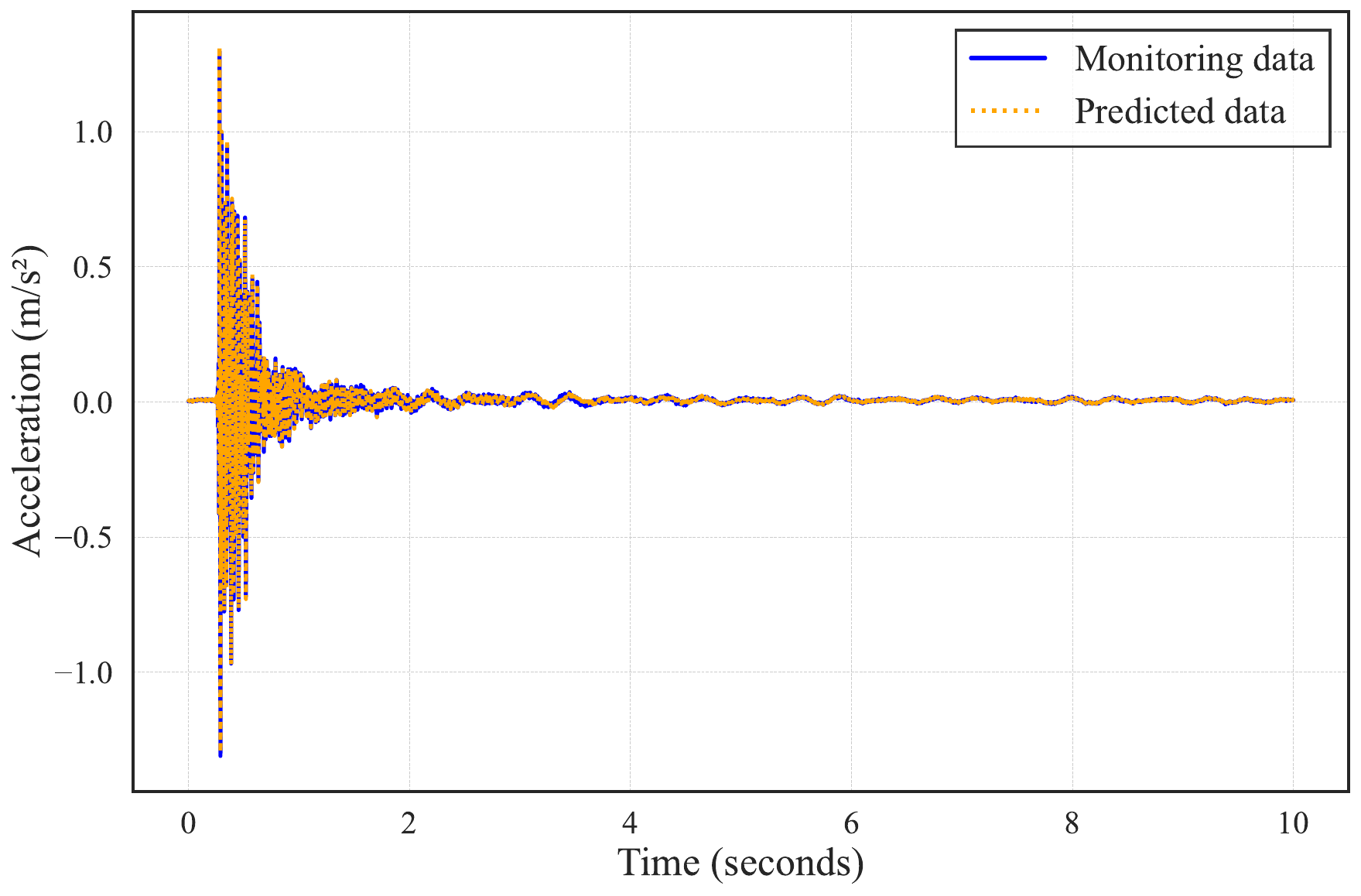}
      \caption{Monitoring Data \( y_{12}(t) \)}
  \end{subfigure}
  \caption{Comparative Analysis of Monitoring and Predicted Data for Various \( y_7 \sim y_{12}\)}
  \label{fig:8}%\ref{fig:8}
\end{figure}

\begin{figure}[!ht]
  \centering
  \vspace{1em}% Space between the two rows of figures
  \begin{subfigure}{0.29\textwidth}
      \centering
      \includegraphics[width=\textwidth]{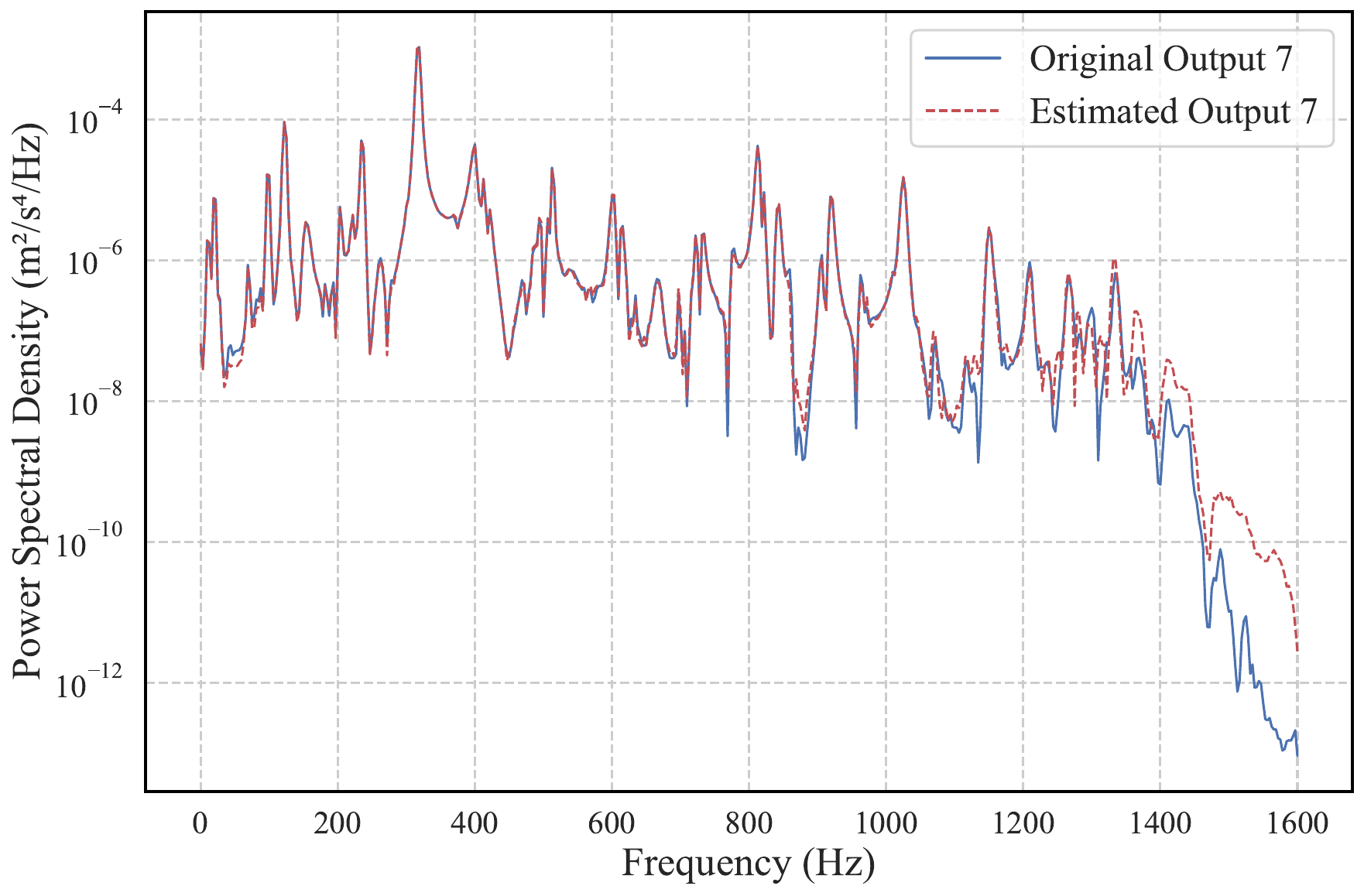}
      \caption{PSD of Monitoring Data \( y_7(t) \)}
  \end{subfigure}
  \hspace{2em}% Space between figures
  \begin{subfigure}{0.29\textwidth}
      \centering
      \includegraphics[width=\textwidth]{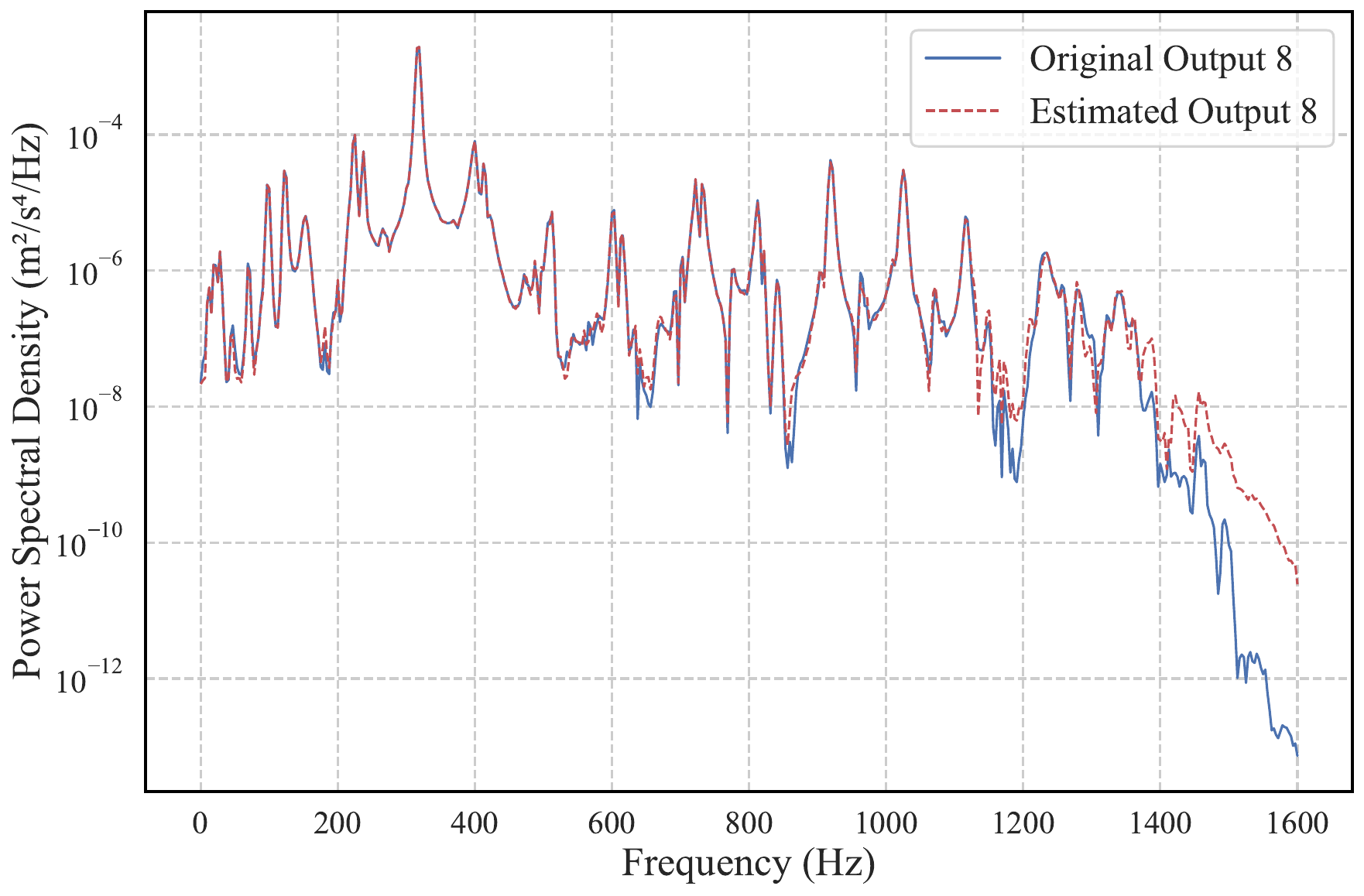}
      \caption{PSD of Monitoring Data \( y_8(t) \)}
  \end{subfigure}
  \hspace{2em}% Space between figures
  \begin{subfigure}{0.29\textwidth}
      \centering
      \includegraphics[width=\textwidth]{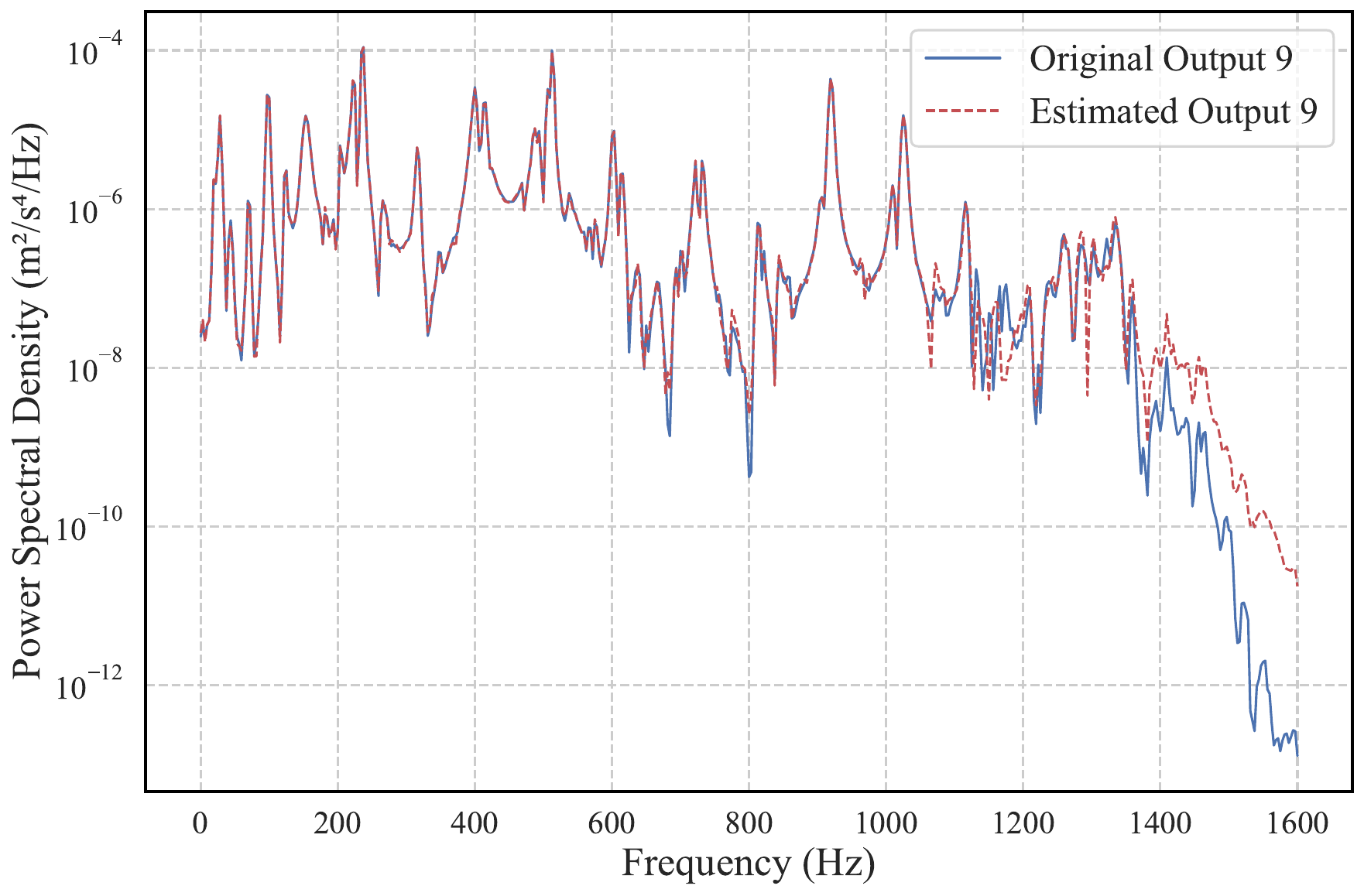}
      \caption{PSD of Monitoring Data \( y_9(t) \)}
  \end{subfigure}

  \vspace{1em}% Space between the two rows of figures
  \begin{subfigure}{0.29\textwidth}
      \centering
      \includegraphics[width=\textwidth]{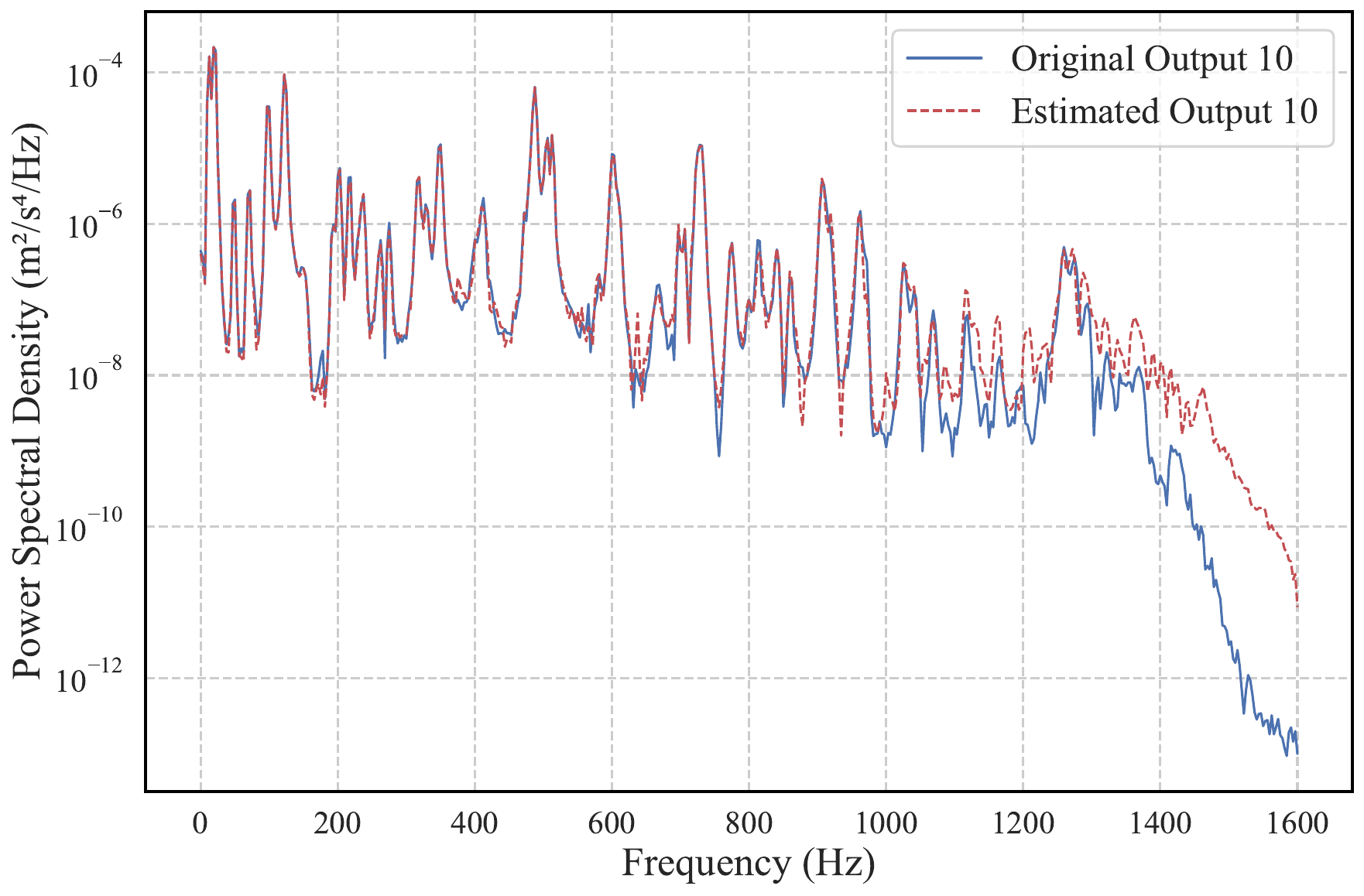}
      \caption{PSD of Monitoring Data \( y_{10}(t) \)}
  \end{subfigure}
  \hspace{2em}% Space between figures
  \begin{subfigure}{0.29\textwidth}
      \centering
      \includegraphics[width=\textwidth]{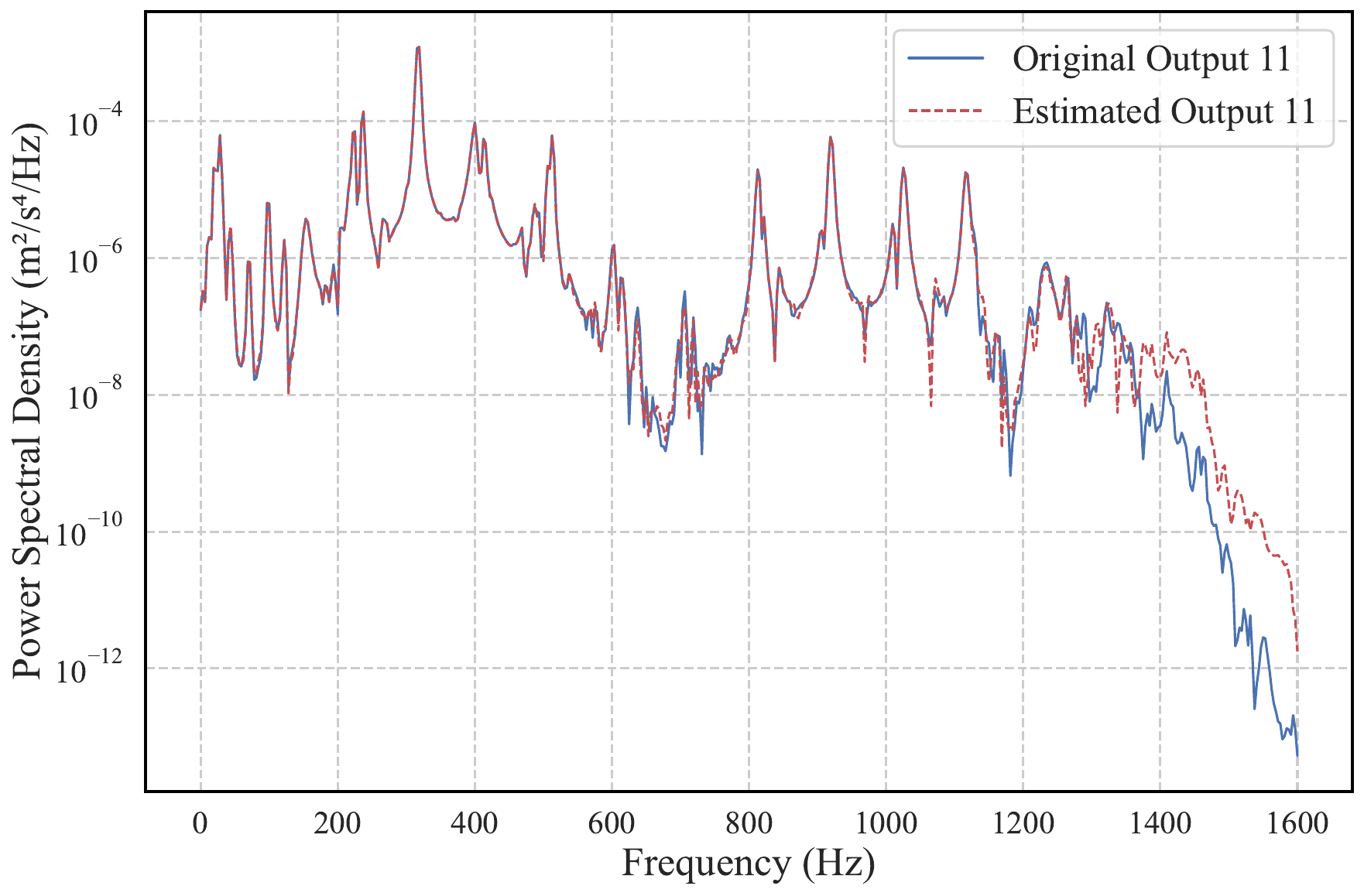}
      \caption{PSD of Monitoring Data \( y_{11}(t) \)}
  \end{subfigure}
  \hspace{2em}% Space between figures
  \begin{subfigure}{0.29\textwidth}
      \centering
      \includegraphics[width=\textwidth]{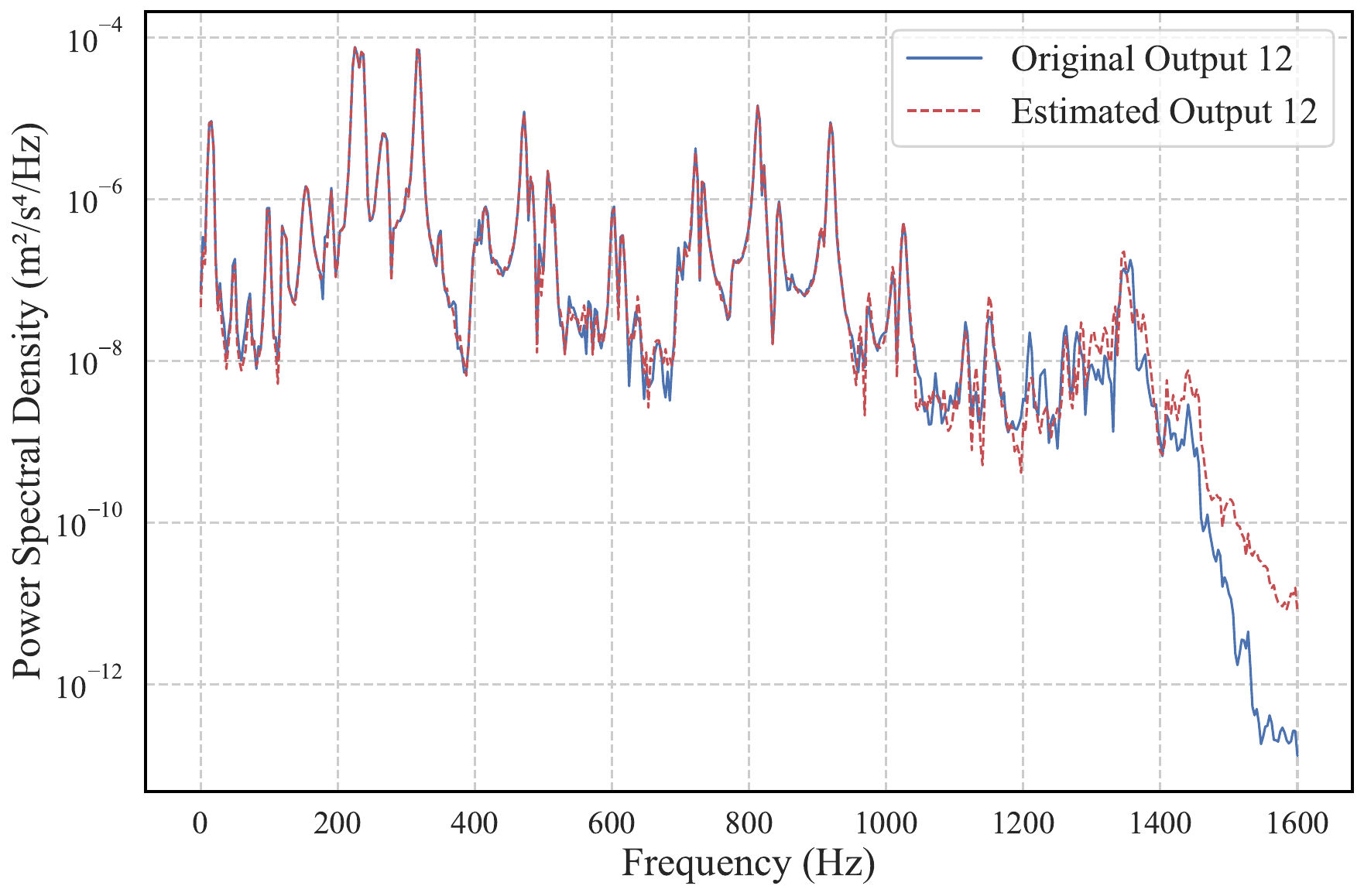}
      \caption{PSD of Monitoring Data \( y_{12}(t) \)}
  \end{subfigure}
  \caption{Comparative Analysis of Power Spectral Density for Various Monitoring Data \( y_7 \sim y_{12}\)}
  \label{fig:10}%\ref{fig:10}
\end{figure}

Furthermore, the NMSE prediction accuracies for APSMC, ERA, and OKID+ERA at \( n = 300 \) in Figure \ref{fig7} are 0.398\%, 1.317\%, and 1.107\%, respectively. 
Figures \ref{fig:15} and \ref{fig:8} illustrate the time-domain prediction results of APSMC, showing an NMSE of 0.398\% and demonstrating near-perfect correspondence. 
The frequency-domain analysis in Figures \ref{fig:9} and \ref{fig:10} further shows that the low-frequency components match closely. Although the high-frequency components 
do not fully align, their magnitude is on the order of \( 10^{-8} \), which is likely due to the absence of signal filtering. It should be noted that during the execution 
of the algorithm, APSMC and ERA did not utilize the input \( u_k \), whereas OKID+ERA did incorporate this input.

\subsection{Convergence Properties of the APSMC Algorithm}

Overall, the convergence steps depicted in Figure \ref{fig6} are somewhat analogous to training a neural network, where repeated training continues until 
the loss function converges. Therefore, in practice, manual tuning or the use of adaptive algorithms is necessary. Similarly, the results in Table \ref{tab:impact_data} 
resemble the process of evaluating the generalization performance of a neural network, where the model's effectiveness is validated using test data. 
In fact, if a neural network consists of only one neuron with a linear activation function, it can be considered equivalent to a state-space model.

The optimization process of the proposed method is similar to stochastic gradient descent (SGD). However, unlike SGD, where samples are randomly selected 
for training, this method sequentially selects samples. Therefore, for the scaled model data, if a highly accurate initial model is constructed using the 
entire time series data with ERA, a larger learning rate may cause the optimization results to oscillate around the optimal point, leading to decreased NMSE
 prediction accuracy. 
 
\begin{figure}[!ht]
  \centering
  \begin{subfigure}{0.4\textwidth}
      \centering
      \includegraphics[width=\textwidth]{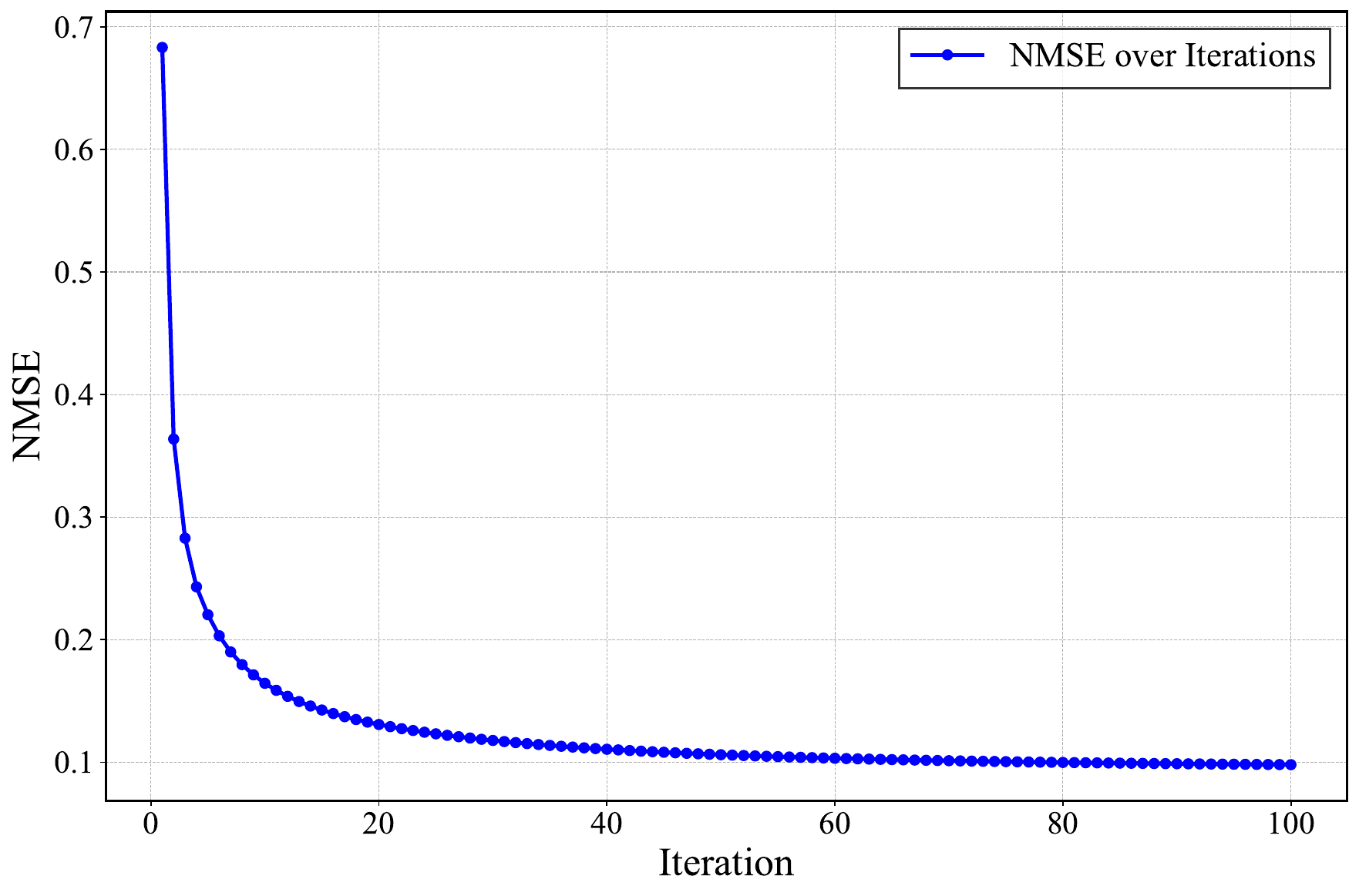}
      \caption{First 100 iterations}\label{fig11}
  \end{subfigure}
  \hspace{2em}% Space between figures
  \begin{subfigure}{0.4\textwidth}
      \centering
      \includegraphics[width=\textwidth]{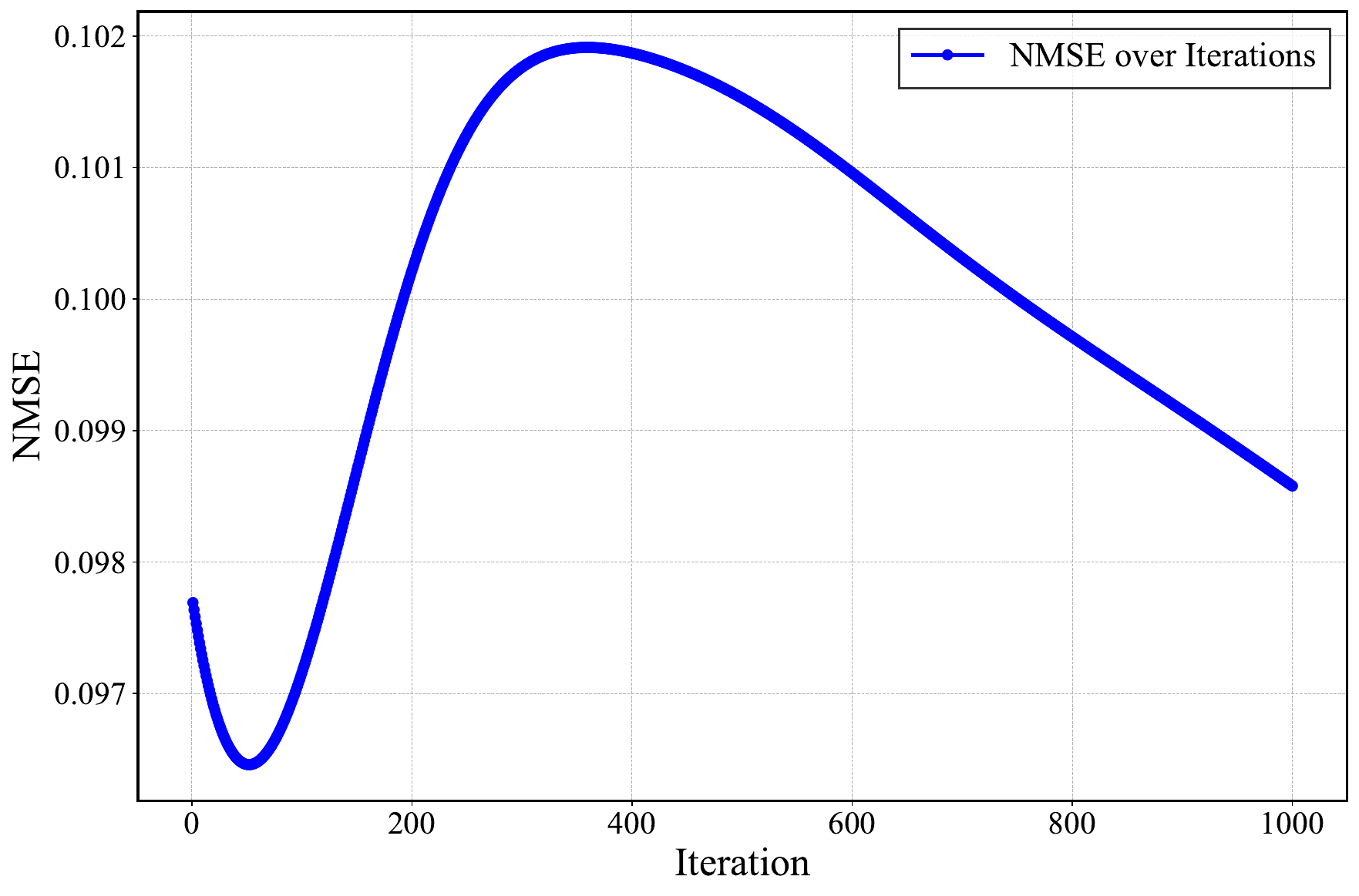}
      \caption{Iterations 101 to 1101}\label{fig12}
  \end{subfigure}
  \caption{Convergence curves for model order 50}\label{fig13}
\end{figure}

 Figure \ref{fig13} illustrates the convergence curves when the initial model is constructed using the ERA method, the observation matrix \( C \) is randomly generated, 
 and the model order \( n \) is 50. After 100 iterations of the APSMC algorithm, the final NMSE value shown in Figure \ref{fig11} is 9.806\%. Figure \ref{fig12} presents 
 the results after running an additional 1000 iterations based on this initial model. 

It is evident that the NMSE in Figure \ref{fig12} exhibits oscillations, with the final NMSE reaching 9.858\%. This behavior occurs because, as the model 
parameters approach their optimal values, the SGD algorithm requires a progressively decreasing learning rate; otherwise, convergence cannot be achieved. 
However, for real-time operation, this is not problematic, as the model parameters themselves are time-varying, and thus, there is no fixed optimal value.
The previous discussions focused on cases where the initial model was relatively accurate. In the following, we briefly discuss scenarios where a randomly 
generated \( A_0 \) may impact the results.

Experimental results indicate that when the initial model \( A_0 \) is significantly inaccurate and the number of sensors is less than the model order, 
the KF fails to provide accurate predictions, preventing the APSMC algorithm from updating to the correct model. In this scenario, 
initializing \( A_0 \) as an identity matrix for optimization typically results in an NMSE similar to that of the ERA model. However, it remains theoretically 
unclear whether initializing \( A_0 \) as an identity matrix guarantees convergence to the optimal solution. 
Conversely, when the number of sensors matches the model order, the APSMC algorithm alone is sufficient for model updating, eliminating the need for a 
KF, and reliably converges to the correct model regardless of the initial value \( A_0 \).

\section{Conclusions}
This paper presents the Adaptive Physics-Constrained System Modeling and Control (APSMC) framework, which integrates a Kalman Filter with physics-informed proximal 
gradient updates. It enables adaptive estimation of time-varying state-space model parameters from noisy input–output data in an online setting, enabling 
real-time tracking of nonlinear structural parameters and effective noise suppression.

Within the stochastic subspace identification framework, it is theoretically shown that as the number of observations approaches infinity, the APSMC 
algorithm converges to the optimal state-space model that satisfies the imposed physical constraints. The method is further extended to accommodate external 
excitations and demonstrates robust performance under arbitrary white noise disturbances.

Numerical simulations involving a nonlinear Duffing oscillator and the seismic response of a frame structure show that the time-varying system matrices identified 
by APSMC correspond to the Jacobians of the underlying nonlinear dynamic systems. With physical constraints imposed, APSMC significantly outperforms traditional 
approaches such as DMDc in capturing nonlinear behavior.

Experimental validation using multiple sets of impact tests on a scaled bridge model confirms the method’s effectiveness. The APSMC algorithm achieves a minimum 
normalized mean square error (NMSE) of 0.398\% under online updating and generalizes well to 19 additional unseen test scenarios.
Compared to existing methods that rely on offline training and are sensitive to noise, APSMC offers the following advantages:
\begin{enumerate}
  \item \textbf{Linearity and physical consistency}: The identified system matrices are approximately linear over short time scales, and the imposed physical 
  constraints ensure interpretability of the matrix structure.
  \item \textbf{Noise robustness}: The method remains stably convergent under white noise disturbances, achieving significantly better prediction performance than 
  unconstrained or globally linear fitting approaches.
  \item \textbf{Low computational complexity}: The proximal-gradient-based incremental optimization strategy supports real-time updates and is suitable for online 
  processing of large-scale monitoring data.
\end{enumerate}

This study provides both theoretical guarantees and practical tools for downstream tasks such as structural health monitoring, online control, and adaptive filtering. 
Future work may extend this framework to large-scale engineering systems involving more complex physical nonlinearities.

\section{Acknowledgments}
This research has been supported by the China National Key R\&D Program (2022YFB26-02103), National Natural Science Foundation of China (General Program, 52378294), 
GuangDong Basic and Applied Basic Research Foundation (2024A1515013224), and Shenzhen Science and Technology Program (GXWD2023112914310001).

\appendix

\section{Proximal Gradient Descent and Matrix Constraints}\label{proximal_gradient}

\subsection{Proximal Gradient Descent}

The proximal gradient method is an efficient optimization algorithm widely applied to convex optimization problems of the general form\cite{combettes2005signal}:
\begin{equation}
  \min_{x} f(x) = g(x) + h(x),
\end{equation}
where \( g(x) \) is convex and differentiable, and \( h(x) \) is convex but potentially non-differentiable. This method is 
particularly advantageous when the proximal operator associated with \( h(x) \) can be computed efficiently. The 
detailed procedure of the proximal gradient method is summarized in Algorithm \ref{algo:proximal_gradient}.

\begin{algorithm}
  \caption{Proximal Gradient Method}\label{algo:proximal_gradient}
  \begin{algorithmic}
    \State \textbf{Input}: Initial point \( x^{(0)} \), step sizes \( \{t_k\} \).
    \State \textbf{Goal}: Minimize \( f(x) = g(x) + h(x) \), with conditions as above.
    \For{\( k = 0, 1, 2, \dots \) until convergence}
      \State Compute gradient: \( \nabla g(x^{(k)}) \).
      \State Update step using proximal operator:
      \State \quad \( x^{(k+1)} = \text{prox}_{t_k h}\left(x^{(k)} - t_k \nabla g(x^{(k)})\right) \).
    \EndFor
    \State \textbf{return}: Sequence \( \{x^{(k)}\} \) converging to solution.
  \end{algorithmic}
\end{algorithm}
The convergence accuracy to optimal solutions in convex optimization problems, where the 
objective function $f(x) = g(x) + h(x)$ combines a $L$-Lipschitz continuous differentiable function $g(x)$\cite{schmidt2011convergence,suzuki2013dual}. The 
choice of an appropriate step size is critical, as it ensures the iterates $x^{(k)}$ 
converge to the optimal solution $x^*$ at a rate of $O(1/\sqrt{k})$, where $k$ is the iteration count.
The proximal operator is formally defined as:
\begin{equation}
  \text{prox}_{t h}(x) = \arg\min_{z}\left\{ \frac{1}{2t}\|x - z\|^2 + h(z) \right\},
\end{equation}
and it simplifies to various common methods depending on \( h(x) \):
\begin{itemize}
  \item When \( h(x) = 0 \), the method reduces to classical gradient descent.
  \item When \( h(x) = I_C \), it becomes projected gradient descent, maintaining feasibility within constraint set \( C \).
\end{itemize}
In this context, the indicator function \( I_C(x) \) is defined as
\begin{equation}
  I_C(x)=\left\{\begin{array}{ll}0 & x \in C \\ \infty & x \notin C\end{array}\right.
\end{equation}
which succinctly expresses the requirement that \( x \) must belong to the set \( C \) \cite{combettes2011proximal}. 
With this definition, the proximal operator associated with \( I_C \) naturally corresponds to a projection onto \( C \):
\begin{align}\label{eqc_{27}}%\eqref{eqc_{27}}
  \operatorname{prox}_{t,I_C}(x) & =\underset{z}{\operatorname{argmin}} \frac{1}{2 t}\|x-z\|_2^2+I_C(z) \\
  & =\underset{z \in C}{\operatorname{argmin}}\|x-z\|_2^2
  \end{align}
Thus, the proximal gradient update step simplifies to a projected gradient step:
\begin{equation}
  x^{(k+1)} = P_C \left( x^{(k)} - t \cdot \nabla g(x^{(k)}) \right),
\end{equation}
where \( P_C(\cdot) \) denotes the projection operator onto the set \( C \), ensuring that each iterate remains feasible with respect to the constraints imposed by \( C \).

\section{Physical Information Matrix Constraints}\label{physics_constrain}

Embedding physical constraints into optimization enhances both the interpretability and robustness of identified models.
Overall, the methods for incorporating physical information can be categorized into constraints for continuous system matrices and discrete system matrices. 

In most physical problems, constraints on the continuous system matrix are more common, especially when the governing equations are known but the parameters are uncertain or imprecise.
However, real-world applications—especially inverse problems based on sampled data—are usually formulated in discrete time. Therefore, this 
section focuses on how physical knowledge can be used to constrain discrete system matrices.

\subsection{Discrete System Physical Information}

In discrete systems, structural constraints can be effectively incorporated through the proximal optimization methods discussed earlier. 
These constraints are generally classified into:
\begin{itemize}
\item \textbf{Hard structural constraints}: enforce strict matrix structures such as circulant, symmetric, upper triangular, or tridiagonal 
forms (denoted $I_{\text{Circulant}}$ to $I_{\text{Tri-Diagonal}}$). These are typically maintained via projected gradient descent to ensure 
structural fidelity\cite{baddooPhysicsinformedDynamicMode2023,chen2024online}.
\item \textbf{Soft constraints}: applied when the exact structure is unknown but desired properties are present. These are imposed through regularization 
terms involving matrix norms, such as the nuclear norm \( \|A_k\|_* \), \( \ell_1 \)-norm \( \|A_k\|_1 \), or Frobenius norm \( \|A_k\|_F^2 \).
\end{itemize}
A summary of constraint types, corresponding optimization formulations, and solution strategies is provided in Table \ref{table:convexity_matrix}. 
The commonly used matrix norms are defined as:
\begin{equation}
    \|A\|_{*} = \sum_{i=1}^{r}\sigma_i(A), \quad \|A\|_1 = \sum_{i=1}^{n} \sum_{j=1}^{m} |a_{ij}|,
\end{equation}
where \( \sigma_i(A) \) denotes the \( i \)-th singular value of \( A \), \( r \) is the number of nonzero singular values of \( A \), and \( a_{ij} \) represents 
the \((i,j)\)-th element of \( A \).

\begin{table}[!h]
    \centering
    \renewcommand{\arraystretch}{1.25}
    \caption{Optimization Problems with Matrix Constraints \cite{chen2024online}}
    \label{table:convexity_matrix}
    \begin{tabular}{lll}
    \toprule
    \textbf{Matrix Structure} & \textbf{Convex Problem} & \textbf{Solution Method} \\
    \midrule
    Implicit Constraint & \( \| Y_k - A_k X_k \|_F^2 \) & Stochastic Gradient Descent \\
    Shift-Invariant     & \( \| Y_k - A_k X_k \|_F^2 + I_{\text{Circulant}} \) & Projected Gradient Descent \\
    Self-Adjoint        & \( \| Y_k - A_k X_k \|_F^2 + I_{\text{Symmetric}} \) & Projected Gradient Descent \\
    Causal              & \( \| Y_k - A_k X_k \|_F^2 + I_{\text{Upper Triangular}} \) & Projected Gradient Descent \\
    Local               & \( \| Y_k - A_k X_k \|_F^2 + I_{\text{Tri-Diagonal}} \) & Projected Gradient Descent \\
    Low-Rank            & \( \| Y_k - A_k X_k \|_F^2 + \lambda \| A_k \|_* \) & Proximal Gradient Method \\
    Sparse              & \( \| Y_k - A_k X_k \|_F^2 + \lambda \| A_k \|_1 \) & Proximal Gradient Method \\
    Norm Shrinkage      & \( \| Y_k - A_k X_k \|_F^2 + \lambda \| A_k \|_F^2 \) & Gradient Descent Method \\
    \bottomrule
    \end{tabular}
\end{table}

The above approach applies primarily to time-invariant linear systems. For discrete nonlinear dynamic systems or systems with time-varying parameters, a similar 
strategy can be employed. Consider the general form of a discrete nonlinear dynamic system:
\begin{align}\label{eqc_{129}}
    x_{k+1} &= f(x_k, u_k),  \\
    y_k &= g(x_k),
\end{align}
where \( f(x_k, u_k) \in \mathbb{R}^n \) is the nonlinear state transition function, and \( g(x_k) \in \mathbb{R}^m \) is the measurement function mapping the 
state \( x_k \) to the observed output \( y_k \).
To address the system’s nonlinearity, a local linearization can be performed around the current state by computing the Jacobian 
matrices \( J_{x} \in \mathbb{R}^{n \times n} \), \( J_{u} \in \mathbb{R}^{n \times l} \), and \( J_g \in \mathbb{R}^{m \times n} \) as:
\begin{equation}
    J_{x} = \frac{\partial f(x_k, u_k)}{\partial x_k}, \quad J_{u} = \frac{\partial f(x_k, u_k)}{\partial u_k}, \quad J_g = \frac{\partial g(x_k)}{\partial x_k}.
\end{equation}
Using this linearization, the nonlinear system in~\eqref{eqc_{129}} can be approximated at each time step by a locally linear model:
\begin{align}
    x_{k+1} &=J_x(x_k, u_k) \cdot x_k + J_u(x_k, u_k) \cdot u_k, \\
    y_k &= J_g(x_k) \cdot x_k,
\end{align}
where \( J_x(x_k, u_k) \), \( J_u(x_k, u_k) \), and \( J_g(x_k) \) are the Jacobian matrices evaluated at the current state $x_k$ and input $u_k$.
At this stage, the matrix \( J_x(x_k, u_k) \) may be subject to specific structural patterns or constraints, as in the linear case. Therefore, appropriate 
constraints can be imposed at each time step, such as:
\begin{equation}
    A_k \in I_C(A_k), \quad \text{or} \quad \| A_k - J_x(\hat{x}_k, u_k) \|_1, \quad \text{or} \quad \| A_k - J_x(\hat{x}_k, u_k) \|_F^2,
\end{equation}
where \( C \) in \( I_C(A_k) \) specifies the structural constraints imposed on the system matrix \( A_k \), serving as a hard constraint, 
which enforces strict compliance with the predefined structural form. In contrast, the terms \( \| \cdot \|_1 \) and \( \| \cdot \|_F^2 \) act as soft constraints, 
which allow a certain degree of deviation from the linearized dynamics. 

These soft constraints introduce flexibility by penalizing, rather than prohibiting, 
discrepancies between \( A_k \) and the Jacobian matrix \( J_x(\hat{x}_k, u_k) \), thereby enabling a trade-off between model accuracy and structural fidelity.

\subsection{Continuous System Physical Information}
In general, physical equations are often derived based on the principles of calculus, and the variables involved are typically continuous in time. 
Consequently, most of the physical information related to dynamic systems is associated with continuous system matrix structures, such as the one 
illustrated in equation~\eqref{eqc_{16}} in the Duffing oscillator numerical example.  
To incorporate such physical information as constraints in the optimization objective, the discrete system matrix \( A_k \) can first be transformed into 
its continuous counterpart \( A_k^c \) using the bilinear transformation:
\begin{equation}
    A_k^c = \frac{2}{\Delta t} \cdot \left( I + A_k \right)^{-1} \cdot \left( A_k - I \right),
\end{equation}
where \( \Delta t \) denotes the time step size. Physical constraints can then be directly imposed on the continuous system matrix \( A_k^c \), such as:
\begin{equation}
    A_k^c \in I_C(A_k^c),
\end{equation}
where \( I_C(A_k^c) \) represents the set of matrices satisfying the desired structural constraints, serving as hard constraints.
For general nonlinear dynamical systems described by:
\begin{align}
   \dot{x} &= f(x,u), \\
   y &= g(x),
\end{align}
where \( f(x,u) \in \mathbb{R}^n \) characterizes the nonlinear system dynamics and \( g(x) \in \mathbb{R}^m \) denotes the measurement function mapping 
the state \( x \) to the observed output \( y \), 
the corresponding continuous-time Jacobian matrices \( J_x^c \in \mathbb{R}^{n\times n} \), \( J_u^c \in \mathbb{R}^{n\times l} \), and \( J_g^c \in \mathbb{R}^{m\times n} \) are 
defined as:
\begin{equation}
    J_x^c = \frac{\partial f(x,u)}{\partial x}, \quad J_u^c = \frac{\partial f(x,u)}{\partial u}, \quad J_g^c = \frac{\partial g(x)}{\partial x}.
\end{equation}
Following the same approach used in the discrete case, appropriate constraints can be imposed on the continuous system matrix \( A_k^c \) at each time step. 
These constraints may take the form:
\begin{equation}
    A_k^c \in I_C(A_k^c), \quad \text{or} \quad \| A_k^c - J_x^c(\hat{x}_k,u_k) \|_1, \quad \text{or} \quad \| A_k^c - J_x^c(\hat{x}_k,u_k) \|_F^2,
\end{equation}
where \( C \) in \( I_C(A_k^c) \) denotes the structural constraints inferred from the Jacobian matrix \( J_x^c \) of the continuous-time system.
Similar to the treatment of discrete system matrices, the terms \( \| \cdot \|_1 \) and \( \| \cdot \|_F^2 \) represent soft constraints imposed on \( A_k^c \).

After analyzing general nonlinear dynamical systems, we now turn to the structural dynamics problems frequently addressed in this study. For such problems, 
the most general form of the linear time-invariant state-space model is expressed as:
\begin{equation}\label{eqc_3}
  \begin{bmatrix}
    \ddot{q} \\
    \dot{q}
  \end{bmatrix} =
  \begin{bmatrix}
    0 & I \\
    -M^{-1}K & -M^{-1}C_1
  \end{bmatrix}
  \begin{bmatrix}
    \dot{q} \\
    q
  \end{bmatrix} +
  \begin{bmatrix}
    0 \\
    M^{-1}
  \end{bmatrix} F,
\end{equation}
where \( M \in \mathbb{R}^{n \times n} \) is the mass matrix, \( C_1 \in \mathbb{R}^{n \times n} \) is the damping matrix, \( K \in \mathbb{R}^{n \times n} \) is 
the stiffness matrix, \( q \in \mathbb{R}^n \) is the displacement vector, and \( F \in \mathbb{R}^n \) is the external force vector.

Typically, the mass matrix \( M \) is diagonal, the stiffness matrix \( K \) is symmetric, and the damping matrix \( C_1 \) is often assumed to follow the Rayleigh 
damping model. Accordingly, the continuous system matrix \( A_k^c \) can generally be constrained to satisfy the following structural form:
\begin{equation}
    A_k^c \in \begin{bmatrix}
        0 & I \\
        \text{Symmetric Matrix} & \text{Symmetric Matrix}
    \end{bmatrix}.
\end{equation}
Moreover, if additional information about the structure of \( -M^{-1}K \) or \( -M^{-1}C_1 \) is available, such as specific patterns or known values of certain 
elements, more refined constraints can be further incorporated.
Even for problems involving nonlinear constitutive relations, the system can still be approximated as linear over a sufficiently short time interval based on a Taylor 
series expansion\cite{chen5097818adaptive}. 

As a result, a nonlinear dynamical system can be regarded as a time-varying linear system, and the aforementioned structural constraints remain 
applicable to nonlinear problems as well.
For instance, in the case of the Duffing oscillator, which exhibits nonlinear characteristics, the Jacobian matrix given in equation \eqref{eqc_{16}} can be regarded 
as a time-varying system matrix of the form shown in equation \eqref{eqc_3}.

Based on this observation, we adopted the constraint form presented in equation \eqref{eqc_{161}}.
After applying physical constraints to the continuous system matrix \( A_k^c \), the matrix can be converted back to its discrete form using the bilinear transformation:
\begin{equation}
    A_k = \left( I + \frac{\Delta t}{2} A_k^c \right) \left( I - \frac{\Delta t}{2} A_k^c \right)^{-1}.
\end{equation}
Subsequently, the Proximal Gradient Descent procedure can be continued.
It is worth noting, however, that unlike direct constraints imposed on discrete matrices, the nonlinear nature of the transformation between continuous 
and discrete matrices may render the resulting optimization problem non-convex. 

Therefore, in convex optimization theory, the Proximal Gradient Descent method does not inherently guarantee convergence to a global optimum. However, 
our numerical experiments demonstrate that, with the integration of suitable physical constraints, the method typically converges more rapidly, exhibits 
enhanced robustness to noisy data, and often still achieves global optimality.

\bibliographystyle{elsarticle-num} 
\bibliography{refers}

\end{document}